\documentclass[twoside,11pt]{article}

\usepackage{jmlr2e}
\usepackage{bm}
\usepackage{amsmath}
\usepackage{rotating}
\usepackage{multicol}
\usepackage{dsfont}
\usepackage{latexsym}
\usepackage{changepage}
\usepackage{multirow}
\usepackage{booktabs}
\usepackage{pbox}
\usepackage[titletoc]{appendix}
\usepackage{float}
\usepackage{caption}
\usepackage[linktoc=page,colorlinks,linkcolor=blue,citecolor=blue,urlcolor=blue]{hyperref}
\usepackage{fancyhdr}
\usepackage{setspace}
\usepackage[top=1in, bottom=1in, left=1in, right=1in]{geometry}
\usepackage{enumitem}
\usepackage[normalem]{ulem}
\usepackage[table]{xcolor}
\usepackage{color}
\usepackage{subcaption}

\usepackage[algoruled,boxed,lined]{algorithm2e}

\SetCommentSty{mycommfont}
\SetKwComment{Comment}{$\triangleright$\ }{}

\newcommand{\dataset}{{\cal D}}

\newcommand{\bv}{\begin{array}}
\newcommand{\ev}{\end{array}}
\newcommand{\bit}{\begin{itemize}}
\newcommand{\eit}{\end{itemize}}
\newcommand{\ben}{\begin{enumerate}}
\newcommand{\een}{\end{enumerate}}
\newcommand{\beq}{\begin{equation}}
\newcommand{\eeq}{\end{equation}}
\newcommand{\bvq}{\begin{eqnarray}}
\newcommand{\evq}{\end{eqnarray}}


\newcounter{parentnumber}

\newtheorem{assumption}{Assumption}

\newcommand{\N}{\mathcal{N}}

\newcommand{\vZ}{\textbf{Z}}
\newcommand{\vG}{\textbf{G}}
\newcommand{\vY}{\textbf{Y}}
\newcommand{\vX}{\textbf{X}}
\newcommand{\vx}{\textbf{x}}

\newcommand{\vz}{\textbf{z}}

\newcommand{\ind}{\perp\!\!\!\perp}

\newcommand{\vtheta}{\boldsymbol{\theta}}

\ShortHeadings{Estimating Causal Effects Under Interference}{Forastiere , Mealli, Wu and Airoldi} \firstpageno{1}

\begin{document}

\title{Estimating Causal Effects Under Interference Using Bayesian Generalized Propensity Scores}

\author{\name Laura Forastiere$^{~1}$ \email laura.forastiere@yale.edu \\
    \addr $^{1~}$Yale Institute for Network Science\\
       Yale University,
       New Haven, CT 06520, USA
          \AND
       \name Fabrizia Mealli$^{~2}$ \email mealli@disia.unifi.it \\
        \addr $^{2~}$ Department of Statistics, Computer Science, Applications \\
       University of Florence,
       Florence,  50134, Italy
       \AND
       \name Albert Wu$^{~2}$ \email albertwu@g.harvard.edu \\
        \name Edoardo M.\ Airoldi$^{~2}$ \email airoldi@stat.harvard.edu \\
       \addr $^{3~}$Department of Statistics\\
       Harvard University,
       Cambridge, MA 02138, USA}

\editor{}

\maketitle

\begin{abstract}

In most real-world systems units are interconnected and can be
represented as networks consisting of nodes and edges. For
instance, in social systems individuals can have social ties,
family or financial relationships. In settings where some units
are exposed to a treatment and its effect spills over connected
units, estimating both the direct effect of the treatment and
spillover effects presents several challenges. First, assumptions
on the way and the extent to which spillover effects occur along
the observed network are required. Second, in observational
studies, where the treatment assignment is not under the control
of the investigator, confounding and homophily are potential
threats to the identification and estimation of causal effects on
networks. Here, we make two structural assumptions: i)
neighborhood interference, which assumes interference operates
only through a function of the immediate neighbors' treatments ii)
unconfoundedness of the individual and neighborhood treatment,
which rules out the presence of unmeasured confounding variables,
including those driving homophily. Under these assumptions we
develop a new covariate-adjustment estimator for treatment and
spillover effects in observational studies on networks. Estimation
is based on a generalized propensity score that balances
individual and neighborhood covariates across units under
different levels of individual treatment and of exposure to
neighbors' treatment. Adjustment for propensity score is performed
using a penalized spline regression. Inference capitalizes on a
three-step Bayesian procedure which allows to take into account
the uncertainty in the propensity score estimation and avoiding
model feedback. Finally, correlation of interacting units is taken
into account using a community detection algorithm and
incorporating random effects in the outcome model. All these
sources of variability, including variability of treatment
assignment, are accounted for in the posterior distribution of
finite-sample causal estimands. We conducted a simulation study
where we assess the performance of our estimator on different type
of networks, generated from a stochastic block model and a latent
space model or given from the friendship-network of the Add-Health
study.

\end{abstract}


\newpage
\tableofcontents
\newpage

\section{Introduction}
\subsection{Motivation and Background}

In many areas of social, economic and medical sciences,
researchers are interested in assessing not just the association
but the causal relationship between two variables, i.e., exposure
to a condition and an outcome variable that is expected to be
affected by the exposure.  Causality plays a crucial role during
the decision-making process, because the decision maker must know
the consequences of specific actions. Many studies conducted for
assessing the effect of the exposure to a certain observed
condition, as well as many non-experimental or experimental
studies designed for evaluating the impact of public policies and
programs, are actually aiming at inferring causal effects of the
exposure to the observed condition or the implementation of the
program. In both cases, causal conclusions are typically used for
predicting causal consequences of an hypothetical intervention
that manipulates the exposure or implements the public policy or
program eventually by selecting or improving specific components.

Causal inference can be drawn using experimental or
non-experimental studies. In the former, the main challenges lies
in the design, which involves both the sampling design and the
assignment of subjects to different experimental conditions. In
the latter, the main challenge is covariate imbalance across
individual in different conditions. Covariate adjustment methods
are needed to compare similar units in different treatment arms
\citep{Imbens:Rubin:2015, Hernan:Robins:2017}.

Most estimators of causal effects rely on the assumption of no
interference between units, that is, a unit's outcome is assumed
to depend only on the treatment he received. When, instead,
interference is present, common estimators fail to estimate the
causal effect of the treatment. Interference mechanism are common
in many fields, from economics to epidemiology.
For example, \citep{Angelucci:deGiorgi:2009}, recipients of conditional cash transfers may share resources with or purchase goods and services from ineligible households who live in the same area;
 de-worming a group of children may also benefit untreated children by reducing disease transmission \citep{Miguel:Kremer:2004}.

In  the past two decades the literature on causal inference in the presence of interference has rapidly started to grow, with increasingly rapid advances in both areas of experimental design (e.g., \citealt{Hudgens:Halloran:2008, Toulis:Kao:2013, Ugander:2013, Eckles:2014}) and statistical inference. The recently proposed approaches for the estimation of spillover effects can be categorized as dealing with one of the following cases:
randomized experiments on clusters (e.g.,\citealt{Hudgens:Halloran:2008}), randomized experiments on networks (e.g.,\citealt{Rosenbaum:2007, Bowers:2013,
Aronow:2012, Athey:2015, Aronow:Samii:2013}), observational studies on clusters (e.g.,\citealt{ Hong:Raudenbush:2006, TchetgenTchetgen:VanderWeele:2012, Liu:2016}, and observational studies on networks  \citep{VanderLaan:2014, Sofrygin:vanderLaan:2017, Forastiere:2016}.

\subsection{Contributions and Outline of the paper}

Building on recent work by \citet{Forastiere:2016}, we extend the
proposed generalized propensity score estimator to more flexible
functional forms of the outcome model and to incorporate
neighborhood correlation. We develop a Bayesian estimation method
for finite sample causal effects, which relies on a modular technique and the imputation approach
to causal inference. As an alternative to the commonly used frequentist and Fisherian perspectives, this paper pioneers the use of Bayesian inference for the estimation of treatment and spillover effects in the presence of interference.
The proposed Bayesian methodology allows flexible estimation of a large range of causal effects, incorporates different sources of uncertainty and allows taking into account correlation among neighbors using community random effects. In addition,
the modular technique enables preserving robustness to model misspecification.

The remainder of the paper is structured as follows.  In Section \ref{sec:po} we review the potential outcome framework and the imputation-based and Bayesian approach to causal inference. We provide an overview of the propensity score-based methods for observational studies for both binary and continuous treatment.  In Section \label{sec:causalnet} we introduce the problem of causal inference on networks and review the concepts of  individual and neighborhood propensity scores. Section \ref{sec:BayesianGPS} is dedicated to our proposed Bayesian estimator based on the generalized propensity score and its performances are assessed with a simulation study reported in Section \ref{sec:sim}. Section \ref{sec:conc} concludes the paper.

\section{Potential Outcome Framework}
\label{sec:po}
\subsection{Introduction}
The most widely
used statistical framework for causal inference is the potential outcome framework \citep{Rubin:1974, Rubin:1978}, also known as
the Rubin Causal Model (RCM)  \citep{Holland:1986}.
The first component of the RCM are potential outcomes defined as
potential values of the outcome variable that each unit would
experience under each level of the treatment condition. Causal
effects are then defined as comparisons of potential outcomes
under different treatment conditions for the same set of units.
The concept of potential outcomes was first proposed by
\citet{Neyman:1923} in the context of randomized experiments, and
was extended to observational studies by \citet{Rubin:1974}. The
fundamental problem of causal inference under the RCM is that, in
one study, for each unit at most one of the potential outcomes is
observed -- the one corresponding to the treatment to the unit is
actually exposed to --, and the other potential outcomes are
missing. Therefore, unit-level causal effects are not identifiable
without further assumptions. The second component of the RCM is
the assignment mechanism: the process that determines which units
receive which treatments, hence which potential outcomes are
realized and thus can be observed, and which are missing. In
randomized experiment the assignment mechanism is under the
control of the experimenter. In contrast,  in observational
studies the assignment mechanism is the unknown process underlying
the observed distribution of treatment and in general depends on
units' characteristics.
Identification of causal effects relies on specific assumptions on the assignment mechanism.
The last optional component of the RCM is a model for the
potential outcomes and covariates. Incorporating scientific
understanding in a probability distribution allows
formal probability statements about the causal effects

\subsection{Set up and Notation}
Consider a study where we observe a set of N units. Let i be the
indicator of a unit in the sample, with $i=1, \ldots, N$.

The variable the causal effect of which is under investigation can
be the exposure to a certain condition (e.g., environmental
exposure, socio-economic status, behavior) or a certain treatment
or intervention. Throughout we will refer to this variable as
treatment. Let $Z_i\in \mathcal{Z}$ be the treatment variable
indicator for unit i and $Y_i\in \mathcal{Y}$ the observed outcome
that we wish to estimate the effect on. For each unit we also
collect a vector of baseline covariate $\vX_i\in \mathcal{X}$. Let
$\mathbf{O}=(\vX,\vZ,\vY)$ denote the observed data in the sample,
where $\vX=\{\vX_i\}_i^N$ is the collection of baseline covariate
across all units, $\vZ=\{Z_i\}_i^N$ is the treatment vector in the
sample and  $\vY=\{Y_i\}_i^N$ is the the corresponding outcome
vector.

In principle, we should define the potential outcome of each unit
as the potential value of the outcome variable that the unit would
experience under a specific assignment of the whole treatment
vector $\vZ=\vz$. Under this general definition, a potential
outcome is denoted by $Y_i(\vZ=\vz)$ or simply $Y_i(\vz)$. We then
have for each unit $|\mathcal{Z}|^N$ potential outcomes but we can
only observe the one corresponding to the treatment vector that
was actually observed, i.e., $Y_i(\vZ) \,\, \forall i$. A
dimensionality reduction is needed both for the definition and for
the identification of causal estimands as comparisons between
potential outcomes under different treatment conditions.
Restrictions can be given by structural assumptions or by specific
assignment mechanisms in randomized experiments.
\subsection{SUTVA}
The first basic assumption that is typically invoked  is the
stable unit treatment value assumption (SUTVA;
\citealp{Rubin:1980}). There are two components to this
assumption. The first is that there is only one version of each
treatment level possible for each unit (consistency). The second
is the no interference assumption, that is, the treatment of one
unit does not affect the potential outcomes of other units,
formally:

\begin{assumption}{Stable Unit Treatment Value (SUTVA)}
\[
\text{If}\quad Z_i=Z_i' \quad \text{then} \quad Y_i(\vZ)=Y_i(\vZ') \,\,\forall \vZ,\vZ', \forall i
\]
\end{assumption}

Under this assumption we can index potential outcomes of unit i only by the treatment received by unit i, i.e., $Y_i(Z_i=z)$ or simply $Y_i(z)$.  Therefore, under SUTVA, for each unit there exist only one potential outcome for each treatment level, with the observed outcome $Y_i=Y_i(Z_i)$ or $Y_i=\sum_{z \in \mathcal{Z}}I(Z_i=z)Y_i(z)$.

\subsection{Causal Estimands}

Causal estimands are defined as comparisons of potential outcomes
under different treatment levels. Unit-level estimands are
comparisons at the unit level, while average estimands are average
comparisons on the same sets units. A common comparison is the
average difference. For the (finite) population of units, under
SUTVA, this is the SATE or simply ATE:
\[ATE(z,z')=\frac{1}{N}\sum_{i=1}^N (Y_i(z)-Y_i(z'))\].

If the set of $N$ units is considered a sample from a larger
(finite of size $M\geq N$ or infinite) population, then define
population average causal effects, sometimes referred to as
PATE:
\[PATE(z,z')=E[Y_i(z)-Y_i(z')]\].

\subsection{Causal Inference as a Missing Data Problem}
The problem of identifying and estimating causal estimands relies on
the fundamental problem of causal inference \citep{Holland:1986}, that is, the inability
to simultaneously observe all the potential outcomes of the same unit.
In fact,
for each unit, we can observe at most the potential outcome corresponding to the  the treatment to which the unit is exposed, i.e., $Y_i=Y_i(Z_i)$, whereas all the other potential outcomes $Y_i(z)$, with $Z_i\neq z$, are missing. 
The missing potential outcomes are often referred to as `counterfactuals' in the literature.
Therefore, causal inference is inherently a missing data problem and
causal effects are not identifiable without further assumptions, which, in general, allow extrapolations of information on missing potential outcomes.
The central identifying assumptions concern the assignment mechanism

Because potential outcomes of the same unit are never simultaneously observed, we cannot in general estimate individual contrasts in potential outcomes.
We could estimate a marginal contrast in potential outcomes, e.g., $ATE(z,z')$, if we could recover the distribution of $Y_i(z)$ for units $Z_i\neq z$.
This could be done by assuming that the distribution of potential outcomes is independent of the actual treatment received.
This assumption is known as \textit{unconconfoudedness}, \textit{ignorability} or \textit{exchangeability}.
Throughout we will use the term unconconfoudedness.
In the literature we can find alternative expressions of such assumption. Here we report a weak version, which assumes the marginal independence between each potential potential outcome and the treatment receipt.
Formally, we have the following assumption.

 \begin{assumption}{Unconconfoudedness of the Treatment}

 \noindent Given an assignment mechanism, a treatment is unconfounded if it does not depend on the potential outcomes:
 \label{ass:unconf}
\[
Y_i(z) \ind Z_i \,\,\forall z, \forall i
\]
\end{assumption}

This implies that for all individuals treatment assignment does
not depend on the potential outcomes and, in turn, the
distribution of the potential outcomes is independent of the
treatment assignment. The unconfoudedness assumption holds by
design in classical randomized experiments, where the assignment
mechanism is known and the treatment is randomly assigned. Thus,
the randomized assignment mechanism implies that those assigned to
$Z_i = z$ and those assigned to $Z_i \neq z$ are exchangeable. The
independence of the potential outcomes from the treatment
assignment, is more intuitive in the way it is crucial to causal
inference. In fact,
unconfoudedness implies that
$Pr(Y_i(z)|Z_i=z) = Pr(Y(z)|Z_i \neq z)$
and thus allows
imputation of the missing potential outcomes $Y_i(z)$, with $Z_i \neq z$,
from the distribution of observed outcomes in the treatment arm where the treatment received is actually $z$.
This ensures the identifiability of causal effects from the observed data.

Specifically, under Assumption \ref{ass:unconf} $ATE(z,z')$ is identified.


\subsection{Imputation Approach to Causal Inference}
Estimating causal effects requires properly handling the missing potential outcomes.
As a matter of fact, all methods for causal inference can be viewed as imputation
methods. In fact, because any causal estimand
depends on missing potential outcomes, every estimator of causal estimands must explicitly or implicitly estimate or \textit{impute} the missing potential outcomes.

The unconfoudedness assumption is crucial for this imputation: one
can recover the distribution of missing potential outcomes from
the distribution of observed outcomes of units in other treatment
arms.

The imputation approach to causal inference explicitly imputes the missing potential outcomes for every unit in the sample.
This can be done from the empirical distribution of observed outcomes of units in other treatment arms or from a parametric model.

A unit-level causal estimand is a comparison between potential
outcomes corresponding to different treatment levels
$\tau_i=\tau(\vX_i, \{Y_i(z)\}_{z\in \mathcal{Z}})$. A finite
sample causal estimand is a sample moment of the distribution of
$\tau_i$ in the whole sample or in a sub-sample defined by o
bserved treatment and covariates; it can be expressed by
$\tau^{S}=\mu^S(\vX, \vZ, \boldsymbol{\tau})$.
Similarly, a population level causal estimand is a population moment of the population distribution of $\tau_i$ unconditional or conditional on a treatment level and/or covariates; it can be expressed by $\tau^{P}=\mu^P(\vX, \vZ, \boldsymbol{\tau})$.
An imputation-based
method hinges on a stochastic model for all potential outcomes,
both observed and missing:
\[p(\{\vY(z)\}_{z\in \mathcal{Z}}| \vX, \vZ; \boldsymbol{\theta})\]
 Such model generally depends
on some unknown parameters $\boldsymbol{\theta}$. The observed data $\mathcal{O}=(\vX, \vZ, \vY)$ are used to learn about these parameters.
The postulated model with the estimated parameters is then used to
impute the missing potential outcomes, given the observed data,
and to conduct inference for the estimands of interest. Typically,
potential outcomes of different units are assumed independent and
identically distributed. In addition,   potential outcomes of a
unit corresponding to different treatment levels are considered
independent. Therefore, under the unconfoudedness assumption, we
posit a model for each potential outcome
\[p(Y_i(z)| \vX_i; \boldsymbol{\theta})\]
Once parameters are estimated and potential outcomes for all units are imputed, the distribution of causal estimands is drawn. Specifically, population average causal estimands $\tau^P$ are computed as a function of the parameters $\boldsymbol{\theta}$. In contrast, finite sample causal estimands $\tau^S$ are derived using the imputed potential outcomes: first unit-level estimands $\tau_i$ are computed from the imputed set of potential outcomes $\{Y_i(z)\}_{z\in \mathcal{Z}}$, then average causal estimands $\tau^S$ are determined
by sample moments of the distribution of $\tau_i$ in the sample.

\subsection{Bayesian Causal Inference}
\label{sec:bci}
The outline of Bayesian inference
for causal effects was first proposed in  \citet{Rubin:1978}.
From the Bayesian perspective, the observed
outcomes are considered to be realizations of random variables and the unobserved potential outcomes are
unobserved random variables. Thus, missing potential outcomes are considered as unknown
parameters. Bayesian inference for the causal
estimands is obtained from the posterior predictive distribution
of the missing potential outcomes, given a model for potential outcomes and a prior distribution for the parameters.

Let $\vY_i^{mis}=\{Y_i(z)\}_{z\neq Z_i\in \mathcal{Z}}$ be the vector of missing potential outcomes for each unit and $\vY^{mis}$ the corresponding vector in the whole sample.
Given the prior distribution $p(\boldsymbol{\theta})$, under unconfoudedness and common independence assumptions, the posterior predictive distribution of these missing potential outcomes can be written as
\begin{equation}
\label{eq:ppd} p(\vY^{mis}| \vX, \vZ, \vY)= \int \prod_i^N
\prod_{z\neq Z_i\in \mathcal{Z}} p(Y_i(z)| \vX_i,
\boldsymbol{\theta})p(\boldsymbol{\theta}| \vX, \vZ, \vY)
d\boldsymbol{\theta}
\end{equation}

The
inner product of \eqref{eq:ppd} is the likelihood function.
Equation \eqref{eq:ppd} suggests that, under unconfoudedness and
iid units, one only needs to specify the potential outcome distribution
$p(Y_i(z) | \vX_i; \boldsymbol{\theta})$
and the prior distribution
$p(\boldsymbol{\theta})$ to conduct Bayesian inference for causal
effects. Closed-form posterior distribution of the causal estimand
is generally not available. Instead, we can use Markov Chain Monte
Carlo (MCMC) methods such as Gibbs sampler
\citep{Gelfand:Smith:1990}; specifically, one can simulate the
joint posterior distribution of all unknown random variables,
$p(\vY^{mis}, \boldsymbol{\theta}| \vX, \vZ, \vY)$, by
iteratively drawing from the two conditional distributions,
$p(\vY^{mis}| \vX, \vZ, \vY, \boldsymbol{\theta})$ and
$p(\boldsymbol{\theta}|  \vX, \vZ, \vY, \vY^{mis})$. After
obtaining the posterior draws of $\vY^{mis}$ and of the parameters
$\boldsymbol{\theta}$, it is straightforward to extract any
posterior summary, e.g. mean, variance, quantiles, of any causal
estimand.
The posterior distribution of population average estimands $\tau^P$ is derived from posterior draws of the parameters
or functions of the parameters. For
finite-sample average estimands $\tau^S$, one would instead obtain the posterior draws of the missing potential
outcomes $\vY^{mis}$ and then for each draw compute the sample average of interest.



\subsection{Unconfoundedness in Observational Studies}
In randomized
experiments Assumption \ref{ass:unconf} holds by design. Conversely, in observational studies
the investigator does not control the assignment of treatments and cannot ensure that similar subjects receive different levels of treatment.
Therefore,
Assumption
\ref{ass:unconf} cannot be ensured by design.
Moreover, such assumption cannot be directly validated
by the data because it involves missing quantities, that is, the distribution of a potential outcome $Y_i(z)$ cannot be observed in the treatment arms where $Z_i\neq z$ and, hence, we are unable to say whether it would be the same across all treatment arms. In an observational setting we typically relax the
unconditional unconfoudedness assumption by assuming
exchangeability conditional on a set of observed covariates. The
set of covariates that can make the unconfoudedness assumption
more plausible must be chosen based on subject matter knowledge.

The conditional unconfoudedness assumption can be formally stated as follows.

 \begin{assumption}{Conditional Unconconfoudedness of the Treatment}
  \label{ass:cond.unconf}

 \noindent Given an assignment mechanism, the treatment is unconfouded if it does not depend on potential outcomes conditional on an observed covariate set $\vX$:
\[
Y_i(z) \ind Z_i | \vX_i \,\,\forall z\in \mathcal{Z}, \forall i
\]
\end{assumption}
Assumption \ref{ass:cond.unconf} implies that the treatment is as
randomized among units with the same value of the observed
covariates. In other words, we can say that the treatment is
unconfouded conditional on an observed set of covariates if there
are no other unmeasured confounders, that is, unobserved
characteristics that -- by affecting both the exposure to
treatment and the potential outcomes -- are said to confound the
relationship between the treatment and the outcome.

Under conditional unconfoudedness, identification of causal
effects also requires a nonzero probability of all levels of the
treatment for all covariate combinations, also known as the
positivity assumption \citep{Hernan:Robins:2017}.

\subsection{Propensity Score for Binary Treatment}
\label{sec:ps}
Most of causal inference literature is concerned with settings with a binary treatment $Z_i \in \{0,1\}$
In this case, for each unit there are only two potential outcomes $Y_i(0)$ and $Y_i(1)$, one observed and one missing.

Several covariate-adjustment methods, relying on the unconfoudedness assumption, have been proposed.
When the set of covariates needed to make the unconfoudedness assumption more plausible is large and
includes continuous covariates, a simple stratification in groups where units share the same
values of covariates is not possible.
In such settings, propensity score-based methods are particularly useful.
For each unit, the propensity score, denoted by $\phi(\vX_i)$, is defined as the probability of receiving the active treatment given the unit's covariates \citep{Rosenbaum:Rubin:1983}. Formally, we can write
\begin{equation}
\phi(\vX_i)=Pr(Z_i=1 |\vX_i).
\end{equation}
The propensity score $\phi(\vX_i)$ has two important properties: (i) it is a balancing score, that is,
it satisfies $\vX_i \ind Z_i |\phi(\vX_i)$;
(ii) if the treatment assignment is unconfounded given $\vX_i$, then it is also unconfounded given $\phi(\vX_i)$,
that is, $Y_i(z) \ind Z_i |\phi(\vX_i)$.

Therefore, under unconfoundedness,
adjusting for the difference in the propensity scores between treated and control units would
remove all biases due to covariate imbalance, i.e., we can compare the observed outcomes
between treated and control units within groups defined by the value of the propensity score
rather than by the value of covariates. Because
 the propensity score
can be viewed as a scalar summary of the multivariate covariates, the use of the propensity score for covariate adjustment is often more convenient.

In randomized experiments, the propensity score is known and in classical randomized experiments it does not depend on covariates. On the contrary, in observational studies, the
propensity score is typically unknown and must be estimated.
It has been shown that a consistent estimate of the propensity score leads to more efficient estimation of the ATE than the true propensity score \citep{Rosenbaum:1987, Rubin:Thomas:1996, Hahn:1998, Hirano:2003}.

Propensity score methods usually involve two stages: in the first stage (`PS stage'), the propensity
scores  $\phi(\vX_i)$ are determined by estimating the parameters of a model, usually through a logistic regression ($\textrm{logit}(Pr(Z_i=1|\vX_i))=\alpha +\beta_Z^T \vX_i$),
and then by computing the individual probabilities given covariates; in the second stage  (`outcome stage'), estimate the causal
effects based on the estimated propensity scores, through three main alternative strategies:
matching, stratification, weighting or combination of these methods (for a review, see \citealt{Stuart:2010} or \citealt{Austin:2011}).

For matching, arguably the most popular causal inference method, the propensity score
is used as the distance metric for finding matched pairs of treated and control units and the causal
effects are estimated by the average within-pair difference in the observed outcome.
The most common method dor propensity score
matching is one-to-one or pair matching, in which pairs of treated and control subjects are formed
such that matched subjects have similar values of the propensity score. There different sets of options for forming propensity score matched sets. The first choice is between matching with or without replacement.
Matching with replacement allows a given subject to be included in more than one matched set, resulting in closer pairs but higher variance due to the fact that less information is used \citep{Rosenbaum:2002}.
On the contrary, in matching without replacement each unit is included in at most one matched
set. In this case, one must choose between greedy and optimal matching \citep{Rosenbaum:2002}. The greedy matching method is a sequential process where at each step the nearest control unit is selected for
matching to a given treated unit, even if that control unit would better serve as a match for a
subsequent treated unit.
An alternative to greedy matching is optimal matching, in which matches
are formed so as to minimize the total within-pair difference of the propensity score. Different optimization algorithms can be used (e.g., \citealt{Rosenbaum:2002,Hansen:2004}).

When there are large
numbers of control individuals, it is sometimes possible to get multiple good matches for each
treated individual, called ratio matching (e.g., \citealt{Rubin:Thomas:1996}).



For stratification, one stratifies
units into K (usually 5 or 6) subclasses based on quantiles of the propensity scores so that
treated and control units in each subclass have similar covariate distribution, and then calculates
the difference in the average outcome between treated and control units in each subclass. The
stratification estimator is the average of these within-subclass causal estimates weighted by the
size of each subclass \citep{Imbens:Rubin:2015}.

For weighting, the basis is an analogy of the Horvitz-Thompson estimator
in survey sampling (Horvitz and Thompson, 1952):
Therefore, one can define an inverse-probability weight (IPW) for each unit: $\phi(\vX_i)$ for treated
units and $1-\phi(\vX_i)$  for control units, and then the difference of the weighted average outcome
between treatment groups is an unbiased estimator of the average causal effect (e.g., \citealt{Hernan:Robins:2017}).

%
In general, variance estimation for these propensity score-based estimators rely on the common structure of weighted estimators \citep{Imbens:Rubin:2015}. In addition, different empirical formulas have been proposed to account for correlation within matched sets as well as for the uncertainty in the estimation of the propensity score (e.g., \citealt{Abadie:Imbens:2006, Abadie:Imbens:2009}). When researchers want to account for the uncertainty in the propensity score, a bootstrap procedure has been found to outperform other methods \citep{Hill:Reiter:2006}, but standart bootstrap methods can yield invalid inferences when applied to matching estimators \citep{Abadie:Imbens:2008}.

\subsection{Propensity Score for Continuous Treatment}
In many studies the treatment is not binary but units may receive different treatment levels.
In such settings, where the treatment is discrete or continuous, propensity score methods for binary treatment cannot be used. In fact, when the treatment is multivalued there is a propensity score for each treatment level.
Thus, adjusting for the difference in the propensity score corresponding to one specific treatment level $z^{\star}$ does not yield unbiased estimate of causal estimands comparing potential outcomes under different treatment levels, i.e., $ATE(z,z')$ with $z,z \neq z^{\star}$.

In addition, when the treatment is discrete or continuous, we are usually not only interested in comparing specific treatment levels, instead the focus is on assessing the heterogeneity of treatment effects arising from
different amounts of treatment exposure, that is, on estimating an average dose-response function.

Over the last years, propensity score methods have been generalized to
the case of discrete treatments (e.g., \citealt{Imbens:2000, Lechner:2001}) and, more recently,
continuous treatments and arbitrary treatment regimes (e.g., \citealt{Hirano:Imbens:2004, Imai:VanDyk:2004, Flores:2012,
Kluve:2012, Bia:2014}). Here we review the work by
\citet{Hirano:Imbens:2004}, who introduced the concept of the
Generalized Propensity Score (GPS) and use it to estimate an
average dose-response function (aDRF) $\mu(z)=E[Y_i(z)]$.


\subsubsection{Generalized Propensity Score (GPS)}
\label{sec:gps}
The GPS, denoted by $\lambda(z; \vx)$,
is the conditional density of the treatment given the covariates: $\lambda(z; \vX_i)=p(z|\vX_i)$.
Each unit is then characterized by a different density of the treatment.
For each unit the GPS corresponding to the actual treatment to which the unit is exposed, $\Lambda_i=\lambda(Z_i; \vx)$, is the probability for that unit of receiving the treatment he actually received given his characteristics $\vX_i$.
As the classic propensity score, the GPS is a balancing score. In the continuous case, this means that, within strata with the same value of $\lambda(z; \vX_i)$, the probability that $Z_i = z$ does not
depend on the value of $\vX_i$. In other words, for units with the same value of the GPS corresponding to a specific treatment level z, the distribution of covariates is the same between the arm assigned to treatment z and all the other arms.
Furthermore, the unconfoundedness assumption, combined with the
balancing score property, implies that the treatment is unconfounded
given the GPS. Formally,
\begin{equation}
Y_i(z) \ind Z_i| \lambda(z; \vX_i) \,\,\forall z\in \mathcal{Z}, \forall i
\end{equation}
Thus, any bias caused by covariate unbalance  across groups with different treatment
levels can be removed adjusting for the difference in the GPS. Under unconfoundedness of the treatment given
 $\vX_i$, then
\begin{equation}
\label{eq:gps.id}
\mu(z)= E [Y_i(z)]=E[E[Y_i(z)|, \lambda(z, \vX_i)]]=E[E[Y_i| Z_i=z, \lambda(z, \vX_i)]]
\end{equation}
According to Equation \eqref{eq:gps.id}, for each treatment level z, within strata with same GPS at level z,  $ \lambda(z, \vX_i)$, the average potential outcome corresponding to z, $Y_i(z)$, can be estimated using the average  outcome observed for units who were actually exposed to level z of the treatment.
Given the continuous nature of the treatment and the GPS, such stratification is unfeasible.

\citet{Hirano:Imbens:2004}
estimate the DRF using the estimated
GPS by employing a parametric partial mean approach. Specifically,
we posit a model for the treatment $Z_i$
\begin{equation}
\label{eq:Zmodel}
\textrm{h}^{ (Z)}\bigl(Z_i\bigr)\sim f^{(Z)}\biggl(q^{(Z)}\big(\vX_i;\boldsymbol{\beta}^{(Z)}\big), \nu^{(Z)}\biggr)
\end{equation}
where $h^{ (Z)}(\cdot)$ is a link function, $f^{(Z)}$ is a probability density function (pdf), $q^{ (Z)}(\cdot)$ is a flexible function of the covariates depending on a vector of parameters $\boldsymbol{\beta}^{(Z)}$, and $\nu^{(Z)}$ is a scale parameter. For instance, $h^{ (Z)}(\cdot)$ can be the identity function, f is the normal pdf, $q^{ (Z)}(\cdot)$ is a linear combination of the covariates, i.e., $q^{(Z)}\big(\vX_i;\boldsymbol{\beta}^{(Z)}\big)=\beta^{(Z)}_0 + \boldsymbol{\beta}_X^{(Z)T}\vX_i$, and $\nu^{(Z)}$ is the standard deviation of Z.
We also postulate a model for the potential outcomes given the GPS:
\begin{equation}
\label{eq:Ymodel}
\begin{aligned}
&\textrm{h}^{(Y)}\bigl(Y_i(z)\bigr) \sim f^{(Y)}\biggl(q^{(Y)}\big(z, \lambda(z;\vX_i);\boldsymbol{\beta}^{(Y)}\big), \nu^{(Y)}\biggr)
\end{aligned}
\end{equation}
where $h^{ (Y)}(\cdot)$ is a link function, $f^{(Y)}$ is a probability density function (pdf), $q^{ (Y)}(\cdot)$ is a flexible function of the treatment and the GPS depending on a vector of parameters $\boldsymbol{\beta}^{(Y)}$, and $\nu^{(Y)}$ is a scale parameter.
In \citet{Hirano:Imbens:2004} $q^{ (Y)}(\cdot)$  is a cubic polynomial function of the treatment $z$ and the generalized propensity score $\lambda(z;\vX_i)$, including the interaction term.

 \citet{Bia:2014}  replace the parametric approach with a  semiparametric estimator based on
penalized spline techniques. In particular, they use penalized bivariate splines, with radial basis functions of the form $C||(z, \lambda)-(k,k')||$.

Relying on the two models for the treatment (Equation \eqref{eq:Zmodel}) and the outcome (Equation\eqref{eq:Ymodel}), the average DRF is derived using a two-step estimator is used.

\vspace{0.3cm}
\noindent \textbf{Estimating Procedure based on the GPS}\\
We describe here the two-step estimator proposed by \citet{Hirano:Imbens:2004} for the estimation of the average DRF based on the generalized propensity score. The first step involves
parametrically modeling and estimating the GPS. The second step consists of
estimating the average DRF.

\begin{algorithm}[H]
\DontPrintSemicolon
\footnotesize
\KwIn{Dataset $\dataset^{\star}$, PS model, outcome model}
\KwOut{Average DRF $\mu(z)^{\star}, z \in \mathcal{Z}$}
\BlankLine
\SetKwProg{St}{Stage:}{}{}
\textbf{GPS} \St{}{
 \nl Estimate the parameters $\boldsymbol{\beta^{(Z)}}$ and $\nu^{(Z)}$ of the GPS model \;
 \nl \For{$i=1$ to $N$} {
Predict $\widehat{\Lambda}_i=\lambda(Z_i; \vX_i)$
  }

}
\BlankLine

\textbf{Outcome} \St{}{
 \nl Estimate the parameters $\boldsymbol{\beta^{(Y)}}$ and $\nu^{(Y)}$ of the outcome model, given the data $(\vZ^{\star}, \vY^{\star})$ and $\boldsymbol{\widehat{\Lambda}}$ \;
 \nl Impute potential outcomes and Compute average DRF: \;
\For{$z\in \mathcal{Z}$}{
\For{$i=1$ to $N$}{
 Impute the potential outcome $\widehat{Y}_i(z)$:\;
 $\qquad $a. Predict the GPS $\phi(z; \vX_i)$\;
 $\qquad $b. Predict $\widehat{Y}_i(z)$, given z, $\lambda(z; \vX_i)$, and the estimated parameters $\widehat{\boldsymbol{\beta^{(Y)}}}$ and $\widehat{\nu^{(Y)}}$\;
 }
Average the potential outcomes over all units:
    $\widehat{\mu}(z)=\frac{1}{N} \sum_{i=1}^N\widehat{Y}_i(z)$\;
    }
}
\caption{Generalized Propensity Score for Multivalued Treatment}
\label{alg:gps}
\end{algorithm}

\vspace{0.3cm}
\noindent \textbf{Statistical Inference for Population Average DRF}\\
In \citet{Hirano:Imbens:2004}, as well as in \citet{Bia:2014}, the
target causal estimand is the population average dose-response
function $\mu(z)$. In order to assess the sampling variability of
the GPS estimator with respect to the population average DRF,
standard errors and 95\% confidence intervals are derived using
bootstrap methods.

\begin{algorithm}[H]
\DontPrintSemicolon
\footnotesize
\KwIn{Dataset $\dataset$, number of iterations M}
\KwOut{Distribution of average DRF $\mu(z), z \in \mathcal{Z}$}
\BlankLine
\SetKwProg{St}{Stage:}{}{}
\For{$m=1$ to $M$} {
\nl \For{$k=1$ to $N$} {
 Sample $s_k \sim \mathcal{U}(1,N)$
}
\nl Dataset $\dataset^m=\{(\vX_{s_k})_{k=1}^M,(Z_{s_k})_{k=1}^M, (Y_{s_k})_{k=1}^M \}$ \;
\nl Run Algorithm \ref{alg:gps} with $\dataset^m$\;
\Return average DRF $\mu(z)^m, z \in \mathcal{Z}$
  }
\caption{Bootstrap for Generalized Propensity Score for Multivalued Treatment}
\label{alg:gps}
\end{algorithm}

\subsection{Bayesian Propensity Score Adjustment}
\label{sec:bps}
After \citet{Rubin:1985} reflected on the usefulness of propensity scores for Bayesian inference,
only recently has Bayesian estimation of causal effects been combined with propensity score methods.
\citep{Hoshino:2008, McCandless:2009, An:2010, McCandless:2010, Kaplan:Chen:2012, Zigler:2013, McCandless:2012, Zigler:Dominici:2014}.
The first advantage of the Bayesian propensity score approach  is that it allows embedding propensity score adjustment within broader Bayesian modeling strategies, incorporating prior information as well as complex models for hierarchical data, measurement error or missing data \citep{Rubin:1985}.
The second major motivation for using Bayesian inference for propensity scores is the propagation of propensity score uncertainty in the estimation of causal effects. In fact,  traditional frequentist approach accomodate the two-stage nature of propensity score methods with a separate and sequential estimation: the estimated model and the predicted propensity scores from the first stage are treated as fixed and known in the outcome stage.
A limitation of this sequential approach is that
confidence intervals for the treatment effect estimate are usually calculated without acknowledging uncertainty in the estimated propensity scores.
On the contrary, Bayesian methods offer a natural strategy for modeling uncertainty in the propensity scores.

\citet{McCandless:2009} proposed to model the joint distribution
of the data and parameters with the propensity score as a latent
variable. Let $\vtheta^{(Z)}$ and $\vtheta^{(Y)}$ be the vectors
of parameters of the propensity score and the outcome model,
respectively. Markov chain Monte Carlo (MCMC) methods allow to
draw from the posterior distribution for model parameters,
\[p(\boldsymbol{\theta}^{(Z)}, \boldsymbol{\theta}^{(Y)}| \vY, \vZ, \vX)\propto p(\vY, \vZ, \vX | \boldsymbol{\theta}^{(Z)}, \boldsymbol{\theta}^{(Y)}) p(\boldsymbol{\theta}^{(Z)})p(\boldsymbol{\theta}^{(Y)})\] by successively drawing from the full conditional distributions
\[p(\boldsymbol{\theta}^{(Z)}| \vY, \vZ, \vX, \boldsymbol{\theta}^{(Y)})\propto \prod_i^N p(Y_i, Z_i, \vX_i | \boldsymbol{\theta}^{(Z)}, \boldsymbol{\theta}^{(Y)}) p(\boldsymbol{\theta}^{(Z)})\] and \[p(\boldsymbol{\theta}^{(Y)}| \vY, \vZ, \vX,\boldsymbol{\theta}^{(Z)})\propto \prod_i^N p(Y_i, Z_i, \vX_i | \boldsymbol{\theta}^{(Z)}, \boldsymbol{\theta}^{(Y)}) p(\boldsymbol{\theta}^{(Y)})\]
In \citet{McCandless:2009}, as well as in the whole literature on Bayesian propensity score, the focus is on binary treatment.

Within the Bayesian framework, we denote the propensity score of a
binary treatment by $\phi(\vX_i; \boldsymbol{\theta}^{(Z)} )$ to
highlight the dependence on the parameters. With propensity score
adjustment, we do not directly model the dependence of the outcome
of covariates,  but rather we adjust for covariates by modeling
parametrically or semi-parametrically the propensity score
$\phi(\vX_i; \boldsymbol{\theta}^{(Z)} )$. As a consequence, the
outcome model indirectly depends on all parameters, including the
set of parameters of the propensity score models, i.e.,
$\boldsymbol{\theta}^{(Z)}$, and, thus, the likelihood cannot be
factorized into two parts, $p(Y_i |Z_i, \vX_i,
\boldsymbol{\theta}^{(Y)})p(Z_i | \vX_i,
\boldsymbol{\theta}^{(Z)})$, that separately depend on different
sets of parameters. Therefore, the posterior distribution of the
parameters $\boldsymbol{\theta}^{(Z)}$ of the PS stage are in part
informed by the outcome stage.
Because of this phenomenon, referred to as `model feedback' (e.g., \citealt{McCandless:2010, Zigler:2013}),
the joint Bayesian PS estimation has raised some concerns. First, since the propensity score adjustment is meant to approximate the design stage in a randomized experiment it should be done without access to the outcome data \citep{Rubin:2007, Rubin:2008}.
Furthermore, a practical consequence is the propagation of error due to model misspecification. In fact, when the model for the relationship between the outcome and the propensity score is misspecified, then the joint Bayesian approach was shown to provide invalid inferences for $\boldsymbol{\theta}^{(Z)}$, which distorts the balancing property of the propensity score and yields incorrect estimates of the treatment effect \citep{Zigler:2013}.

Various methods described as
`two-step Bayesian' have been recently proposed to `cut the feedback' between the propensity score
and outcome stages \citep{Hoshino:2008, McCandless:2010, Kaplan:Chen:2012}.
These methods  represent a special case of the so-called `modularization' in Bayesian inference \citep{Liu:2009}.
To limit feedback, the general idea is based on an approximate Bayesian technique that
uses the posterior distribution of the PS model as an input when fitting the outcome model.
Specifically, the posterior distribution of the parameters $\boldsymbol{\theta}^{(Z)}$ of the PS model
are not updated from the full conditional, but rather from  the approximate
conditional distribution \[p(\boldsymbol{\theta}^{(Z)}| \vY, \vZ, \vX, \boldsymbol{\theta}^{(Y)})\propto \prod_i^N p(Z_i| \vX_i, \boldsymbol{\theta}^{(Z)}) p(\boldsymbol{\theta}^{(Z)})\] which ignores the likelihood contribution from the outcome.
This restricts the
flow of information between models
during MCMC computation, and is similar in spirit to two-stage estimation \citep{Lunn:2009}.

\begin{algorithm}[H]
\DontPrintSemicolon
\footnotesize
\KwIn{Dataset $\dataset$, PS model, outcome model, priors}
\KwOut{Posterior distribution of causal estimand}
\BlankLine
\For{$m=1$ to $M$}{
\SetKwProg{St}{Stage:}{}{}
\textbf{PS} \St{}{
 \nl Draw the parameters $\boldsymbol{\theta^{(Z)}}\sim p(\boldsymbol{\beta^{(Z)}}|\vX, \vZ) $
 \;
 \nl \For{$i=1$ to $N$} {
Predict $\widehat{\Phi}_i=\phi(\vX_i; \boldsymbol{\theta^{(Z)}})$
  }
}
\BlankLine

\textbf{Outcome} \St{}{
 \nl Draw the parameters of the outcome model $\textrm{h}^{(Y)}\bigl(Y_i(z)\bigr) \sim f^{(Y)}\biggl(q^{(Y)}\big(z, \widehat{\Phi}_i;\boldsymbol{\theta}^{(Y)}\big), \nu^{(Y)}\biggr)$: $\boldsymbol{\beta^{(Y)}}\sim p(\boldsymbol{\beta^{(Y)}}|\vY, \vZ, \vX, \boldsymbol{\widehat{\Phi}}) $ and $\nu^{(Y)}\sim p(\nu^{(Y)}|\vY, \vZ, \vX, \boldsymbol{\widehat{\Phi}})$
 \;
  \nl \For{$i=1$ to $N$} {
 Impute potential outcomes $Y^{mis}_i=Y_i(1)(1-Z_i) +Y_i(0)Z_i$ from the posterior predictive distribution
  $p(Y_i^{mis}| Z_i, \vX_i, \widehat{\Phi}_i, \boldsymbol{\beta^{(Y)}}, \nu^{(Y)})$}
 \nl Compute the causal estimand $ATE=\frac{1}{N}\sum_{i=1}^N (Y_i-Y_i^{mis})(2Z_i-1)$ \;
}
}
\caption{Bayesian Two-Stage Propensity Score for Binary Treatment}
\label{alg:2stagePS}
\end{algorithm}

\section{Causal Inference under Interference on Network Data}
\label{sec:causalnet}
In this section we describe the problem of interference on network data in observational studies.
The presence of interference on networks poses two major challenges: i)  spillover effects of a unit's treatment on other units' outcomes, including through a contagion mechanism, and
homophily, that is, the tendency of units with similar characteristics of forming ties, which creates a dependence structure among interacting units in both pre-treatment characteristics and in the outcome.

In addition, in observational studies, where the treatment is not
randomized, homophily can generate correlation among neighbors'
treatment due to similar propensity of taking the treatment given
similar covariates, as well as peer influence in the treatment
uptake.

We will first define the problem of interference on networks and
review common assumptions of interference on neighborhood. Then,
we will define causal estimands of interest, review identifying
assumptions, and finally  propose a novel Bayesian estimator.

\subsection{Network Data}
\label{sec:netnotation}

Consider a network $\N$ of N units, indexed by i, with adjacency matrix $\mathbf{A}$, where element $A_{ij}>0$ represents the presence of a tie between unit i and unit j.  Ties are assumed to be fixed and known.
Recall that for each unit we measure a vector of covariates $\vX_i$, a treatment variable $Z_i$, and an outcome variable $Y_i$. Here we focus on binary treatments $Z_i\in\{0,1\}$.
The adjacency matrix A defines for each unit $i$ the set of units
that have a direct tie with $i$. We refer to this set as
neighborhood of unit $i$, denoted by $\N_i=\{j: A_{ij}>0\}$, and
to the units belonging to this set as neighbors of unit $i$. In
real-world applications, neighbors can be geographical neighbors,
or friends, partners or collaborators. Let $N_i=\sum_{j\neq
i}I(A_{ij}>0)$ denote the number of neighbors of unit it, referred
to as \textit{degree} of unit i. The complement of $\N_i$ in $\N$,
excluding i, is denoted by  $\N_{-i} \backslash \N_i$.

For each unit $i$, the partition $(i, \N_i, \N_{-i} \backslash
\N_i)$ defines the following partitions of the treatment and
outcome vectors: $(Z_i,\vZ_{\mathcal{N}_i},\vZ_{\mathcal{N}_{-i}
\backslash \N_i})$ and $(Y_i,\vY_{\N_i},\vY_{\N_{-i} \backslash
\N_i})$.
With non-network data, $\vX_i$ typically includes individual
characteristics or cluster-level characteristics representing
contextual factors (e.g., demographic or socio-economic factors)
or contextual covariates (e.g., geographical factors or presence
of infrastructures). On the contrary, in network data $\vX_i$
might also include variables describing the network. In
particular, it might contain variables representing  of the
neighborhood $\N_i$, including the topology but also the
distribution of individual-level characteristics, and it can
contain network properties at node-level representing the position
of unit's neighborhood in the graph (e.g., centrality,
betweenness, number of shared neighbors, ... ).

\subsection{Neighborhood interference}


In general, the potential outcome for unit $i$ depends on the
entire treatment assignment vector $\vZ$, i.e.,  $Y_i(\vZ=\vz)$.
The no interference assumption, or SUTVA, restricts the dependency
to only the treatment received by unit $i$, i.e., $Y_i(Z_i=z)$. On
the contrary, under interference  the potential outcome for unit
$i$ depends on the treatment received by other units. However, if
each outcome depends on the whole treatment vector, then for each
treatment vector $\vZ$ each unit would be observed under the same
treatment $\vZ$. Therefore, the data would not provide any
information on missing potential outcomes under different
treatment conditions. It is clear that the definition of causal
effects as well as their estimation requires assumption that
restricts the number of potential outcomes for each unit.
Depending on the type of interference $Y_i$ can depend on the
treatment received by a specific group of units. In network data,
the adjacency matrix provides information on the interaction
between units and thus on the flow of the treatment effect.
Oftentimes, we have reasons to assume that the outcome of a unit
only depends on the treatment received by the neighbors, that is,
by the units that individual is in direct contact with. In such
case, the potential outcome could be written as
$Y_i(\vZ_{\N_i}=\vz_{\N_i})$. This assumption excludes spillover
effects of the treatment received by higher order connections.
Nevertheless, when the number of neighbors is substantial then
under each treatment vector $\vZ$ the variability on the treatment
vector $\vZ_{\N_i}$ across units is still very low, and
information on the $2^{N_i}-1$ missing potential outcomes would be
hard to extrapolate.

For this reason, we introduce the concept of \textit{exposure mapping}.
In general terms, we define an exposure mapping as a function that maps a treatment vector $\vz$, the adjacency matrix $ \mathbf{A}$ and unit-level characteristics  $\mathbf{X}$ to an exposure value denoted by $G_i$: $G_i=g(\vz, \mathbf{A}, \vX_i,  \vX_{-i})$, with  $g : \mathcal{Z}^N \times  \mathcal{A}^{N^2}  \times  \mathcal{X}^N  \rightarrow \mathcal{G}_i$.
\citet{Hudgens:Halloran:2008} consider the `partial interference' assumption, that allows units to be affected only by the treatment received by units belonging to the same clusters.
This can be expressed by a specific exposure mapping function that only depends on group indicators.
In network data, a special case is the function $g_{\N}(\vz_{\N_i}, \mathbf{A}_i, \vX_i, \vX_{\N_i})$, which receives as input
only the treatment vector in the neighborhood, the unit's row of the adjacency matrix, the unit's covariates,
and the covariates of units in the neighborhood.
Given this definition, we can formalize the neighborhood interference assumption.

\begin{assumption}[Neighborhood Interference]
\label{ass:SUTNVA}
Given a function $g_{\N}: \mathcal{Z}^{N_i}\times \mathcal{A}^{N_i}\times \mathcal{X}^{N_i+1}\rightarrow \mathcal{G}_i$, $\forall i \in \N$,
$ \forall \, \vZ_{\mathcal{N}_{-i}},\vZ'_{\mathcal{N}_{-i}} $ and
$\forall \, \vZ, \vZ' \in \mathcal{Z}^N:
 g_{\N}(\vZ_{\N_i}, \mathbf{A}_i, \vX_i, \vX_{\N_i})=g_{\N}(\vZ'_{\N_i}, \mathbf{A}_i, \vX_i, \vX_{\N_i})$,
 then
 $Y_i(\vZ)=Y_i(\vZ')$.
\end{assumption}
%

Assumption \ref{ass:SUTNVA} rules out the dependence of the
potential outcomes of unit $i$ from the treatment received by
units outside his neighborhood, i.e.,
$\vZ_{\mathcal{N}_{-i}\backslash \N_i}$, but allows $Y_i$ to
depend on the treatment received by his neighbors, i.e.,
$\vZ_{\mathcal{N}_i}$. Moreover, this dependence  is assumed to be
through a specific exposure mapping function $g_{\N}(\cdot)$. This
formulation is similar to the `exposure mapping' introduced by
\citet{Aronow:Samii:2013} and the one in \citet{VanderLaan:2014}.
In Assumption  \ref{ass:SUTNVA} the function $g_{\N}(\cdot)$ is
assumed to be known and well-specified.
When analyzing network data, we must use
substantive knowledge of the subject matter and judgment about the mechanism of interference to fix the exposure mapping function.
We refer to $G_i=g_{\N}(\vZ_{\N_i}, \mathbf{A}_i, \vX_i, \vX_{\N_i})$ as the \textit{neighborhood treatment}, and, by contrast, we refer to $Z_i$ as the \textit{individual treatment}.
In general, we can write $G_i=\sum_{j\in \N_i}A_{ij}w(\vX_i,\vX_j)Z_j$, where $A_{ij}$ is the element of the adjacency matrix, which could be binary or weighted, and $w(\vX_i,\vX_j)$ is a function of unit-level characteristics of the interacting units.
This means that we assume that the extent to which the treatment of unit j affects the outcome of unit i
depends on the level on interaction between the two units, encoded in $A_{ij}$, and on their similarity in terms of characteristics.
In the simplest case, $G_i$ can be the number or the proportion of treated neighbors, i.e., $G_i=\sum_{j\in \N_i}Z_j$ or $G_i=\frac{\sum_{j\in \N_i}Z_j}{N_i}$, respectively.
The domain of $G_i$ depends on how the function $g_{\N}(\cdot)$ is defined. For example, if we consider the simple number of treated neighbors, then $\mathcal{G}_i=\{0,1,\ldots, N_i\}$.
We denote the overall domain by $\mathcal{G}=\bigcup_i\mathcal{G}_i$.

Under Assumption \ref{ass:SUTNVA}, potential outcomes
can be indexed just by the the individual treatment and the
neighborhood treatment, i.e., $Y_i(Z_i=z,G_i=g)$, which can be simplified to
$Y_i(z,g)$.
The potential
outcome $Y_i(z,g)$ represents the outcome that unit i would exhibit
under individual treatment $Z_i=z$ and if exposed to the value $g$ of a function $g_{\N}(\cdot)$
of the
treatment vector of his neighbors, $\vZ_{\mathcal{N}_i}$.

A potential
outcome $Y_i(z,g)$ is defined only for a subset of nodes where
$G_i$ can take on value $g$. We denote this subset by $V_g=\{i\! :g\in\mathcal{G}_i\}$.
For instance, in the case where $G_i$ is the number of treated
neighbors, $V_g$ is the set of nodes with degree $N_i\geq g$, that
is, with at least $g$ neighbors. It is worth noting that each unit
can belong to different subsets $V_g$, depending on the
cardinality of $\mathcal{G}_i$.

\subsection{Causal estimands: Treatment and Spillover effects}
\label{sec:effects}




We define here the causal estimands of interest under the
neighborhood interference. We focus on finite-sample causal
estimands, that is, estimands that are defined on the network at
hand. The advantage of this type of estimands is that their
definition does not require the specification of the sampling
mechanism from a larger population, which can be difficult to
conceptualize in network settings. We first define the average
potential outcome $Y_i(z,g)$ in a set of units $V$ as:
\begin{equation}
\qquad \mu(z,g; V)=\frac{1}{|V|}\sum_{i\in V} Y_i(z,g) \qquad z\in{0,1}, g\in \mathcal{G}
\end{equation}
 $V$ is a set of units, possibly defined by covariates, including individual or network characteristics. In order for the potential outcomes
 to be well-defined, $V$ must be the set (or a subset of the set) of units where $g$ is a possible value, i.e., $V \subseteq V_g$.
We can view $ \mu(z,g;V)$ as an average dose-response function
(ADRF), which measures the heterogeneity of potential outcomes
arising from a variation in the bivariate treatment, i.e., the
binary individual treatment, and the neighborhood treatment, which
is a discrete or continuous variable.

We can now define causal estimands as comparisons between average potential outcomes.
We define the average (individual) \textit{treatment effect}
at neighborhood level $g\in \mathcal{G}$ by
\begin{equation}
\label{eq:taug}
\tau(g; V)=\mu(1,g; V)-\mu(0,g; V)
\end{equation}
which denotes the average causal effect of the individual treatment when the neighborhood treatment is set to level $g$. Again here $V \subseteq V_g$.
 Instead of fixing the neighborhood treatment, we can consider an hypothetical intervention that assigns the neighborhood treatment to unit i based on a probability distribution $\pi^{\star}(g;\vX_{\N_i})$. Thus, we define the average \textit{treatment effect} $\tau(\pi^{\star}, V)$ given the neighborhood treatment assignment $\pi^{\star}(g;\vX_{\N_i})$ by the average effect of the individual treatment marginalized over the probability distribution of the neighborhood treatment, that is
\begin{equation}
\label{eq:tau} \tau(\pi^{\star}, V)=\int \bigg(\mu(1,g;
V)-\mu(0,g; V) \bigg)\pi^{\star}(g;\vX_{\N_i}) dg.
\end{equation}
We now define
the causal effects of the neighborhood treatment, often referred
to as spillover effects or peer effects. We define the
 \textit{average spillover effect} of having the neighborhood treatment set
to level $g$ versus $g'$, when the unit is under the individual
treatment $z$, by
\begin{equation}
\label{eq:deltagz}
\delta(g, g', z; V)=\mu(z,g; V)-\mu(z,g'; V)
\end{equation}
Notice that $V$ must be a subset of units belonging to both $V_g$
and $V_{g'}$, i.e., $V \subseteq V_g\cap V_{g'}$. Finally, define
the \textit{average spillover effect} of intervention
$\pi^{\star}$ vs $\pi'$ by
\begin{equation}
\label{eq:Delta}
\Delta(\pi^{\star}, \pi', z; V)=\int \mu(z,g; V)\pi^{\star}(g;\vX_{\N_i}) dg-\int \mu(z,g; V) \pi'(g;\vX_{\N_i}) dg
\end{equation}

Hypothetical Intervention $\pi^{\star}(g;\vX_{\N_i})$ vs
$\pi'(g;\vX_{\N_i})$ can be given by real experiments with
assignment mechanism $p(\vZ=\vz)$, which reflects into the
probability distribution of the neighborhood treatment, or it can
be directly defined as a probability distribution of $G_i$ given
covariates (see also Papadogeorgou et al., 2018).

The treatment effects $\tau(g; V)$ in \eqref{eq:taug} and spillover effects $\delta(g, g', z; V)$ in \eqref{eq:deltagz} are average comparisons of potential outcomes under fixed values of the individual and neighborhood treatment. Conversely, in the average treatment and spillover effects in \eqref{eq:tau} and \eqref{eq:Delta}, the individual treatment is kept fixed while the neighborhood treatment is drawn from the probability of hypothetical interventions.

\subsection{Identifying Assumption: Unconfoundedness}
\label{sec:unconf}

Because the causal effects of interest depend on the comparison between two quantities $\mu(z,g; V)$ with different values of the individual and neighborhood treatments, identification results can focus on the identification of the ADRF $\mu(z,g; V)$.
\begin{assumption}[Unconfoundedness of Individual and Neighborhood Treatment]
\label{ass: Totunconf}
\[
Y_i(z,g) \ind Z_i,G_i \mid  \vX_i \qquad \forall z\in \{0,1\}, g
\in \mathcal{G}_i, \forall i \in V.
\]
This assumption states that the individual and neighborhood treatments are independent of the potential outcomes of unit $i$, conditional on the vector of covariates $\vX_i$.
\end{assumption}
Assumption \ref{ass: Totunconf} states that the vector $\vX_i$
contains all the potential confounders of the relationship between
the individual and the neighborhood treatment and the potential
outcomes for each unit i. The plausibility of this assumption
depends on how the vector $\vX_i$ is defined in relation to the
probability distribution of the treatment and to the network
structure. Assumption \ref{ass: Totunconf} rules out the presence
of latent variables (not included in $\vX_i$) that affect both the
probability of taking the treatment and/or the value of
neighborhood treatment and the outcome. Neighborhood covariates,
that is, the topology of the neighborhood or individual-level
covariates among neighbors, are potential confounders only if the
affect the outcome of the unit.

In addition, in principle the assumption does not rule out the
presence of homophily, that is tendency of individuals who share
similar characteristics to form ties. In fact, homophily does not
violate the unconfoudedness assumption in the cases where
characteristics driving the homophily mechanism i) are included in
$\vX_i$, ii) even if unobserved they do not affect the outcomes
iii) they correspond to  the treatment variable, that
is people who share the same treatment/exposure variable tend to
form ties. The only situation where homophily is a threat to
identification is when variables underlying the network formation
process  are not included in $\vX_i$ and affect the outcome.


\citet{Forastiere:2016} show that under Assumption \ref{ass: Totunconf} the ADRF $\mu(z,g; V)$ is identified from the observed data, and estimation can be conducted by taking the average of the observed outcomes of units with $Z_i=z$ and $G_i=g$ within groups of units defined by covariates $\vX_i$.

\section{Bayesian Generalized Propensity Score Estimator for Causal Effects under Neighborhood Interference}
\label{sec:BayesianGPS}
Building on the generalized propensity score estimator proposed by
\citet{Forastiere:2016}, here we develop a new Bayesian
semi-parametric estimator for the ADRF $\mu(z,g;V)$, and in turn
for the causal estimands in Section \ref{sec:effects}. The idea is
to combine results on the generalized propensity score for
multivalued treatment proposed by \citet{Hirano:Imbens:2004}
(Section \ref{sec:gps}) and extended  by \citet{Forastiere:2016}
to interference settings, with the two-step Bayesian propensity
score estimator for a binary treatment without interference (see
Section \ref{sec:bps}). The proposed estimator lies within the
Bayesian imputation approach to causal inference reviewed in
Section \ref{sec:bci}. In addition we will replace the parametric
partial mean approach of \citet{Hirano:Imbens:2004} with a
semiparametric technique based on penalized Bayesian multivariate
splines. To take into account the dependence in the outcome, we
also include random effects in the outcome model, with groups
defined by a community detection algorithm. This Bayesian approach
allows us to easily quantify the unceratinty due to the assignment
of $Z$ and $G$ and to the inherent variability of the (missing)
potential outcomes.

\subsection{Individual and Neighborhood Propensity Scores}

Under the unconfoudedness assumption (Assumption \ref{ass: Totunconf}) the ADRF $\mu(z,g; V)$  could be estimated by taking the average of the observed outcomes
 within cells defined by covariates.
Nevertheless, the presence of continuous covariates or a large number of covariates poses some challenges in the estimation.
Under SUTVA, propensity score-based estimators are common solutions (see Section \ref{sec:ps}).
Conversely, under the neighborhood interference assumption, \citet{Forastiere:2016} propose a new propensity score-based estimator, based on the adjustment for the so-called individual and neighborhood propensity scores.

The \textit{individual propensity score}, denoted by $\phi(z;\vX_i^z)$, is the probability of having
the individual treatment at level $z$ conditional on covariates
$\vX^z_i$, i.e., $P(Z_i=z|\vX^z_i=\vx^z)$.
Similarly, the \textit{neighborhood
propensity score},
denoted by $\lambda(g; z; \vX_i^g)$,
is the probability of having the neighborhood treatment at level
$g$ conditional on a specific
value $z$ of the individual treatment and on the vector of
covariates $\vX^g_i$, i.e., $P(G_i=g| Z_i=z, \vX^g_i=\vx^g)$.
$\vX^z_i\in \mathcal{X}^z\subset  \mathcal{X}$ is the subset of
covariates affecting the individual treatment, and
$\vX^g_i\in \mathcal{X}^g\subset  \mathcal{X}$ is the subset
of covariates affecting the neighborhood treatment.
Typically,  $\vX^z_i$ should include individual characteristics
and $\vX^g_i$ is likely to include neighborhood characteristic.
Nevertheless, $\vX^z_i$ and $\vX^g_i$ could also coincide
and both include all kind of covariates.

\citet{Forastiere:2016}  show that the individual and neighborhood
propensity scores satisfy the balancing and unconfoundedness
properties. In particular, if Assumption \ref{ass: Totunconf}
holds given $\vX_i$, then the unconfoundedness assumption holds
conditional on the two propensity scores separately, i.e., $
Y_i(z,g) \ind Z_i, G_i | \lambda(g; z; \vX_i^g), \phi(z;\vX_i^z),
\forall z\in \{0,1\}, g \in \mathcal{G}_i $. This property allows
deriving a covariate-adjustment method that separately adjusts for
the individual propensity score $\phi(z;X_i^z)$ and for the
neighborhood propensity score $\lambda(g; z; X_i^g)$.
Because $\phi(z;\vX_i^z)$ is the propensity score of a binary treatment, we can always adjusts for the propensity score $\phi(\vX_i^z)=\phi(\; \vX_i^z)$.
\citet{Forastiere:2016} propose the use of a subclassification
approach on the individual propensity score $\phi(1;x^z)$ and,
within subclasses that are approximately homogenous in
$\phi(1;x^z)$, a model-based approach for the neighborhood
propensity score, similar to the one in
\citet{Hirano:Imbens:2004}. Here we replace the frequentist
subclassification and generalized propensity score-based estimator
with a semiparametric Bayesian approach.

\subsection{Propensity Scores and Outcome Models}

\subsubsection{Individual and Neighborhood Propensity Scores Models}
We first posit a model for the binary individual treatment $Z_i$
\begin{equation}
\begin{aligned}
\label{eq:ZZmodel}
&Z_i\sim Ber(\phi(1;\vX^z_i))\\
\textrm{h}^{ (Z)}&\bigl(\phi(1;\vX^z_i)\bigr)= \boldsymbol{\beta}^{(Z)T}\vX_i^z
\end{aligned}
\end{equation}
where $h^{ (G)}(\cdot)$ is the logit or probit link function,
and a model
for the neighborhood treatment $G_i$
\begin{equation}
\label{eq:Gmodel}
\textrm{h}^{ (G)}\bigl(G_i\bigr)\sim f^{(G)}\biggl(q^{(G)}\big(\vX_i;\boldsymbol{\beta}^{(G)}\big), \nu^{(G)}\biggr)
\end{equation}
where again $h^{ (G)}(\cdot)$ is a link function, $f^{(G)}$ is a probability density function (pdf), $q^{ (G)}(\cdot)$ is a flexible function of the covariates depending on a vector of parameters $\boldsymbol{\beta}^{(G)}$, and $\nu^{(G)}$ is a scale parameter.

\subsubsection{Outcome Model with Penalized Splines and Random Effects}
We now postulate a model for the potential outcomes given $\phi(1;\vX_i^z)$ and $\lambda(g; z; \vX_i^g)$:
\begin{equation}
\label{eq:YYmodel}
\begin{aligned}
&\textrm{h}^{(Y)}\bigl(Y_i(z,g)\bigr) \sim f^{(Y)}\biggl(q^{(Y)}\big(z, g,\phi(1;\vX^z_i),\lambda(g; z; \vX_i^g);\boldsymbol{\beta}^{(Y)}\big), \nu^{(Y)}\biggr)
\end{aligned}
\end{equation}
where as usual $h^{ (Y)}(\cdot)$ is a link function, $f^{(Y)}$ is a probability density function (pdf), and $\nu^{(Y)}$ is a scale parameter.
The key feature here is  $q^{ (Y)}(\cdot)$, which we
model semiparametrically using a set of penalized spline
basis functions. Splines yield several advantages that include
flexibility as well as interpretability via representations that
use a compact set of basis functions and coefficients.
In particular, the predictor $q^{ (Y)}(z, \mathbf{V}(z,g)_i)$, where $\mathbf{V}(z,g)_i=[g, \phi(1;\vX^z_i), \lambda(g; z; \vX_i^g)]^T$,
can be written in the mixed model representation \citep{Ruppert:2003}:
\begin{equation}
\label{eq:splines}
\begin{aligned}
&q^{(Y)}\big(z, \mathbf{V}_i(z,g))=\boldsymbol{\beta}^{(Y)T}_V \mathbf{V}'_i(z,g)+\boldsymbol{\beta}^{(Y)T}_{Vz} \mathbf{V}'_i(z,g)z+ \boldsymbol{b}_U^{T}\mathbf{U}_i(z,g) + \boldsymbol{b}_{Uz}^{T}\mathbf{U}_i(z,g)z + u_j\\
&u_j\sim \mathcal{N}(0, \Sigma_u) \qquad
\boldsymbol{b}_{U} \sim  \mathcal{N}(0, \sigma^2_{b_u}I_K) \qquad \boldsymbol{b}_{Uz} \sim  \mathcal{N}(0, \sigma^2_{b_{uz} }I_K)
\end{aligned}
\end{equation}
where $\mathbf{V}'_i(z,g)=[1, g, \phi(1;\vX^z_i), \lambda(g; z; \vX_i^g), \lambda(g; z; \vX_i^g)g]^T$, such that the first two terms of Equation \eqref{eq:splines} represent the linear predictor with interactions,
and $\mathbf{U}_i(z,g)$ contains spline basis functions. In particular, we use
multivariate smoothing splines with
radial basis functions of the form
\begin{equation}
\label{eq:splines_basis}
\mathbf{U}_{ik}(z,g)= C(|| \mathbf{V}_i(z,g)- \mathbf{k}_k||)\Omega^{-1/2} ;\qquad \Omega=\big[C(||\mathbf{k}_k-\mathbf{k}_{k'} ||)\big]_{1\leq k,k', \leq K}\end{equation}
where $||\cdot||$ is the euclidean norm and $C(\cdot)$ is a basis function.
Here our choice goes to thin plate splines of the form
\begin{equation}
\label{eq:radial_basis}
C(|| \mathbf{r}||)=\begin{cases}
|| \mathbf{r}||^{2m-|\mathbf{r}|}\\
||\mathbf{r}||^{2m-|\mathbf{r}|} log(\mathbf{r})
\end{cases}\end{equation}
where m is an integer satisfying $2m - |\mathbf{r}| > 0$, that controls the order of the spline and its smoothness
\citep{Ruppert:2003, Wood:2003}. The default is to use the smallest integer satisfying that condition.
The advantage of radial basis functions in multivariate smoothing is that they are rotational invariant.
The distribution of the coefficients $\boldsymbol{b}_{U}$ and $\boldsymbol{b}_{Uz}$ is a mixed model representation of penalties. The variances $\sigma^2_{b_{u} }$ and $\sigma^2_{b_{uz} }$ are indeed the parameters controlling the degree of smoothness. A large value of these parameters , that is,
a strong roughness penalty, leads to a smoother fit, while a
small value (close to zero) leads to an irregular fit and
essentially interpolation of the data.
A key component in fitting splines is the choice of the
number and the placement of knots (K). We address this
issue by using penalized splines; wherein we choose
a large enough number of knots that are sufficient to capture
the local nonlinear features present in the data and control for
overfitting by using a penalization on the basis coefficients.
Knots are first placed on data locations.
For large datasets we randomly subsample a maximum number of data locations (the default maximum number is 2000).
Then a truncated eigen-decomposition is used to achieve a rank reduction \citep{Wood:2003, Wood:2017}.

In Equation \ref{eq:YYmodel}, the outcome model depends on the individual treatment, the neighborhood treatment and both the individual and the neighborhood propensity scores.
An alternative approach is to replace the model-based adjustment for the individual propensity score
with a matching approach. The idea is to match units on the individual propensity score $\phi(1;\vX^z_i)$ to create a matched sample where covariates $\vX_i^z$ are balanced across the treated group ($Z_i=1$) and the control group ($Z_i=0$). Adjustment for the neighborhood propensity score is then handled, as previously, by a model-based generalized propensity score method applied to the matched samples.
For matching with replacement or variable ratio matching weights need to be incorporated into the analysis \citep{Dehejia:Wahba:1999, Hill:2004, Stuart:2010}. When matching with replacement, individuals receive a frequency weight that reflects the number of matched sets they belong to. 
When using variable ratio matching, units receive a weight that is proportional to the actual number of units matched to them.
In a Bayesian framework, weights can be incorporated by weighting the scale parameter. Therefore, we would assume a model for the outcome as in \eqref{eq:YYmodel}, with $V_i=[g, \lambda(g; z; \vX_i^g)]$, $\mathbf{V}'_i=[1, g, \lambda(g; z; \vX_i^g), \lambda(g; z; \vX_i^g)g]^T$ and $\nu^{(Y)}$ scaled by the matching weights.

Finally, the last term in Equation \eqref{eq:splines}, $u_j$,
$j=1,\ldots,J$, is the random effect for community j, with $j=1,
\ldots, J$. We include this term to take into account any
dependence in the outcome data between a unit and his neighbors.
In principle, each unit belongs to the neighborhoods $\N_k$ of all
units $k \in \N_i$ in his own neighborhood. Such overlapping
nature of neighborhoods complicates the estimation of the
correlation structure.
We propose an alternative dependence structure by identifying larger non-overlapping clusters incorporating the neighborhoods of multiple units. By defining random effects $u_j$ at such cluster-level, we assume the presence of latent random variable which is shared by all units belonging to the same cluster.
Clusters are defined using a community detection algorithm that
identifies groups of nodes that are heavily connected among
themselves, but sparsely connected to the rest of the network.
This definition of communities enables taking into account both
the ego-alter correlation and a broader cluster correlation
structure.

\subsubsection{Community Detection}
\label{sec:comdet}

Unfolding the communities in real networks is widely used to determine the structural properties of these networks. Community detection or clustering algorithms aim at finding groups of related nodes that are densely interconnected and have fewer connections with the rest of the network. These groups of nodes are called communities or clusters.
As communities are often associated with important structural characteristics of a complex system, community detection is a common first step in the understanding and analysis of networks. The search for communities that optimize a given quantitative performance criterion is typically an NP-hard problem, so in most cases one must rely on approximate algorithms to identify community structure.
The problem of how to find communities in networks has been extensively studied and a substantial amount of work has been done on developing clustering algorithms (an overview can be found in \citep{Schaeffer:2007, Lancichinetti:Fortunato:2009, Fortunato:2010}).
In the simulation section (Section \ref{sec:sim}), we will descrive the specific methods used for our simulations.

\subsubsection{Priors}
Within the Bayesian framework we should posit  a prior distribution for the parameter vector
\[\boldsymbol{\theta}=\big[\boldsymbol{\theta^{(Z)}}, \boldsymbol{\theta^{(G)}}, \boldsymbol{\theta^{(Y)}}\big]\]
where $\boldsymbol{\theta^{(Z)}}=\boldsymbol{\beta^{(Z)}}$, $\boldsymbol{\theta^{(G)}}=[\boldsymbol{\beta^{(G)}}, \nu^{(G)}]$, $\boldsymbol{\theta^{(Y)}}=[\boldsymbol{\beta^{(Y)}}, \nu^{(Y)},  \Sigma_u, \sigma^2_{b_{u}}, \sigma^2_{b_{uz}}]$.
In particular, we recommend the use of weakly informative priors to provide moderate regularization and help stabilize computation.

We assume a multivariate normal prior for all regression coefficients:
\begin{equation}
\boldsymbol{\beta^{(Z)}} \sim \mathcal{N}(\boldsymbol{\eta}^{Z}, K^Z) \qquad  \boldsymbol{\beta^{(G)}} \sim \mathcal{N}(\boldsymbol{\eta}^{G}, K^G)\qquad  \boldsymbol{\beta^{(Y)}} \sim \mathcal{N}(\boldsymbol{\eta}^{Y}, K^Y)
\end{equation}

The priors on the scale parameters $\nu^{(G)}$ and $\nu^{(Y)}$ should depend on the neighborhood treatment distribution $f^{(G)}$ and on the outcome distribution $f^{(Y)}$, respectively. However, a general prior distribution could be
\begin{equation}
\nu^{(G)} \sim \textrm{Exp}(\gamma_{\nu^G }) \qquad  \nu^{(Y)} \sim \textrm{Exp}(\gamma_{\nu^Y })
\end{equation}

The random effect covariance matrix $\Sigma_u$ is decomposed into a diagonal matrix of standard deviations and the correlation matrix \citep{McElreath:2016}:
\[\Sigma_u=\textrm{diag}(\boldsymbol{\sigma}_u) \Omega_u\textrm{diag}(\boldsymbol{\sigma}_u)\]
For the correlation matrix we use a prior distribution called LKJ, whose density is proportional to the determinant of the correlation matrix raised to the power of a positive regularization parameter minus one:
\[\Omega_u\sim LKJ(\zeta) \qquad p(\Omega_u)\propto \textrm{det}(\Omega_u)^{\zeta-1}, \zeta>0\]
so $\zeta=1$ leads to a uniform distribution on correlation matrices, while the magnitude of correlations between components decreases as $\zeta \rightarrow \infty$.
The standard deviations $\boldsymbol{\sigma}_u$ are in turn decomposed into the product of a simplex vector $\boldsymbol{\pi}_u)$ and the trace of the covariance matrix:
\[\textrm{diag}(\boldsymbol{\sigma}_u)= \textrm{tr}(\Sigma_u) \boldsymbol{\pi}_u=J e^2\boldsymbol{\pi}_u\]
where the trace (total variance) is the product of the order of the matrix and the square of a scale parameter and the element $\pi_j$ of the simplex vector is the proportion of the total variance attributable to the corresponding random effect $u_j$.
For the scale parameter $e$ we posit a Gamma prior, with shape and scale parameters both set to 1. For the simplex vector $\boldsymbol{\pi}_u$ we use a symmetric Dirichlet prior, which has a single concentration parameter $\chi>0$.

For the smoothing parameters we posit the following prior distribution:
\begin{equation}
\sigma^2_{b_{u} } \sim \textrm{Exp}(\gamma_{b_{u} }) \qquad  \sigma^2_{b_{uz} } \sim \textrm{Exp}(\gamma_{b_{uz} })
\end{equation}


\subsection{Three-Step Estimating Procedure}
Here we propose a three-step Bayesian estimator that extends the `two-step Bayesian' estimator proposed by
\citet{Hoshino:2008}, \citet{McCandless:2010} and \citet{Kaplan:Chen:2012}  to the neighborhood interference setting, with the individual and the neighborhood propensity score.

The three steps refer to the posterior distributions of the
parameters of the individual propensity score, the neighborhood
propensity score and the outcome models. Since the outcome model
involves the individual and neighborhood propensity scores, in
principle the outcome model indirectly depends on all parameters,
including the set of parameters of the two propensity score
models, i.e., $\boldsymbol{\theta}^{(Z)}$, and
$\boldsymbol{\theta}^{(G)}$. Therefore, the posterior distribution
of these parameters should in part be informed by the outcome
stage. To restrict the flow of information between models during
MCMC computation and, hence, to avoid `model feedback', we take a
three-step approach which approximate the joint posterior
distribution by drawing the parameters of the propensity score
models from the approximate conditional distributions
\[p(\boldsymbol{\theta}^{(Z)}| \vY, \vG, \vZ, \vX, \boldsymbol{\theta}^{(Y)}, \boldsymbol{\theta}^{(G)})\propto \prod_i^N p(Z_i| \vX_i, \boldsymbol{\theta}^{(Z)}) p(\boldsymbol{\theta}^{(Z)})\]
and
\[p(\boldsymbol{\theta}^{(G)}| \vY, \vG, \vZ, \vX, \boldsymbol{\theta}^{(Y)},\boldsymbol{\theta}^{(Z)})\propto \prod_i^N p(G_i| Z_i, \vX_i, \boldsymbol{\theta}^{(G)}) p(\boldsymbol{\theta}^{(G)})\]
which ignore the likelihood contribution from the outcome and, hence,  do not depend neither on $\boldsymbol{\theta}^{(Y)} )$ nor on the parameters of the neighborhood or individual  propensity score model, respectively.
The posterior distributions of the individual and the neighborhood propensity score models are then used as
 an input when deriving the posterior distribution of the parameters of the outcome model:
 \[p(\boldsymbol{\theta}^{(Y)}| \vY, \vG, \vZ, \vX,\boldsymbol{\theta}^{(Z)}, \boldsymbol{\theta}^{(G)})\propto \prod_i^N p(Y_i, G_i, Z_i, \vX_i | \boldsymbol{\theta}^{(Z)}, \boldsymbol{\theta}^{(G)},\boldsymbol{\theta}^{(Y)}) p(\boldsymbol{\theta}^{(Y)})\]

After the posterior distribution of all the parameters $\vtheta$ is drawn,
the posterior distribution of our finite-sample average dose-response function ADRF $\mu(z,g; V)$ is obtained by drawing from the posterior predictive distribution the potential outcomes $Y_i(z,g)$, for each value of z and g and for each unit $i \in V$. Then for each draw the ADRF $\mu(z,g; V)$ is computed by taking the average of the imputed potential outcomes $Y_i(z,g)$ over all units of the set V.
Causal estimands are simply computed from the ADRF as comparisons of average potential outcomes at different levels.


We describe below all the steps of the algorithm.
\newpage

\begin{algorithm}[H]
\DontPrintSemicolon
\footnotesize
\KwIn{Dataset $\dataset$, Adjacency Matrix $\mathbf{A}$, Z model, G model, Y model, priors, Matching=FALSE}
\KwOut{Posterior distribution of ADRF $\mu(z,g)$}
\BlankLine
\SetKwProg{St}{Stage:}{}{}

\textbf{Community Detection} \St{}{
\nl Run a \textit{Community Detection algorithm} on $\mathbf{A}$ $\Longrightarrow$ community indicators $C_i \in \{1,\ldots,J\}, \forall i \in \N$
}
\BlankLine
\nl Initialize parameters $\vtheta^{(0)}=[\vtheta^{(Z)(0)}, \vtheta^{(G)(0)}, \vtheta^{(Z)(0)}]$\;
\For{$m=1$ to $M$}{
\textbf{PS} \St{}{
\nl Define $\vX_i^z \in \vX_i$ \tcc*[h] {including individual, network and neighborhood characteristics} \;
 \nl Draw the parameters $\boldsymbol{\beta^{(Z)(m)}}\sim p(\boldsymbol{\beta^{(Z)}}|\vX^z, \vZ) $  \;
  \nl \For{$i=1$ to $N$} {
Predict $\widehat{\Phi}^{(m)}_i=\phi(1; \vX^z_i; \boldsymbol{\beta^{(Z)(m)}})$
  }

\nl  \If{Matching=TRUE}{
 Run a \textit{Matching algorithm} with distance metric=$\widehat{\lambda}_i$ $\Longrightarrow$ matched sets $\mathcal{S}_k, \,k=1,\ldots, K$ \;
 Given $S_{ik}=I(i\in \mathcal{S}_k)$, define weights as $w_i=\frac{1}{\sum_{k=1}^K S_{ik}}$, set $\mathcal{M}=\{i : \sum_{k=1}^K S_{ik}>0\}$, and define the matched sample as
 $\dataset^{\star}=\{(\vX_{i})_{i \in \mathcal{M}},(Z_{i})_{i \in \mathcal{M}}, (Y_{i})_{i \in \mathcal{M}} \}$
 }
\Else{ Set $w_i=1, \forall i \in \N$; $\mathcal{M} = \N$ ; $\dataset^{\star}= \dataset$}
}
\BlankLine

\textbf{GPS} \St{}{
\nl Compute $G_i=g_{\N}(\vZ_{\N_i}, \mathbf{A}_i, \vX_i, \vX_{\N_i}), \forall i \in \N$ \;
\nl Define $\vX_i^g \in \vX_i$ \tcc*[h] {including individual, network and neighborhood characteristics} \;
 \nl Draw the parameters $\boldsymbol{\beta^{(G)(m)}}\sim p(\boldsymbol{\beta^{(G)}}|\vX^g, \vZ, \vG) $ and $\nu^{(G)(m)}\sim p(\nu^{(G)}|\vX^g, \vZ, \vG)$  \;
 \nl \For{$i=1$ to $N$} {
Predict $\widehat{\Lambda}^{(m)}_i=\lambda(G_i;Z_i;\vX^g_i; \boldsymbol{\beta^{(G)(m)}}, \nu^{(G)(m)})$
  }
}
\BlankLine
\textbf{Outcome} \St{}{
\nl  \If{Matching=TRUE}{
Define $\mathbf{V}^{(m)}(Z_i,G_i)_i=[G_i, \widehat{\Lambda}^{(m)}_i]^T$ \;
 }
\Else{ Define $\mathbf{V}_i^{(m)}(Z_i,G_i)=[G_i, \widehat{\Phi}^{(m)}_i, \widehat{\Lambda}^{(m)}_i]^T$}
\nl Compute spline basis functions $\mathbf{U}_i^{(m)}(Z_i, G_i)$ as in \eqref{eq:splines_basis} and \eqref{eq:radial_basis} \;
 \nl Draw the parameters $\vtheta^{(Y)(m)}$ of the outcome model in  \eqref{eq:YYmodel} and \eqref{eq:splines}
 using the Gibbs sampler algorithm \ref{alg:Ygibbs}\;
 \nl \For{$z=0,1$} {
  \nl \For{$g\in \mathcal{G}$} {
  \nl \For{$i=1$ to $N$} {
 Impute potential outcomes $\widehat{Y}_i(z,g)$:\;
 $\qquad $a. Predict the neighborhood GPS $\lambda(g; z; \vX^g_i; \boldsymbol{\beta^{(G)(m)}}, \nu^{(G)(m)})$\;
 $\qquad $b. Define $\mathbf{V}^{(m)}_i(z,g)$ and compute $\mathbf{U}^{(m)}_i(z,g)$\;
  $\qquad $c. Predict $\widehat{Y}_i(z,g)$, given z,g, $\mathbf{V}^{(m)}_i(z,g)$, $\mathbf{U}^{(m)}_i(z,g)$, the random effects $\mathbf{u}^{(m)}$, and $\phantom{.} \,\,\, \qquad $ the  parameters ${\boldsymbol{\beta^{(Y)(m)}}}$, $\boldsymbol{b}^{(m)}_{U}$, $\boldsymbol{b}^{(m)}_{Uz}$ and ${\nu^{(Y)(m)}}$\;
 }
 Average the potential outcomes over all units:
    $\widehat{\mu}(z,g)=\frac{1}{N} \sum_{i=1}^N\widehat{Y}_i(z,g)$\;

}
}
}
}
\caption{Bayesian Three-Step Generalized Propensity Score}
\label{alg:3stagePS}
\end{algorithm}

\newpage
\begin{algorithm}[H]
\DontPrintSemicolon
\footnotesize
\KwIn{Dataset $\dataset$, Adjacency Matrix $\mathbf{A}$, Y model, priors}
\KwOut{Posterior distribution of $\vtheta^{(Y)(m)}$}
 \nl Draw the parameters $\vtheta^{(Y)(m)}$ of the outcome model in  \eqref{eq:YYmodel} and \eqref{eq:splines}:\;
$\quad $a. Draw the random effects $u^{(m)}_j \sim p(u_j| \Sigma^{m-1}_u)$  \;
$\quad $b. Draw the coefficients $\boldsymbol{b}^{(m)}_{U}\sim p(\boldsymbol{b}_{U}|\sigma^{2(m-1)}_{b_{u} })$ and $\boldsymbol{b}^{(m)}_{Uz}\sim p(\boldsymbol{b}_{Uz}|\sigma^{2(m-1)}_{b_{uz} })$ \;
$\quad $c.  Draw $\Sigma^{(m)}_u \sim p(\Sigma_u | \vY, \vZ, \mathbf{V}^{(m)}(\vZ,\vG),\mathbf{U}^{(m)}(\vZ,\vG), \mathbf{u}^{(m)}, \boldsymbol{\beta^{(Y)(m-1)}}, \boldsymbol{\nu^{(Y)(m-1)}}, \boldsymbol{b}^{(m-1)}_{U}, \boldsymbol{b}^{(m-1)}_{Uz})$ \;
$\quad $d. Draw the smoothing parameters:
$\phantom{.} \qquad \sigma^{2(m)}_{b_{u}} \sim p( \sigma^{2(m)}_{b_{u}} | \vY, \vZ, \mathbf{V}^{(m)}(\vZ,\vG),\mathbf{U}^{(m)}(\vZ,\vG), \mathbf{u}^{(m)}, \boldsymbol{b}^{(m)}_{U}, \boldsymbol{b}^{(m)}_{Uz}, \boldsymbol{\beta^{(Y)(m-1)}}, \boldsymbol{\nu^{(Y)(m-1)}})$ \; $\phantom{.} \,\, \quad $ and $\sigma^{2(m)}_{b_{uz} } \sim p( \sigma^{2(m)}_{b_{uz}} | \vY, \vZ, \mathbf{V}^{(m)}(\vZ,\vG),\mathbf{U}^{(m)}(\vZ,\vG), \mathbf{u}^{(m)}, \boldsymbol{b}^{(m)}_{U}, \boldsymbol{b}^{(m)}_{Uz}, \boldsymbol{\beta^{(Y)(m-1)}}, \boldsymbol{\nu^{(Y)(m-1)}})$ \;
$\quad $e. Draw $\boldsymbol{\beta^{(Y)(m)}}\sim p(\boldsymbol{\beta^{(Y)}}|\vY, \vZ, \mathbf{V}^{(m)}(\vZ,\vG),\mathbf{U}^{(m)}(\vZ,\vG), \mathbf{u}^{(m)},  \boldsymbol{b}^{(m)}_{U}, \boldsymbol{b}^{(m)}_{Uz}, \boldsymbol{\nu^{(Y)(m-1)}})$ \;
$\phantom{.} \!\quad $f. Draw $\nu^{(Y)(m)}\sim p(\nu^{(Y)}|\vY, \vZ, \mathbf{V}^{(m)}(\vZ,\vG),\mathbf{U}^{(m)}(\vZ,\vG), \mathbf{u}^{(m)},  \boldsymbol{b}^{(m)}_{U}, \boldsymbol{b}^{(m)}_{Uz},  \boldsymbol{\beta^{(Y)(m)}})$
 \;
\caption{Gibbs Sampler for Parameters of the Outcome Model}
\label{alg:Ygibbs}
\end{algorithm}

\subsection{Posterior Predictive Checks}
\label{ref:ppv}
The unbiasedness of our proposed estimator relies on the structural Assumptions \ref{ass:SUTNVA} and \ref{ass: Totunconf} and on the correctness of the posited model for the individual and neighborhood propensity scores (Equations \eqref{eq:ZZmodel} and \eqref{eq:Gmodel}) and for the outcome (Equation \eqref{eq:YYmodel}). It is worth noting that Assumptions \ref{ass:SUTNVA} can be seen as both a structural and a modeling assumption because, in addition to ruling out the presence of interference from units outside the neighborhood of unit i, it also assumes a specific function $g_{\N}(\cdot)$ through which interference takes place.
As a method for model checking we propose the use of posterior predictive checks, where
the posterior predictive distribution is compared to the observed outcome distribution \citep{Gelman:1996}.
This approach simply relies on the intuitive idea that if a model is a good fit to the data, then replicated data predicted from that model should look similar to the observed data.
To simplify the comparison, a summary discrepancy measure $T(\cdot)$
is generally used.

In practice, a posterior predictive check can be incorporated in our estimator by sampling from the
the posterior predictive distribution of the potential outcomes corresponding to the observed outcomes:
\begin{equation}
p(Y(Z_i,G_i)| \vY, \vG, \vZ, \vX)=\int p(Y(Z_i,G_i)| \vY, \vG, \vZ, \vX,\vtheta) p(\vtheta| \vY, \vG, \vZ, \vX) d\theta
\end{equation}
After the models have been specified (Equations \eqref{eq:ZZmodel}, \eqref{eq:Gmodel} and \eqref{eq:YYmodel}), together with the prior distribution $p(\vtheta)$, and the posterior distribution of the parameters has been sampled, we draw from the posterior predictive distribution many replicate data sets, that is, for each unit we draw outcomes corresponding to the observed individual and neighborhood treatments, i.e.,  $Y_i^{rep}=Y_i(Z_i, G_i)$. We then calculate a test statistic $T(\mathbf{Y}^{rep})$ for each replicate data set and we compare the distribution of $T(\mathbf{Y}^{rep})$ with the test stastistic applied to the observed data $T(\mathbf{Y}^{obs})$.
Parameter uncertainty is explicitly accounted for because the data realizations are generated from parameter values randomly drawn from the posterior distribution.
 Posterior predictive P-values are defined as posterior probability that the test statistic applied to replicated data, generated under the posited model with parameter sampled from the posterior distribution, exceeds the value of the test statistic applied to the observed data:
 \begin{equation}
\mathrm{p}\text{-value}=Pr(T(\mathbf{Y}^{rep})\geq T(\mathbf{Y}^{obs})| \vY, \vG, \vZ, \vX)
\end{equation}
 Posterior predictive P-values  can easily be calculated as  the proportion of data replications under the specified model resulting in a value of the test statistics that exceeds the observed value.


\section{Simulation Study}
\label{sec:sim}

\subsection{Data Generating Process and Estimators}
\label{sec: DGP}
To assess the performance of our proposed estimator we conducted a simulation study. The study consists of 12 (3 x 2 x 2) scenarios, characterized by a different data generating process given by the network type, the outcome functional form and the outcome correlation structure. Here we describe the general structure of the simulation scenarios. Details of the data generating models can be found in
Appendix \ref{app: DGP}.

With regard to the network type we generated the network from i) a stochastic block model, ii) a latent cluster position model, iii) the Add Health dataset.
Each network consists of 1000 nodes. For the first two cases, where the network was generated using a formation model, before generating the adjacency matrix two covariates -- one continuous and one binary -- were drawn for each unit.
The two formation models differ by presence of homophily, that is the probability of forming links depending on covariates. In the stochastic block model \citep{Holland:1983} the probability of link between two units only depends on their community membership, whereas in the latent cluster position network \citep{Handcock:Raftery:2007} it also depends on the similarity in their characteristics. Finally, in the third case, we extracted 1000 students from the Add Health dataset \citep{Harris:Udry:2017} and we used race, grade and sex as covariates.

Regarding the outcome functional form, for each network type we simulated the outcome from a normal distribution with the predictor $q^{(Y)}(\cdot)$ being a) a linear function of the individual treatment, the neighborhood treatment, the individual propensity score and neighborhood propensity score, b) a non-linear function of the two treatments and the two propensity scores, resulting in a sigmoid function of the neighborhood treatment.

Finally, for each scenario defined by the network type and the outcome functional form we generated the outcome with and without correlation between units belonging to the same community, that is with and without the inclusion of a community random intercept.


%

For each scenario, we ran 500 simulations where, given the sample
of 1000 nodes and its adjacency  matrix, for each unit i: 1) we
generated the individual treatment $Z_i$ from an individual
propensity score model (see Appendix \ref{app: DGP}); 2) we
computed the neighborhood treatment $G_i$ as the proportion of
treated neighbors; 3) we generated the outcome $Y_i$ from a model
defined by the specific scenario (linear or non-linear and with or
without a community intercept). The posterior variance of the
estimator should capture the variability of the treatment and
outcome stochastic models conditional on a given network.
Therefore, in order for the variance across simulations to
reproduce the posterior variance of the estimator in principle the
treatment and the outcome should be repeatedly drawn on the same
network of units.
 When the network was sampled from a formation model,
 we generated 5 different networks and we ran 100 simulations for each network, in order
 to assess the performance of the estimator beyond a specific sampled network.

In all 12 scenarios the average dose-response functions, and in turn the causal treatment and spillover effects, were estimated using the proposed estimator with different options.
The first option refers to the adjustment for the difference in the individual propensity score. This was performed using a model-based approach, that is by including the individual propensity score in the outcome model, or with a matching approach. The second variant of the estimator was the use of a linear outcome model, i.e. $q^{(Y)}(\cdot)$ including only the first linear part,  or the use of penalized splines, i.e. $q^{(Y)}(\cdot)$ including both the linear part and the non-linear part with spline basis functions. Finally, we investigated the performance of the estimator with and without the inclusion of a community random intercept. Since the scope of the simulation study was to assess under
different
scenarios the frequentist performances of our proposed estimator with different modeling assumptions,
in the case where community correlation structure was taken into account we defined random intercepts on the known communities as provided by the data generating process. In this way we could disentangle the performance of our Bayesian estimator based on the individual and neighborhood propensity score from the performance of community detection algorithms.


\subsection{Results}
We investigated the frequentist performance of the Bayesian generalized propensity score estimator in estimating the average dose-response function and the causal treatment and spillover effects.
Given the posterior distribution of the ADRF and the causal effects for each replication, we then computed across the 500 replication the average bias and mean square error of the posterior mean and the coverage of the 95\% credible intervals. In the tables below we report the bias, RMSE and coverage for the treatment and spillover effects. The performance of the estimator with respect to the ADRF can be seen in the figures.
The figures represent posterior distribution of ADRF averaged across 500 simulations. The posterior mean and 95\% credible intervals are plotted against the true average dose-response function.

\begin{figure}[t!]
        \begin{adjustwidth}{-0.8cm}{0cm}
    \begin{subfigure}[t]{0.45\textwidth}
        \centering
        \includegraphics[width=1.25\textwidth]{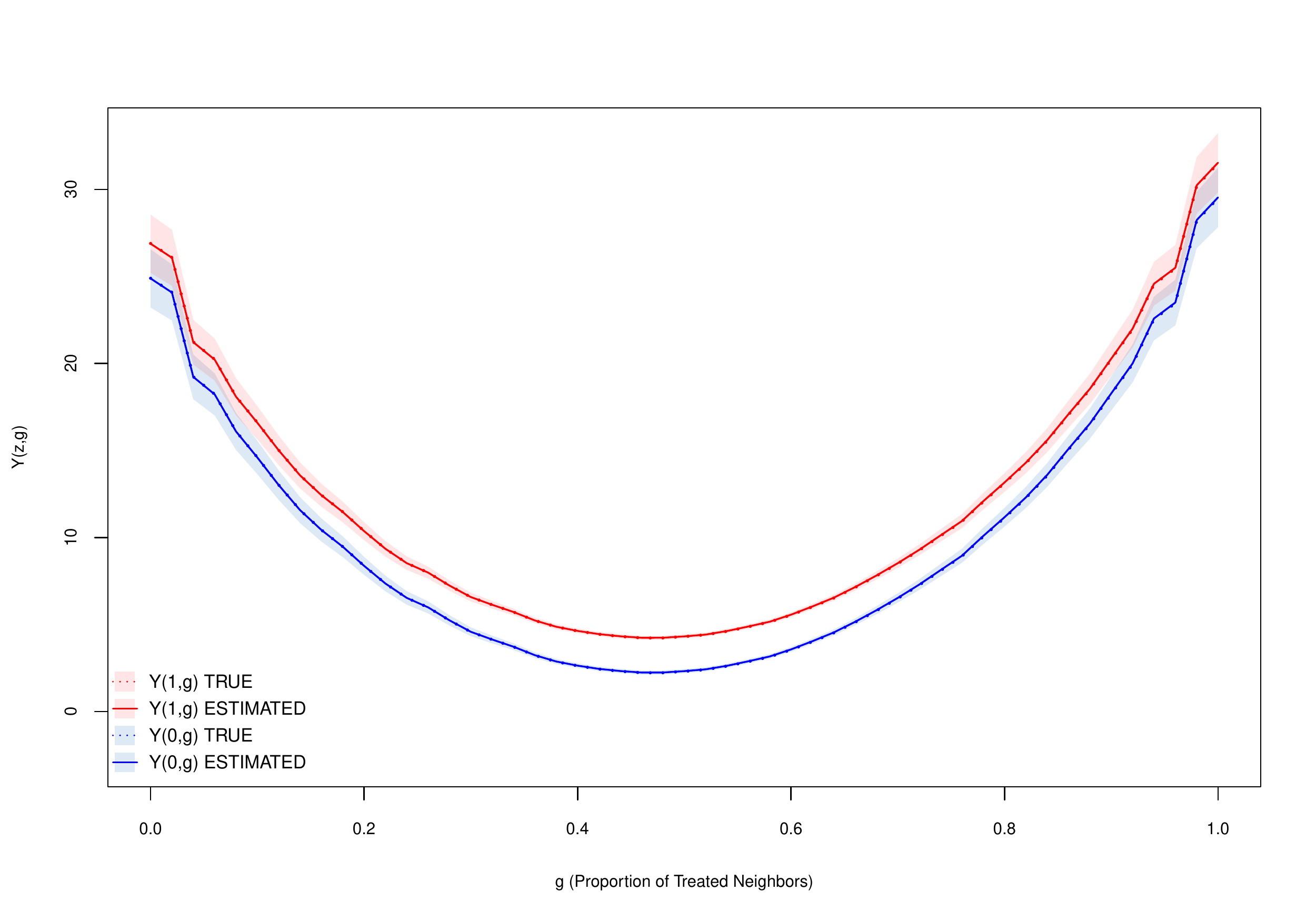}
        \caption{Linear DGP - Linear Model \phantom{0.3cm}}
    \end{subfigure}%
        ~~~~~~~~~~
    \begin{subfigure}[t]{0.45\textwidth}
        \centering
        \includegraphics[width=1.25\textwidth]{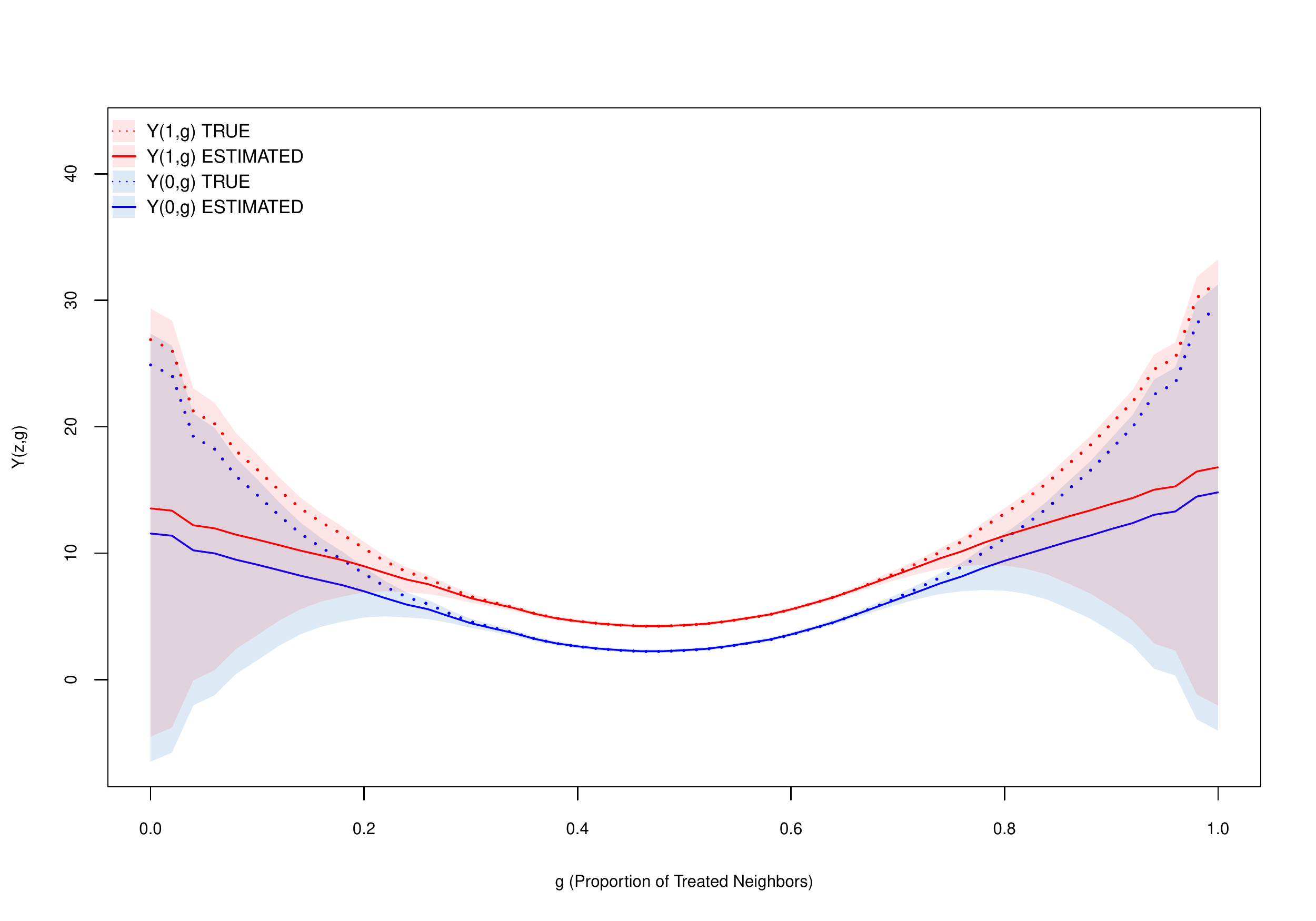}
        \caption{Linear DGP - Splines \phantom{0.3cm}}
    \end{subfigure}
     \end{adjustwidth}
        \begin{adjustwidth}{-0.8cm}{0cm}

   \begin{subfigure}[t]{0.45\textwidth}
        \centering
        \includegraphics[width=1.25\textwidth]{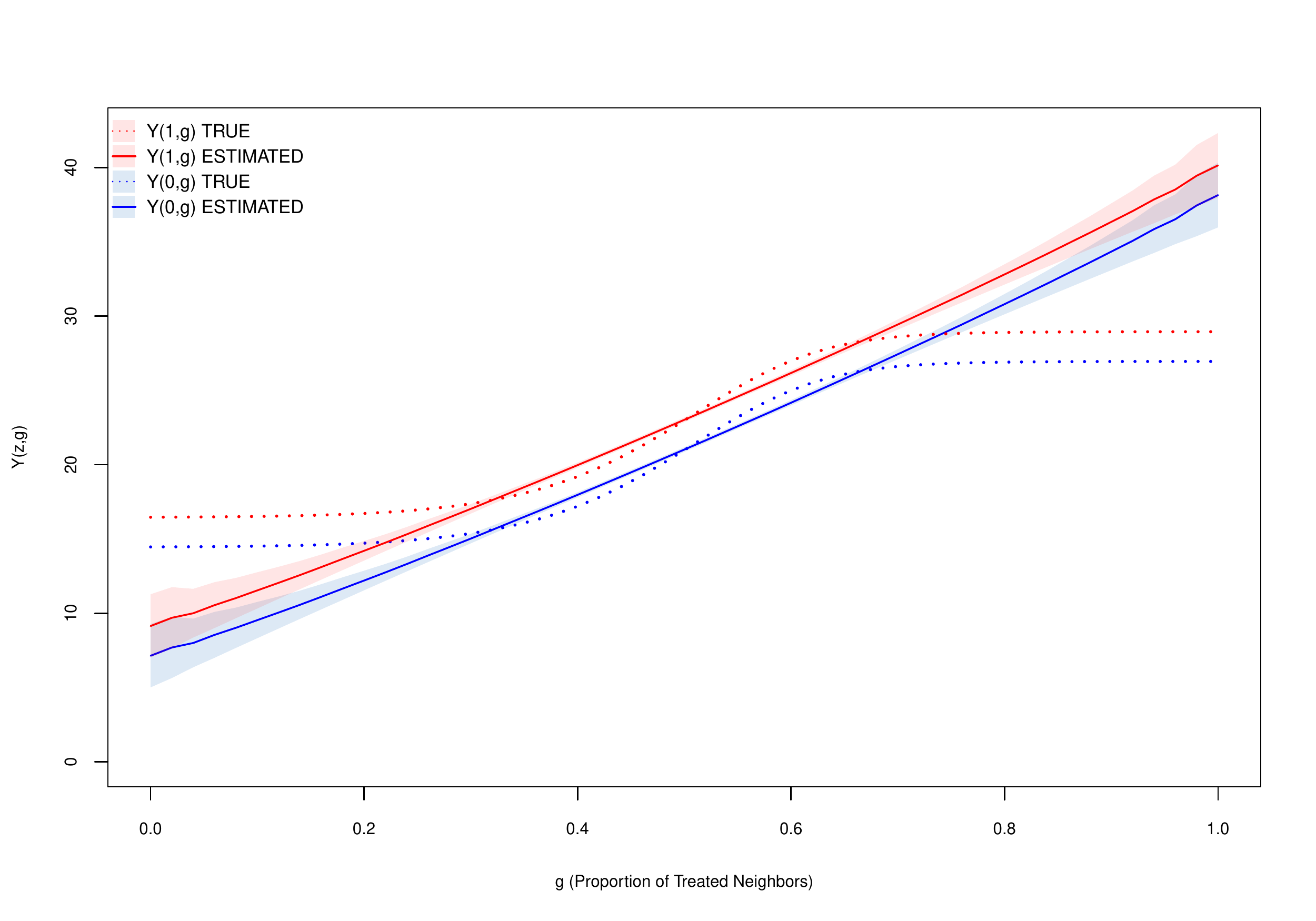}
        \caption{Non-Linear DGP - Linear Model }
    \end{subfigure}%
        ~~~~~~~~~~
    \begin{subfigure}[t]{0.45\textwidth}
        \centering
        \includegraphics[width=1.25\textwidth]{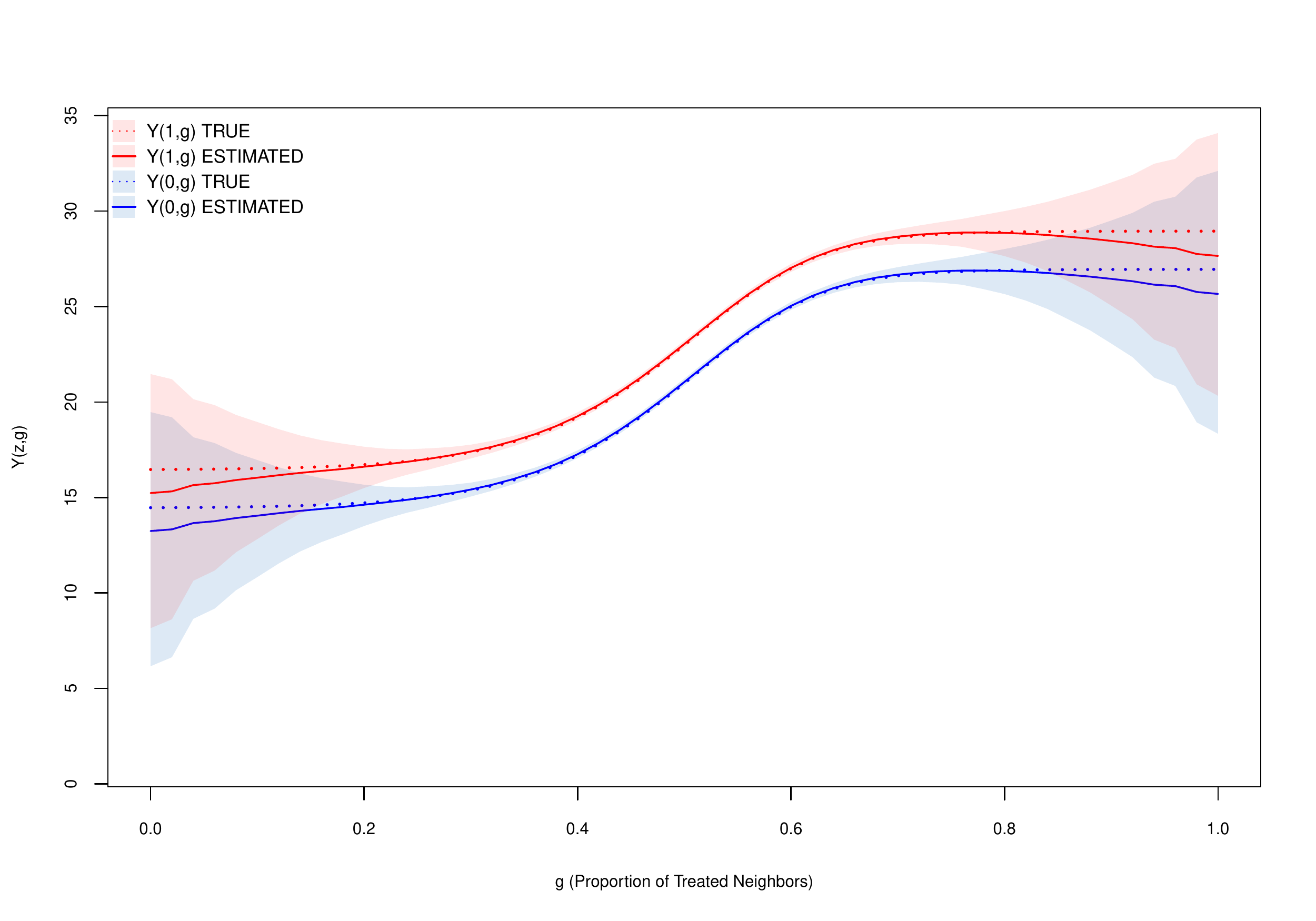}
        \caption{Non-Linear DGP - Splines}
    \end{subfigure}
     \end{adjustwidth}
  \caption{Estimated and true ()ADRF $\mu(z,g; V_g)$ under the scenario where the network is generated from a stochastic block model and the outcome is drawn from a linear model (top) or a non-linear model (bottom) with random effects. Average posterior mean (solid line) and 95 \% credible intervals of $\mu(0,g; V_g)$ (blue) and $\mu(1,g; V_g)$ (red) of linear (left) and splines-based (right) estimator with random effects are represented with the true ADRFs (dotted line). }
\label{fig: sbm}
\end{figure}

In Table \ref{tab:sbm_lin_re} we report the
performance of our estimator for treatment effects $\tau(g, V_g)$, with $g\in\{0,0.1, \ldots, 0.9,1\}$,  and spillover effects  $\delta(g, g-0.1, z, V_g \cup V_{g-0.1})$, with $g\in\{0.1, \ldots, 0.9,1\}$ and $z \in \{0,1\}$,  under the scenario where the network is generated from a stochastic block model and the outcome is drawn from a linear model with random effects.
The estimator that makes use of a linear model, reflecting the data generating model, performs well in terms of bias, mean square error and coverage (close to the nominal level) for all treatment and spillover effects.
For treatment effects the bias and mean square error are constant across different levels of the neighborhood treatment g. On the contrary the mean square error and coverage of the estimator w.r.t spillover effects vary with g and get respectively higher and smaller as g departs from its median value.
In these regions the overall performance of the estimator show severe deterioration
when the linear model is replaced with penalized splines. In fact we can see in Figure \ref{fig: sbm} that splines result in large confidence intervals for the average dose-response function for values of the neighborhood treatment that are not found in less than 2\% of the sample, that is, below the 1st percentile and above the 99th percentile.
This deterioration in areas where data points decrease is a
typical drawback of the use of smoothing splines. Nevertheless,
the splines-based estimator w.r.t. treatment effects still
performs well in the entire range of g, with the bias and mean
square error staying constant. When correlation is not taken into
account, that is, the estimator does not include community random
intercepts, the root MSE of both linear and splines-based
estimators is doubled, with a slight increase in the bias and a
higher increase in the variance. The coverage remains close to the
nominal level, but we do see a small decrease for the linear
estimator. This confirms the robustness of the Bayesian estimation
approach, which generally has very good frequentist properties.
Specifically, because the Bayesian approach does not rely on large
sample approximations and specific estimation of the variance, it
tends to be robust to (small) model missispecifications. In Table
\ref{tab:sbm_lin_nore}, where network is again generated from a
stochastic block model and the outcome is drawn from a linear
model without random effects, we can see that the mean square
error is insensitive to the inclusion of random intercepts in the
estimator and is in general slightly lower but very close to the
one that results from generating correlated outcomes and
estimating the causal effects taking the correlation into account.
Again we see that for the linear estimator the coverage is
slightly higher when we try to estimate a correlation that is not
present. This is not the case for the splines-based estimator.
This means that in general including a random intercept when it is
not needed is not harmful.

\begin{figure}[t!]
        \begin{adjustwidth}{-0.8cm}{0cm}
    \begin{subfigure}[t]{0.45\textwidth}
        \centering
        \includegraphics[width=1.25\textwidth]{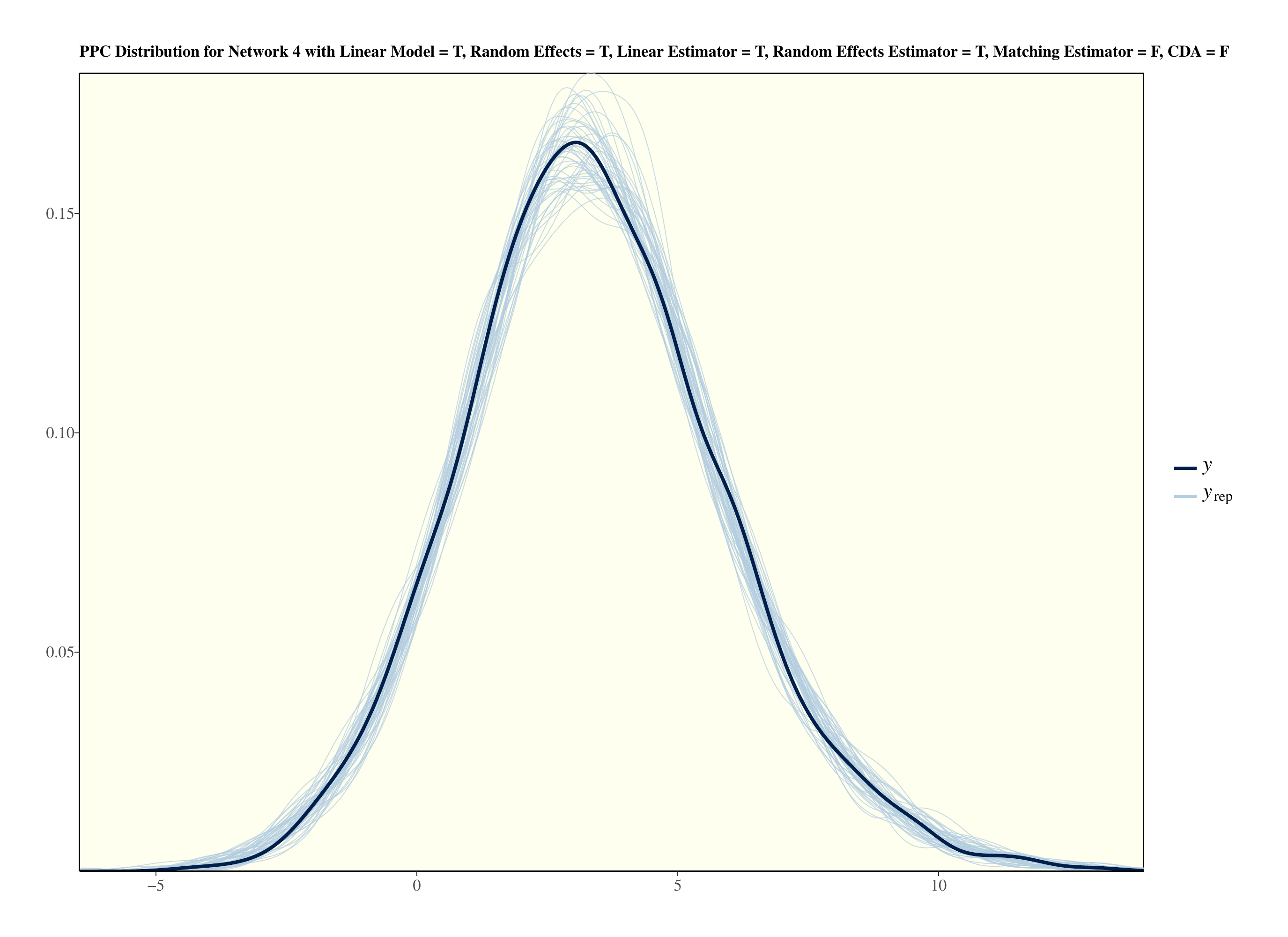}
        \caption{Linear DGP - Linear Model \phantom{0.3cm}}
    \end{subfigure}%
        ~~~~~~~~~~
    \begin{subfigure}[t]{0.45\textwidth}
        \centering
        \includegraphics[width=1.25\textwidth]{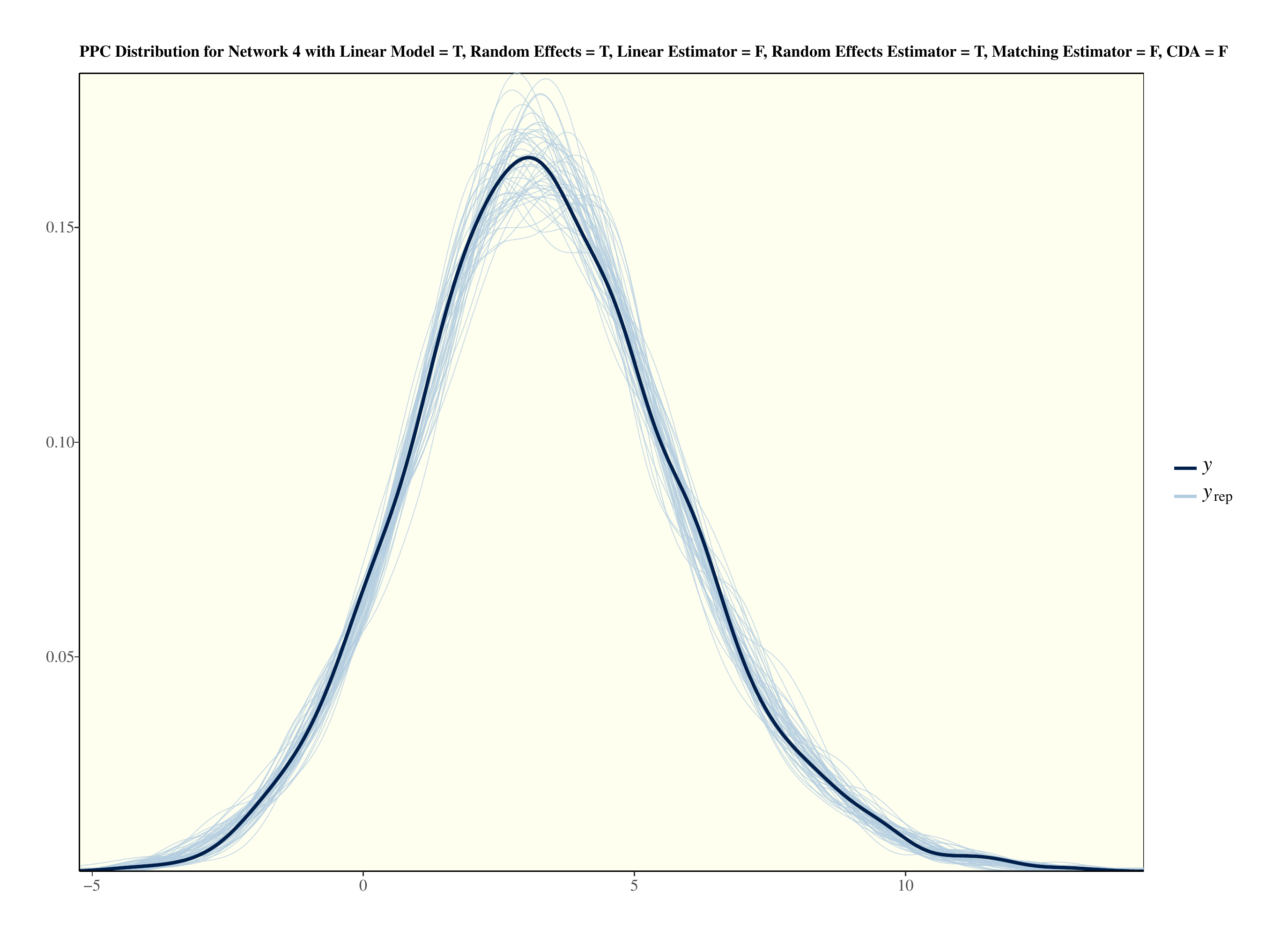}
        \caption{Linear DGP - Splines \phantom{0.3cm}}
    \end{subfigure}
     \end{adjustwidth}
        \begin{adjustwidth}{-0.8cm}{0cm}

   \begin{subfigure}[t]{0.45\textwidth}
        \centering
        \includegraphics[width=1.25\textwidth]{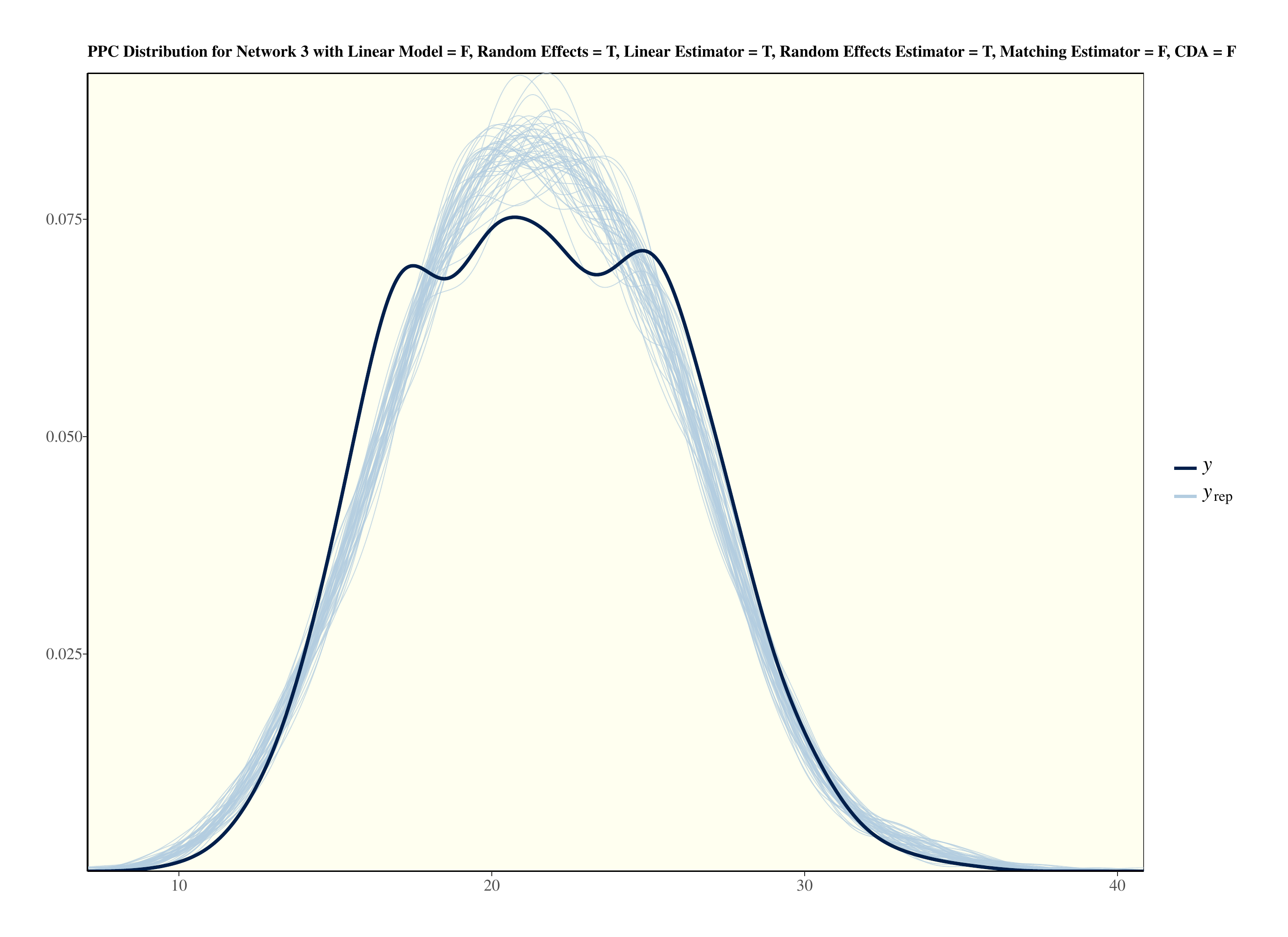}
        \caption{Non-Linear DGP - Linear Model }
    \end{subfigure}%
        ~~~~~~~~~~
    \begin{subfigure}[t]{0.45\textwidth}
        \centering
        \includegraphics[width=1.25\textwidth]{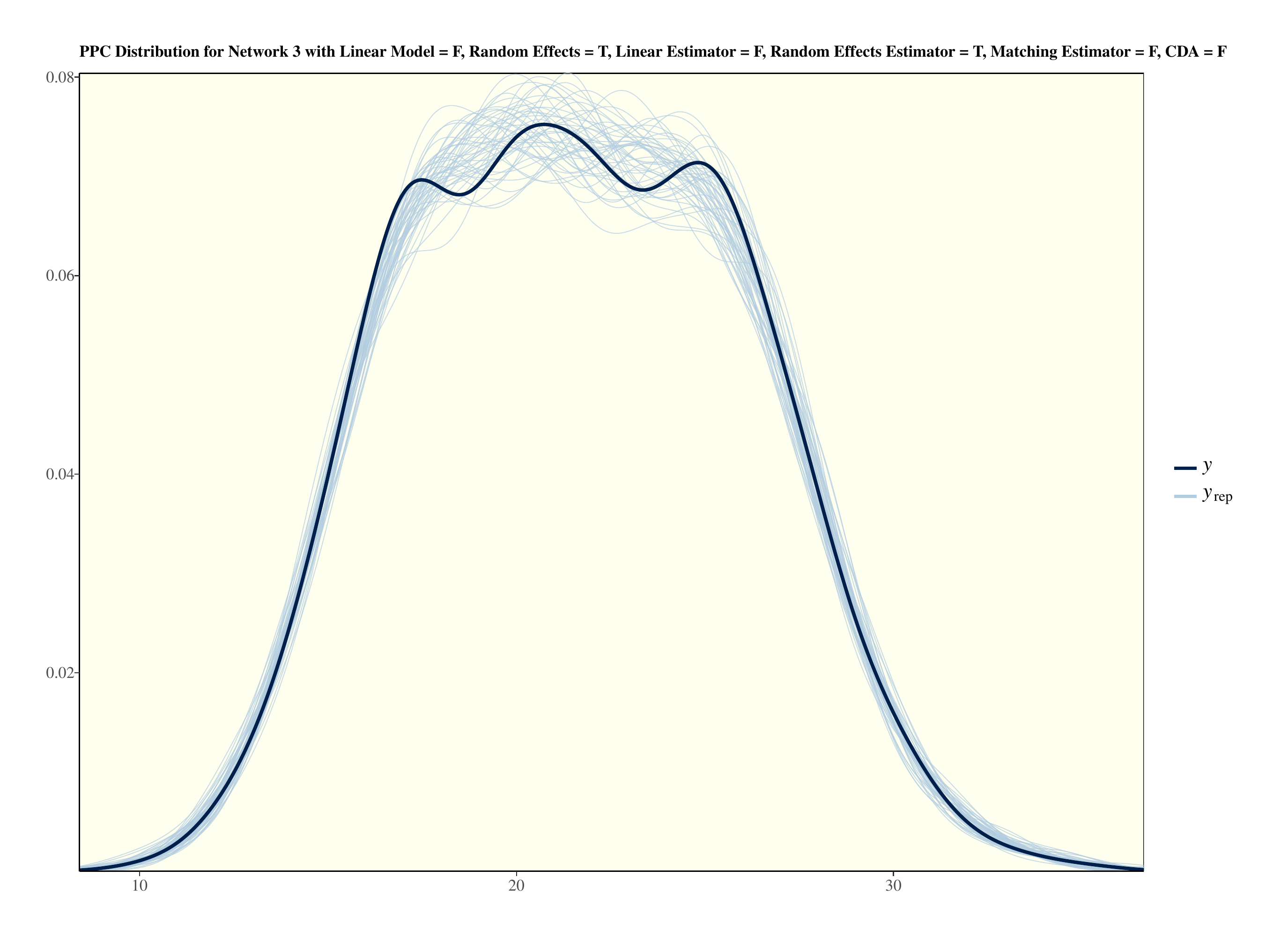}
        \caption{Non-Linear DGP - Splines}
    \end{subfigure}
     \end{adjustwidth}
  \caption{Posterior predictive checks under the scenario where the network is generated from a stochastic block model and the outcome is drawn from a linear model (top) or a non-linear model (bottom) with random effects. Draws from the posterior predictive distribution (blue lines) are represented with the observed distribution (black line). }
\label{fig: sbm_ppc}
\end{figure}

Figure \ref{fig: sbm_ppc} reports an example of posterior predictive checks, where 
draws from the posterior predictive distribution are plotted against the observed outcome distribution.
We could also obtain the posterior predictive distribution of a test statistic (e.g. mean, median, standard deviation, quantiles) and compare it with its observed value.
The figure shows that, as expected, the linear estimator gives rise to a posterior predictive distribution which does not match the observed one when the data generating model is non-linear.
Moreover, the posterior predictive distribution resulting from a splines-based estimator has more variability than the one resulting from the linear estimator.
 
\begin{table}[ht!]
\small
\def\firstrowcolor{}
\def\secondrowcolor{}
\begin{center}
\begin{adjustwidth}{0cm}{0cm}
\begin{tabular}{c<{\hspace{-\tabcolsep}\,\,\,\,}|>{\hspace{-\tabcolsep}}c<{\hspace{-\tabcolsep}}>{\hspace{-\tabcolsep}}c<{\hspace{-\tabcolsep}}>{\hspace{-\tabcolsep}\,}c<{\hspace{-\tabcolsep}}>{\hspace{-\tabcolsep}\,}c<{\hspace{-\tabcolsep}}>{\hspace{-\tabcolsep}\,}c<{\hspace{-\tabcolsep}}>{\hspace{-\tabcolsep}\,}c<{\hspace{-\tabcolsep}}>{\hspace{-\tabcolsep}\,}c<{\hspace{-\tabcolsep}}>{\hspace{-\tabcolsep}\,}c<{\hspace{-\tabcolsep}}>{\hspace{-\tabcolsep}\,}c<{\hspace{-\tabcolsep}}>{\hspace{-\tabcolsep}\,}c<{\hspace{-\tabcolsep}}>{\hspace{-\tabcolsep}}c<{\hspace{-\tabcolsep}}>{\hspace{-\tabcolsep}}c}
 &\multicolumn{6}{c}{{Linear}}&\multicolumn{6}{c}{{Splines}}\\
          &\multicolumn{3}{c}{{RE}}&\multicolumn{3}{c}{{NO RE}}&\multicolumn{3}{c}{{RE}}&\multicolumn{3}{c}{{NO RE}}\\
\cmidrule(lr){2-4}\cmidrule(lr){5-7}\cmidrule(lr){8-10}\cmidrule(lr){11-13}
          &Bias& RMSE&Coverage&Bias& RMSE &Coverage&Bias& RMSE&Coverage&Bias& RMSE&Coverage\\
\hline
$\tau(0, V_0)$&  -0.004&0.083 &0.960&0.001&0.173&0.948&-0.017&0.085&0.960&-0.028&0.177&0.964\\
$\tau(0.1, V_{0.1})$&-0.004  &0.083 &0.962&0.001&0.173&0.942&-0.017&0.085&0.958&-0.028&0.177&0.960\\
$\tau(0.2, V_{0.2})$&  -0.004&0.083 &0.966&0.001&0.173&0.948&-0.017&0.085&0.966&-0.028&0.177&0.962\\
$\tau(0.3, V_{0.3})$&  -0.004& 0.083&0.964&0.001&0.173&0.942&-0.017&0.085&0.958&-0.028&0.177&0.966\\
$\tau(0.4, V_{0.4})$&  -0.004& 0.083&0.968&0.001&0.173&0.942&-0.017&0.085&0.958&-0.028&0.177&0.962\\
$\tau(0.5, V_{0.5})$&  -0.004& 0.083&0.962&0.001&0.173&0.942&-0.017&0.085&0.956&-0.028&0.177&0.962\\
$\tau(0.6, V_{0.6})$&  -0.004& 0.083&0.962&0.001&0.173&0.944&-0.017&0.085&0.960&-0.028&0.177&0.964\\
$\tau(0.7, V_{0.7})$&  -0.004& 0.083&0.964&0.001&0.173&0.946&-0.017&0.085&0.954&-0.028&0.177&0.960\\
$\tau(0.8, V_{0.8})$&  -0.004& 0.083&0.962&0.001&0.173&0.942&-0.017&0.085&0.960&-0.028&0.177&0.966\\
$\tau(0.9, V_{0.9})$&  -0.004& 0.083&0.960&0.001&0.173&0.946&-0.017&0.085&0.958&-0.028&0.177&0.970\\
$\tau(1, V_{1})$&  -0.004& 0.083&0.964&0.001&0.173&0.944&-0.017&0.085&0.960&-0.028&0.177&0.964\\[0.5cm]
\hline
$\delta(0.1,0,0)$&  0.006&0.379 &0.944&0.019&0.801&0.944&7.826&8.778&0.806&8.363&9.173&0.898\\
$\delta(0.2,0.1,0)$&  0.006& 0.237&0.942&0.014&0.501&0.936&4.125&4.696&0.818&4.467&4.983&0.902\\
$\delta(0.3,0.2,0)$&  0.005& 0.153&0.942&0.012&0.322&0.936&1.272&1.575&0.840&1.440&1.789&0.904\\
$\delta(0.4,0.3,0)$&  0.005& 0.090&0.960&0.010&0.191&0.928&1.124&0.209&0.938&0.147&0.318&0.966\\
$\delta(0.5,0.4,0)$& 0.005 &0.042 &0.998&0.008&0.089&0.934&0.013&0.054&1.000&0.019&0.111&0.998\\
$\delta(0.6,0.5,0)$& 0.005 & 0.042&0.992&0.007&0.090&0.940&-0.005&0.056&0.994&0.006&0.110&0.994\\
$\delta(0.7,0.6,0)$&  0.005& 0.093&0.972&0.005&0.198&0.944&0.165&0.284&0.920&-0.240&0.418&0.954\\
$\delta(0.8,0.7,0)$&  0.005& 0.150&0.962&0.003&0.318&0.954&-1.597&1.900&0.856&-1.848&2.187&0.864\\
$\delta(0.9,0.8,0)$&  0.005& 0.236&0.952&0.001&0.500&0.956&-4.600&5.139&0.846&-4.993.&5.484&0.858\\
$\delta(1,0.9,0)$&  0.005& 0.380&0.950&-0.004&0.805&0.952&-8.350&9.274&0.856&-8.940&9.724&0.868\\[0.5cm]
\hline
$\delta(0.1,0,1)$&  0.006&0.379 &0.944&0.019&0.801&0.944&7.826&8.778&0.806&8.363&9.173&0.898\\
$\delta(0.2,0.1,1)$&  0.006& 0.237&0.948&0.014&0.501&0.936&4.125&4.696&0.818&4.467&4.983&0.902\\
$\delta(0.3,0.2,1)$&  0.005& 0.153&0.946&0.012&0.322&0.936&1.272&1.575&0.840&1.440&1.789&0.904\\
$\delta(0.4,0.3,1)$&  0.005& 0.090&0.958&0.010&0.191&0.928&1.124&0.209&0.938&0.147&0.318&0.966\\
$\delta(0.5,0.4,1)$& 0.005 &0.042 &0.994&0.008&0.089&0.934&0.013&0.054&1.000&0.019&0.111&0.998\\
$\delta(0.6,0.5,1)$& 0.005 & 0.042&0.996&0.007&0.090&0.940&-0.005&0.056&0.994&0.006&0.110&0.994\\
$\delta(0.7,0.6,1)$&  0.005& 0.093&0.966&0.005&0.198&0.944&0.165&0.284&0.920&-0.240&0.418&0.954\\
$\delta(0.8,0.7,1)$&  0.005& 0.150&0.950&0.003&0.318&0.954&-1.597&1.900&0.856&-1.848&2.187&0.864\\
$\delta(0.9,0.8,1)$&  0.005& 0.236&0.946&0.001&0.500&0.956&-4.600&5.139&0.846&-4.993.&5.484&0.858\\
$\delta(1,0.9,1)$&  0.005& 0.380&0.946&-0.004&0.805&0.952&-8.350&9.274&0.856&-8.940&9.724&0.868\\
\end{tabular}
\end{adjustwidth}
\end{center}
\caption{Performance of Bayesian GPS estimator for treatment effects $\tau(g, V_g)$, with $g\in\{0,0.1, \ldots, 0.9,1\}$,  and spillover effects  $\delta(g, g-0.1, z, V_g \cup V_{g-0.1})$, with $g\in\{0.1, \ldots, 0.9,1\}$ and $z \in \{0,1\}$,  under the scenario where the network is generated from a stochastic block model and the outcome is drawn from a linear model with random effects.}
\label{tab:sbm_lin_re}
\end{table}%

\begin{table}[ht!]
\small
\def\firstrowcolor{}
\def\secondrowcolor{}
\begin{center}
\begin{adjustwidth}{0cm}{0cm}
\begin{tabular}{c<{\hspace{-\tabcolsep}\,\,\,\,}|>{\hspace{-\tabcolsep}}c<{\hspace{-\tabcolsep}}>{\hspace{-\tabcolsep}}c<{\hspace{-\tabcolsep}}>{\hspace{-\tabcolsep}\,}c<{\hspace{-\tabcolsep}}>{\hspace{-\tabcolsep}\,}c<{\hspace{-\tabcolsep}}>{\hspace{-\tabcolsep}\,}c<{\hspace{-\tabcolsep}}>{\hspace{-\tabcolsep}\,}c<{\hspace{-\tabcolsep}}>{\hspace{-\tabcolsep}\,}c<{\hspace{-\tabcolsep}}>{\hspace{-\tabcolsep}\,}c<{\hspace{-\tabcolsep}}>{\hspace{-\tabcolsep}\,}c<{\hspace{-\tabcolsep}}>{\hspace{-\tabcolsep}\,}c<{\hspace{-\tabcolsep}}>{\hspace{-\tabcolsep}}c<{\hspace{-\tabcolsep}}>{\hspace{-\tabcolsep}}c}
 &\multicolumn{6}{c}{{Linear}}&\multicolumn{6}{c}{{Splines}}\\
          &\multicolumn{3}{c}{{RE}}&\multicolumn{3}{c}{{NO RE}}&\multicolumn{3}{c}{{RE}}&\multicolumn{3}{c}{{NO RE}}\\
\cmidrule(lr){2-4}\cmidrule(lr){5-7}\cmidrule(lr){8-10}\cmidrule(lr){11-13}
          &Bias& RMSE&Coverage&Bias& RMSE &Coverage&Bias& RMSE&Coverage&Bias& RMSE&Coverage\\
\hline
$\tau(0, V_0)$&  -0.002&0.076 &0.978&-0.002&0.076&0.950&-0.014&0.077&0.974&-0.013&0.077&0.974\\
$\tau(0.1, V_{0.1})$&-0.002  &0.076 &0.976&-0.002&0.076&0.950&-0.014&0.077&0.976&-0.013&0.077&0.974\\
$\tau(0.2, V_{0.2})$&  -0.002&0.076 &0.968&-0.002&0.076&0.950&-0.014&0.077&0.974&-0.013&0.077&0.976\\
$\tau(0.3, V_{0.3})$&  -0.002& 0.076&0.980&-0.002&0.076&0.950&-0.014&0.077&0.978&-0.013&0.077&0.982\\
$\tau(0.4, V_{0.4})$&  -0.002& 0.076&0.980&-0.002&0.076&0.950&-0.014&0.077&0.980&-0.013&0.077&0.976\\
$\tau(0.5, V_{0.5})$&  -0.002& 0.076&0.974&-0.002&0.076&0.950&-0.014&0.077&0.980&-0.013&0.077&0.978\\
$\tau(0.6, V_{0.6})$&  -0.002& 0.076&0.980&-0.002&0.076&0.950&-0.014&0.077&0.980&-0.013&0.077&0.976\\
$\tau(0.7, V_{0.7})$&  -0.002& 0.076&0.982&-0.002&0.076&0.950&-0.014&0.077&0.974&-0.013&0.077&0.976\\
$\tau(0.8, V_{0.8})$&  -0.002& 0.076&0.978&-0.002&0.076&0.950&-0.014&0.077&0.972&-0.013&0.077&0.976\\
$\tau(0.9, V_{0.9})$&  -0.002& 0.076&0.968&-0.002&0.076&0.950&-0.014&0.077&0.976&-0.013&0.077&0.976\\
$\tau(1, V_{1})$&  -0.002& 0.076&0.980&-0.002&0.076&0.950&-0.014&0.077&0.976&-0.013&0.077&0.974\\[0.5cm]
\hline
$\delta(0.1,0,0)$&  0.009&0.365&0.944&0.008&0.364&0.952&7.386&8.434&0.800&7.375&8.451&0.794\\
$\delta(0.2,0.1,0)$&  0.008& 0.227&0.942&0.007&0.227&0.946&3.890&4.500&0.820&3.886&4.514&0.798\\
$\delta(0.3,0.2,0)$&  0.008& 0.146&0.942&0.007&0.146&0.948&1.212&1.527&0.830&1.215&1.537&0.840\\
$\delta(0.4,0.3,0)$&  0.007& 0.086&0.960&0.007&0.086&0.946&0.129&0.227&0.914&0.130&0.227&0.920\\
$\delta(0.5,0.4,0)$& 0.007 &0.040 &0.998&0.007&0.040&0.938&0.011&0.054&0.994&0.012&0.054&0.998\\
$\delta(0.6,0.5,0)$& 0.006 & 0.042&0.992&0.007&0.042&0.946&0.005&0.057&0.996&0.004&0.056&0.996\\
$\delta(0.7,0.6,0)$&  0.007& 0.092&0.972&0.007&0.091&0.944&-0.156&0.294&0.906&-0.157&0.296&0.896\\
$\delta(0.8,0.7,0)$&  0.005& 0.146&0.962&0.007&0.146&0.938&-1.498&1.851&0.750&-1.150&1.865&0.756\\
$\delta(0.9,0.8,0)$&  0.005& 0.229&0.952&0.006&0.229&0.942&-4.325&4.930&0.730&-4.322.&4.946&0.738\\
$\delta(1,0.9,0)$&  0.004& 0.369&0.950&0.005&0.368&0.942&-7.873&8.910&0.738&-7.861&8.926&0.740\\[0.5cm]
\hline
$\delta(0.1,0,1)$&  0.009&0.365 &0.944&0.008&0.364&0.952&7.386&8.434&0.802&7.375&8.451&0.792\\
$\delta(0.2,0.1,1)$&  0.008& 0.228&0.948&0.007&0.227&0.946&3.890&4.500&0.818&3.886&4.514&0.806\\
$\delta(0.3,0.2,1)$&  0.008& 0.146&0.946&0.007&0.146&0.948&1.212&1.527&0.822&1.215&1.537&0.834\\
$\delta(0.4,0.3,1)$&  0.007& 0.086&0.958&0.007&0.086&0.946&0.129&0.227&0.922&0.130&0.227&0.918\\
$\delta(0.5,0.4,1)$& 0.007 &0.040&0.994&0.007&0.040&0.938&0.011&0.054&0.996&0.012&0.054&0.996\\
$\delta(0.6,0.5,1)$& 0.007 & 0.042&0.996&0.007&0.042&0.946&0.005&0.057&0.996&0.004&0.056&0.998\\
$\delta(0.7,0.6,1)$&  0.006& 0.091&0.966&0.007&0.091&0.944&-0.156&0.294&0.910&-0.157&0.296&0.902\\
$\delta(0.8,0.7,1)$&  0.005& 0.146&0.950&0.007&0.146&0.938&-1.498&1.851&0.750&-1.150&1.865&0.762\\
$\delta(0.9,0.8,1)$&  0.005& 0.229&0.946&0.006&0.229&0.942&-4.325&4.930&0.726&-4.322.&4.946&0.732\\
$\delta(1,0.9,1)$&  0.004& 0.369&0.946&0.005&0.368&0.942&-7.873&8.910&0.736&-7.861&8.926&0.734\\
\end{tabular}
\end{adjustwidth}
\end{center}
\caption{Performance of Bayesian GPS estimator for treatment effects $\tau(g, V_g)$, with $g\in\{0,0.1, \ldots, 0.9,1\}$,  and spillover effects  $\delta(g, g-0.1, z, V_g \cup V_{g-0.1})$, with $g\in\{0.1, \ldots, 0.9,1\}$ and $z \in \{0,1\}$,  under the scenario where the network is generated from a stochastic block model and the outcome is drawn from a linear model without random effects.}
\label{tab:sbm_lin_nore}
\end{table}%

Tables \ref{tab:sbm_nolin_re} and \ref{tab:sbm_nolin_nore} report the results
 under the scenarios where the network is still generated from a stochastic block model but the outcome is drawn from a
  non-linear model, respectively with and without random effects. In this case the use of a linear model fails to recover the average dose-response function. This results in low performances of the linear estimator w.r.t. the spillover effects. Nevertheless, the linear estimator is still able to yield unbiased estimates of the treatment effects with coverage close to the nominal rate.  On the contrary, the splines-based estimator applied to the non-linear case is able to recover the average dose-response function and to unbiasedly estimate both treatment and spillover effects. In the critical regions of low and high values of the neighborhood treatment,  the performance of the splines-based estimator seems to improve when such estimator is applied to non-linear data rather than linear data. In fact,  here the bias and mean square error of this estimator w.r.t spillover effects at low or high values of g is reduced by a factor of 8-10 compared to the linear case.

\begin{table}[ht!]
\small
\def\firstrowcolor{}
\def\secondrowcolor{}
\begin{center}
\begin{adjustwidth}{0cm}{0cm}
\begin{tabular}{c<{\hspace{-\tabcolsep}\,\,\,\,}|>{\hspace{-\tabcolsep}}c<{\hspace{-\tabcolsep}}>{\hspace{-\tabcolsep}}c<{\hspace{-\tabcolsep}}>{\hspace{-\tabcolsep}\,}c<{\hspace{-\tabcolsep}}>{\hspace{-\tabcolsep}\,}c<{\hspace{-\tabcolsep}}>{\hspace{-\tabcolsep}\,}c<{\hspace{-\tabcolsep}}>{\hspace{-\tabcolsep}\,}c<{\hspace{-\tabcolsep}}>{\hspace{-\tabcolsep}\,}c<{\hspace{-\tabcolsep}}>{\hspace{-\tabcolsep}\,}c<{\hspace{-\tabcolsep}}>{\hspace{-\tabcolsep}\,}c<{\hspace{-\tabcolsep}}>{\hspace{-\tabcolsep}\,}c<{\hspace{-\tabcolsep}}>{\hspace{-\tabcolsep}}c<{\hspace{-\tabcolsep}}>{\hspace{-\tabcolsep}}c}
 &\multicolumn{6}{c}{{Linear}}&\multicolumn{6}{c}{{Splines}}\\
          &\multicolumn{3}{c}{{RE}}&\multicolumn{3}{c}{{NO RE}}&\multicolumn{3}{c}{{RE}}&\multicolumn{3}{c}{{NO RE}}\\
\cmidrule(lr){2-4}\cmidrule(lr){5-7}\cmidrule(lr){8-10}\cmidrule(lr){11-13}
          &Bias& RMSE&Coverage&Bias& RMSE &Coverage&Bias& RMSE&Coverage&Bias& RMSE&Coverage\\
\hline
$\tau(0, V_0)$&  0.001&0.101 &0.982&0.010&0.197&0.946&-0.010&0.082&0.976&-0.017&0.188&0.966\\
$\tau(0.1, V_{0.1})$&0.001  &0.101 &0.984&0.010&0.197&0.946&-0.010&0.082&0.974&-0.017&0.188&0.970\\
$\tau(0.2, V_{0.2})$& 0.001 &0.101 &0.976&0.010&0.197&0.946&-0.010&0.082&0.976&-0.017&0.188&0.964\\
$\tau(0.3, V_{0.3})$& 0.001 & 0.101&0.974&0.010&0.197&0.946&-0.010&0.082&0.976&-0.017&0.188&0.966\\
$\tau(0.4, V_{0.4})$& 0.001 & 0.101&0.976&0.010&0.197&0.946&-0.010&0.082&0.972&-0.017&0.188&0.970\\
$\tau(0.5, V_{0.5})$& 0.001 & 0.101&0.978&0.010&0.197&0.946&-0.010&0.082&0.972&-0.017&0.188&0.966\\
$\tau(0.6, V_{0.6})$& 0.001 & 0.101&0.976&0.010&0.197&0.946&-0.010&0.082&0.974&-0.017&0.188&0.968\\
$\tau(0.7, V_{0.7})$& 0.001 & 0.101&0.974&0.010&0.197&0.946&-0.010&0.082&0.974&-0.017&0.188&0.966\\
$\tau(0.8, V_{0.8})$& 0.001 & 0.101&0.972&0.010&0.197&0.946&-0.010&0.082&0.972&-0.017&0.188&0.970\\
$\tau(0.9, V_{0.9})$& 0.001 & 0.101&0.978&0.010&0.197&0.946&-0.010&0.082&0.976&-0.017&0.188&0.968\\
$\tau(1, V_{1})$& 0.001 & 0.101&0.978&0.010&0.197&0.946&-0.010&0.082&0.976&-0.017&0.188&0.974\\[0.5cm]
\hline
$\delta(0.1,0,0)$&  2.345&2.742 &0.144&2.349&2.807&0.346&0.751&1.195&0.996&0.664&1.594&1.000\\
$\delta(0.2,0.1,0)$&  2.464&2.611 &0.008&2.466&2.638&0.054&0.373&0.731&0.994&0.368&1.051&0.998\\
$\delta(0.3,0.2,0)$&  2.171&2.234 &0.000&2.172&2.246&0.002&0.142&0.392&0.988&0.177&0.624&0.998\\
$\delta(0.4,0.3,0)$&  1.111& 1.149&0.000&1.112&1.155&0.004&0.017&0.164&0.984&0.042&0.294&0.984\\
$\delta(0.5,0.4,0)$&  -0.731&0.739 &0.000&-0.731&0.746&0.002&0.002&0.107&0.972&0.002&0.185&0.992\\
$\delta(0.6,0.5,0)$&  -0.865&0.876 &0.000&-0.866&0.883&0.000&-0.026&0.114&0.980&-0.075&0.206&0.984\\
$\delta(0.7,0.6,0)$&  1.649& 1.694&0.000&1.647&1.683&0.000&0.022&0.177&0.972&0.097&0.328&0.978\\
$\delta(0.8,0.7,0)$&3.084  &3.146 &0.000&3.082&3.147&0.000&-0.091&0.387&0.992&-0.019&0.324&0.992\\
$\delta(0.9,0.8,0)$&  3.491& 3.617&0.000&3.488&3.626&0.004&0.462&0.799&0.994&-0.398&1.060&1.000\\
$\delta(1,0.9,0)$&3.799  & 4.086&0.022&3.794&4.117&0.102&-0.792&1.236&0.996&-0.657&1.596&1.000\\[0.5cm]
\hline
$\delta(0.1,0,0)$&  2.345&2.742 &0.146&2.349&2.807&0.346&0.751&1.195&0.996&0.664&1.594&1.000\\
$\delta(0.2,0.1,0)$&  2.464&2.611 &0.008&2.466&2.638&0.054&0.373&0.731&0.994&0.368&1.051&0.998\\
$\delta(0.3,0.2,0)$&  2.171&2.234 &0.000&2.172&2.246&0.002&0.142&0.392&0.988&0.177&0.624&0.998\\
$\delta(0.4,0.3,0)$&  1.111& 1.148&0.000&1.112&1.155&0.004&0.017&0.164&0.984&0.042&0.294&0.984\\
$\delta(0.5,0.4,0)$&  -0.731&0.739 &0.000&-0.731&0.746&0.002&0.002&0.107&0.972&0.002&0.185&0.992\\
$\delta(0.6,0.5,0)$&  -0.865&0.876 &0.000&-0.866&0.883&0.000&-0.026&0.114&0.980&-0.075&0.206&0.984\\
$\delta(0.7,0.6,0)$&  1.649& 1.694&0.000&1.647&1.683&0.000&0.022&0.177&0.972&0.097&0.328&0.978\\
$\delta(0.8,0.7,0)$&3.084  &3.146 &0.000&3.082&3.147&0.000&-0.091&0.387&0.992&-0.019&0.324&0.992\\
$\delta(0.9,0.8,0)$&  3.491& 3.617&0.000&3.488&3.626&0.004&0.462&0.799&0.994&-0.398&1.060&1.000\\
$\delta(1,0.9,0)$&3.799  & 4.086&0.022&3.794&4.117&0.102&-0.792&1.236&0.996&-0.657&1.596&1.000\\
\end{tabular}
\end{adjustwidth}
\end{center}
\caption{Performance of Bayesian GPS estimator for treatment effects $\tau(g, V_g)$, with $g\in\{0,0.1, \ldots, 0.9,1\}$,  and spillover effects  $\delta(g, g-0.1, z, V_g \cup V_{g-0.1})$, with $g\in\{0.1, \ldots, 0.9,1\}$ and $z \in \{0,1\}$,  under the scenario where the network is generated from a stochastic block model and the outcome is drawn from a non-linear model with random effects.}
\label{tab:sbm_nolin_re}
\end{table}%

\begin{table}[ht!]
\small
\def\firstrowcolor{}
\def\secondrowcolor{}
\begin{center}
\begin{adjustwidth}{0cm}{0cm}
\begin{tabular}{c<{\hspace{-\tabcolsep}\,\,\,\,}|>{\hspace{-\tabcolsep}}c<{\hspace{-\tabcolsep}}>{\hspace{-\tabcolsep}}c<{\hspace{-\tabcolsep}}>{\hspace{-\tabcolsep}\,}c<{\hspace{-\tabcolsep}}>{\hspace{-\tabcolsep}\,}c<{\hspace{-\tabcolsep}}>{\hspace{-\tabcolsep}\,}c<{\hspace{-\tabcolsep}}>{\hspace{-\tabcolsep}\,}c<{\hspace{-\tabcolsep}}>{\hspace{-\tabcolsep}\,}c<{\hspace{-\tabcolsep}}>{\hspace{-\tabcolsep}\,}c<{\hspace{-\tabcolsep}}>{\hspace{-\tabcolsep}\,}c<{\hspace{-\tabcolsep}}>{\hspace{-\tabcolsep}\,}c<{\hspace{-\tabcolsep}}>{\hspace{-\tabcolsep}}c<{\hspace{-\tabcolsep}}>{\hspace{-\tabcolsep}}c}
 &\multicolumn{6}{c}{{Linear}}&\multicolumn{6}{c}{{Splines}}\\
          &\multicolumn{3}{c}{{RE}}&\multicolumn{3}{c}{{NO RE}}&\multicolumn{3}{c}{{RE}}&\multicolumn{3}{c}{{NO RE}}\\
\cmidrule(lr){2-4}\cmidrule(lr){5-7}\cmidrule(lr){8-10}\cmidrule(lr){11-13}
          &Bias& RMSE&Coverage&Bias& RMSE &Coverage&Bias& RMSE&Coverage&Bias& RMSE&Coverage\\
\hline
$\tau(0, V_0)$&  -0.001&0.105 &0.960&-0.002&0.109&0.956&-0.009&0.080&0.978&-0.009&0.080&0.974\\
$\tau(0.1, V_{0.1})$&-0.001  &0.105 &0.958&-0.002&0.109&0.956&-0.009&0.080&0.974&-0.009&0.080&0.972\\
$\tau(0.2, V_{0.2})$&  -0.001&0.105 &0.956&-0.002&0.109&0.956&-0.009&0.080&0.976&-0.009&0.080&0.974\\
$\tau(0.3, V_{0.3})$&  -0.001& 0.105&0.980&-0.002&0.109&0.956&-0.009&0.080&0.976&-0.009&0.080&0.972\\
$\tau(0.4, V_{0.4})$&  -0.001& 0.105&0.960&-0.002&0.109&0.956&-0.009&0.080&0.972&-0.009&0.080&0.974\\
$\tau(0.5, V_{0.5})$&  -0.001& 0.105&0.958&-0.002&0.109&0.956&-0.009&0.080&0.972&-0.009&0.080&0.978\\
$\tau(0.6, V_{0.6})$&  -0.001& 0.105&0.960&-0.002&0.109&0.956&-0.009&0.080&0.974&-0.009&0.080&0.976\\
$\tau(0.7, V_{0.7})$&  -0.001& 0.105&0.964&-0.002&0.109&0.956&-0.009&0.080&0.972&-0.009&0.080&0.966\\
$\tau(0.8, V_{0.8})$&  -0.001& 0.105&0.958&-0.002&0.109&0.956&-0.009&0.080&0.972&-0.009&0.080&0.980\\
$\tau(0.9, V_{0.9})$&  -0.001& 0.105&0.956&-0.002&0.109&0.956&-0.009&0.080&0.976&-0.009&0.080&0.976\\
$\tau(1, V_{1})$&  -0.001& 0.105&0.954&-0.002&0.109&0.956&-0.009&0.080&0.974&-0.009&0.080&0.972\\[0.5cm]
\hline
$\delta(0.1,0,0)$&  2.348&2.699 &0.146&2.352&2.703&0.162&0.819&1.271&1.000&0.811&1.269&1.000\\
$\delta(0.2,0.1,0)$&  2.465&2.595 &0.002&2.467&2.598&0.010&0.416&0.761&0.998&0.412&0.761&0.998\\
$\delta(0.3,0.2,0)$&  2.171&2.227 &0.000&2.172&2.229&0.000&0.136&0.371&0.990&0.133&0.371&0.988\\
$\delta(0.4,0.3,0)$&  1.110& 1.144&0.000&1.111&1.145&0.000&0.021&0.170&0.970&0.020&0.169&0.972\\
$\delta(0.5,0.4,0)$&  -0.731&0.739 &0.000&-0.731&0.739&0.000&0.006&0.103&0.980&0.006&0.102&0.976\\
$\delta(0.6,0.5,0)$&  -0.867&0.876 &0.000&-0.867&0.876&0.000&-0.023&0.111&0.976&-0.023&0.111&0.974\\
$\delta(0.7,0.6,0)$&  1.645& 1.685&0.000&1.645&1.684&0.000&0.007&0.177&0.984&0.007&0.177&0.976\\
$\delta(0.8,0.7,0)$&3.080  &3.134 &0.000&3.079&3.133&0.000&-0.117&0.387&0.986&-0.113&0.386&0.988\\
$\delta(0.9,0.8,0)$&  3.485& 3.596&0.000&3.484&3.595&0.000&-0.500&0.832&0.994&-0.497&0.830&0.996\\
$\delta(1,0.9,0)$&3.792  & 4.044&0.012&3.790&4.043&0.022&-0.864&1.314&0.996&-0.857&1.312&1.000\\[0.5cm]
\hline
$\delta(0.1,0,0)$&  2.348&2.699 &0.142&2.352&2.703&0.162&0.819&1.271&1.000&0.811&1.269&1.000\\
$\delta(0.2,0.1,0)$&  2.465&2.595 &0.008&2.467&2.598&0.010&0.416&0.761&0.998&0.412&0.761&0.998\\
$\delta(0.3,0.2,0)$&  2.171&2.227 &0.000&2.172&2.229&0.000&0.136&0.371&0.980&0.133&0.371&0.988\\
$\delta(0.4,0.3,0)$&  1.110& 1.144&0.000&1.111&1.145&0.000&0.021&0.170&0.970&0.020&0.169&0.972\\
$\delta(0.5,0.4,0)$&  -0.731&0.739 &0.000&-0.731&0.739&0.000&0.006&0.103&0.980&0.006&0.102&0.976\\
$\delta(0.6,0.5,0)$&  -0.867&0.876 &0.000&-0.867&0.876&0.000&-0.023&0.114&0.976&-0.023&0.111&0.974\\
$\delta(0.7,0.6,0)$&  1.645& 1.685&0.000&1.645&1.684&0.000&0.007&0.177&0.972&0.007&0.177&0.976\\
$\delta(0.8,0.7,0)$&3.080  &3.134 &0.000&3.079&3.133&0.000&-0.117&0.387&0.992&-0.113&0.386&0.988\\
$\delta(0.9,0.8,0)$&  3.485& 3.596&0.000&3.484&3.595&0.004&-0.500&0.832&0.994&-0.497&0.830&0.996\\
$\delta(1,0.9,0)$&3.792  & 4.044&0.014&3.790&4.043&0.022&-0.864&1.314&0.996&-0.857&1.312&1.000\\[0.5cm]
\end{tabular}
\end{adjustwidth}
\end{center}
\caption{Performance of Bayesian GPS estimator for treatment effects $\tau(g, V_g)$, with $g\in\{0,0.1, \ldots, 0.9,1\}$,  and spillover effects  $\delta(g, g-0.1, z, V_g \cup V_{g-0.1})$, with $g\in\{0.1, \ldots, 0.9,1\}$ and $z \in \{0,1\}$,  under the scenario where the network is generated from a stochastic block model and the outcome is drawn from a non-linear model without random effects.}
\label{tab:sbm_nolin_nore}
\end{table}%


Tables \ref{tab:ls_lin_re}, \ref{tab:ls_lin_nore}, \ref{tab:ls_nolin_re} and \ref{tab:ls_nolin_nore}
report the results of the scenarios where the network is generated from a latent cluster model, which which is characterized by the presence of homophily along the observed covariates. We essentially see the same patterns shown in Tables
 \ref{tab:sbm_lin_re}, \ref{tab:sbm_lin_nore}, \ref{tab:sbm_nolin_re} and \ref{tab:sbm_nolin_nore}.
 This means that the adjustment for the individual and neighborhood propensity score is able to disentangle the spillover effects from the mechanism of homophily when this is driven by observed characteristics. In addition to this results, which follows from the theory and is expressed by the uncounfoudedness assumption conditional on the two propensity scores \citep{Forastiere:2016}, our empirical results show that the performance of our estimator is not affected by the presence of homophily in the network.
The only difference is in a better performance of the estimators w.r.t the spillover effects at low and high values of g, that is, the decrease in performance happens at a lower rate than what we see with the stochastic block model network. This is presumably due to the distribution of the neighborhood treatment which is more spread for the latent cluster model with more data at the tails of the distribution.


\begin{table}[ht!]
\small
\def\firstrowcolor{}
\def\secondrowcolor{}
\begin{center}
\begin{adjustwidth}{0cm}{0cm}
\begin{tabular}{c<{\hspace{-\tabcolsep}\,\,\,\,}|>{\hspace{-\tabcolsep}}c<{\hspace{-\tabcolsep}}>{\hspace{-\tabcolsep}}c<{\hspace{-\tabcolsep}}>{\hspace{-\tabcolsep}\,}c<{\hspace{-\tabcolsep}}>{\hspace{-\tabcolsep}\,}c<{\hspace{-\tabcolsep}}>{\hspace{-\tabcolsep}\,}c<{\hspace{-\tabcolsep}}>{\hspace{-\tabcolsep}\,}c<{\hspace{-\tabcolsep}}>{\hspace{-\tabcolsep}\,}c<{\hspace{-\tabcolsep}}>{\hspace{-\tabcolsep}\,}c<{\hspace{-\tabcolsep}}>{\hspace{-\tabcolsep}\,}c<{\hspace{-\tabcolsep}}>{\hspace{-\tabcolsep}\,}c<{\hspace{-\tabcolsep}}>{\hspace{-\tabcolsep}}c<{\hspace{-\tabcolsep}}>{\hspace{-\tabcolsep}}c}
 &\multicolumn{6}{c}{{Linear}}&\multicolumn{6}{c}{{Splines}}\\
          &\multicolumn{3}{c}{{RE}}&\multicolumn{3}{c}{{NO RE}}&\multicolumn{3}{c}{{RE}}&\multicolumn{3}{c}{{NO RE}}\\
\cmidrule(lr){2-4}\cmidrule(lr){5-7}\cmidrule(lr){8-10}\cmidrule(lr){11-13}
          &Bias& RMSE&Coverage&Bias& RMSE &Coverage&Bias& RMSE&Coverage&Bias& RMSE&Coverage\\
\hline
$\tau(0, V_0)$&  0.002&0.077&0.982&0.008&0.176&0.948&-0.011&0.078&0.980&-0.023&0.178&0.978\\
$\tau(0.1, V_{0.1})$&0.002  &0.077 &0.980&0.008&0.176&0.948&-0.011&0.078&0.976&-0.023&0.178&0.980\\
$\tau(0.2, V_{0.2})$&  0.002&0.077 &0.982&0.008&0.176&0.948&-0.011&0.078&0.976&-0.023&0.178&0.980\\
$\tau(0.3, V_{0.3})$&  0.002& 0.077&0.982&0.008&0.176&0.948&-0.011&0.078&0.976&-0.023&0.178&0.978\\
$\tau(0.4, V_{0.4})$&  0.002& 0.077&0.982&0.008&0.176&0.948&-0.011&0.078&0.980&-0.023&0.178&0.974\\
$\tau(0.5, V_{0.5})$&  0.002& 0.077&0.982&0.008&0.176&0.948&-0.011&0.078&0.978&-0.023&0.178&0.974\\
$\tau(0.6, V_{0.6})$&  0.002& 0.077&0.982&0.008&0.176&0.948&-0.011&0.078&0.978&-0.023&0.178&0.970\\
$\tau(0.7, V_{0.7})$&  0.002& 0.077&0.982&0.008&0.176&0.948&-0.011&0.078&0.974&-0.023&0.178&0.974\\
$\tau(0.8, V_{0.8})$&  0.002& 0.077&0.980&0.008&0.176&0.948&-0.011&0.078&0.970&-0.023&0.178&0.980\\
$\tau(0.9, V_{0.9})$&  0.002& 0.077&0.984&0.008&0.176&0.948&-0.011&0.078&0.978&-0.023&0.178&0.978\\
$\tau(1, V_{1})$&  0.002& 0.077&0.980&0.008&0.176&0.948&-0.011&0.078&0.974&-0.023&0.178&0.976\\[0.5cm]
\hline
$\delta(0.1,0,0)$&  -0.018&0.273 &0.956&-0.041&0.801&0.944&5.128&5.860&0.840&5.727&6.384&0.922\\
$\delta(0.2,0.1,0)$& -0.011& 0.165&0.962&-0.025&0.501&0.936&2.207&2.591&0.850&2.525&2.926&0.930\\
$\delta(0.3,0.2,0)$&  -0.007& 0.109&0.974&-0.017&0.322&0.936&0.599&0.786&0.876&0.724&0.985&0.938\\
$\delta(0.4,0.3,0)$&  -0.004& 0.069&0.974&-0.010&0.191&0.928&0.064&0.133&0.958&0.085&0.227&0.980\\
$\delta(0.5,0.4,0)$& -0.002 &0.032&1.000&-0.004&0.089&0.934&0.005&0.046&0.998&0.008&0.093&1.000\\
$\delta(0.6,0.5,0)$& 0.001 & 0.033&0.998&0.002&0.090&0.940&0.004&0.049&1.000&-0.001&0.091&1.000\\
$\delta(0.7,0.6,0)$&  0.004& 0.071&0.982&0.009&0.198&0.944&-0.072&0.154&0.968&-0.112&0.251&0.964\\
$\delta(0.8,0.7,0)$&  0.005& 0.110&0.970&0.014&0.318&0.954&-0.777&0.972&0.808&-0.960&1.199&0.916\\
$\delta(0.9,0.8,0)$&  0.011& 0.179&0.968&0.025&0.500&0.956&-2.843&3.241&0.776&-3.253&3.650&0.896\\
$\delta(1,0.9,0)$&  0.016& 0.262&0.960&0.037&0.805&0.952&-5.348&6.010&0.786&-5.967&6.554&0.902\\[0.5cm]
\hline
$\delta(0.1,0,1)$&  -0.018&0.273 &0.948&-0.041&0.801&0.944&5.128&5.860&0.840&5.727&6.384&0.922\\
$\delta(0.2,0.1,1)$&  -0.011& 0.165&0.942&-0.025&0.501&0.936&2.207&2.591&0.852&2.524&2.926&0.928\\
$\delta(0.3,0.2,1)$&  -0.007& 0.109&0.948&-0.017&0.322&0.936&0.599&0.786&0.866&0.724&0.985&0.934\\
$\delta(0.4,0.3,1)$&  -0.004& 0.069&0.952&-0.010&0.191&0.928&0.064&0.133&0.968&0.085&0.227&0.980\\
$\delta(0.5,0.4,1)$& -0.002 &0.032 &0.944&-0.004&0.089&0.934&0.005&0.046&0.998&0.008&0.093&1.000\\
$\delta(0.6,0.5,1)$& 0.001 & 0.033&0.962&0.002&0.090&0.940&0.004&0.049&0.998&-0.001&0.091&0.998\\
$\delta(0.7,0.6,1)$&  0.004& 0.071&0.954&0.009&0.198&0.944&-0.072&0.154&0.972&-0.112&0.251&0.962\\
$\delta(0.8,0.7,1)$&  0.005& 0.110&0.960&0.014&0.318&0.954&-0.777&0.972&0.806&-0.960&1.199&0.922\\
$\delta(0.9,0.8,1)$&  0.011& 0.179&0.960&0.025&0.500&0.956&-2.843&3.241&0.780&-3.253&3.650&0.898\\
$\delta(1,0.9,1)$&  0.016& 0.262&0.960&0.037&0.805&0.952&-5.348&6.010&0.786&-5.967&6.554&0.902\\
\end{tabular}
\end{adjustwidth}
\end{center}
\caption{Performance of Bayesian GPS estimator for treatment effects $\tau(g, V_g)$, with $g\in\{0,0.1, \ldots, 0.9,1\}$,  and spillover effects  $\delta(g, g-0.1, z, V_g \cup V_{g-0.1})$, with $g\in\{0.1, \ldots, 0.9,1\}$ and $z \in \{0,1\}$,  under the scenario where the network is generated from a latent cluster model and the outcome is drawn from a linear model with random effects.}
\label{tab:ls_lin_re}
\end{table}%

\begin{figure}[t!]
        \begin{adjustwidth}{-0.8cm}{0cm}
    \begin{subfigure}[t]{0.45\textwidth}
        \centering
        \includegraphics[width=1.25\textwidth]{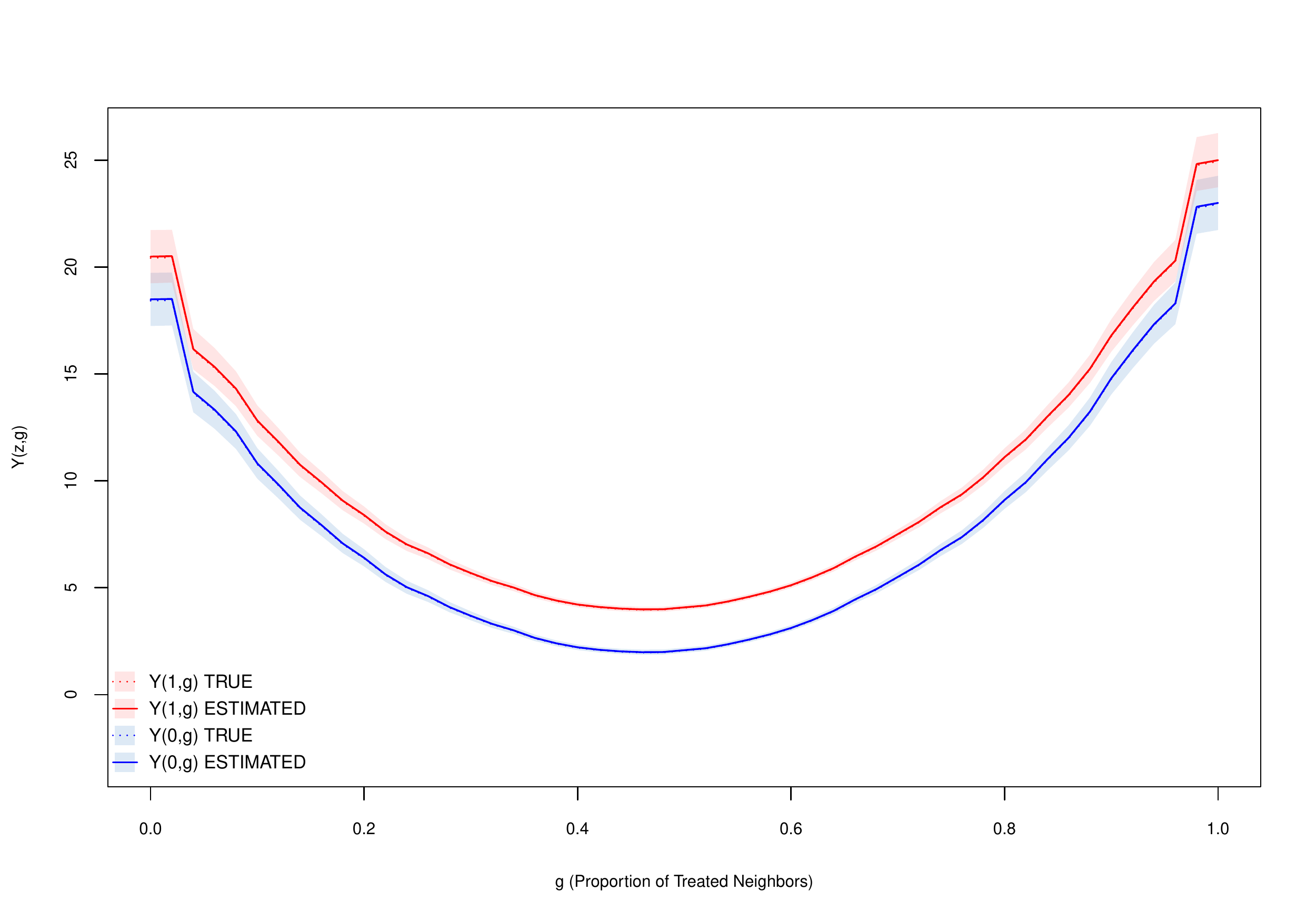}
        \caption{Linear DGP - Linear Model \phantom{0.3cm}}
    \end{subfigure}%
        ~~~~~~~~~~
    \begin{subfigure}[t]{0.45\textwidth}
        \centering
        \includegraphics[width=1.25\textwidth]{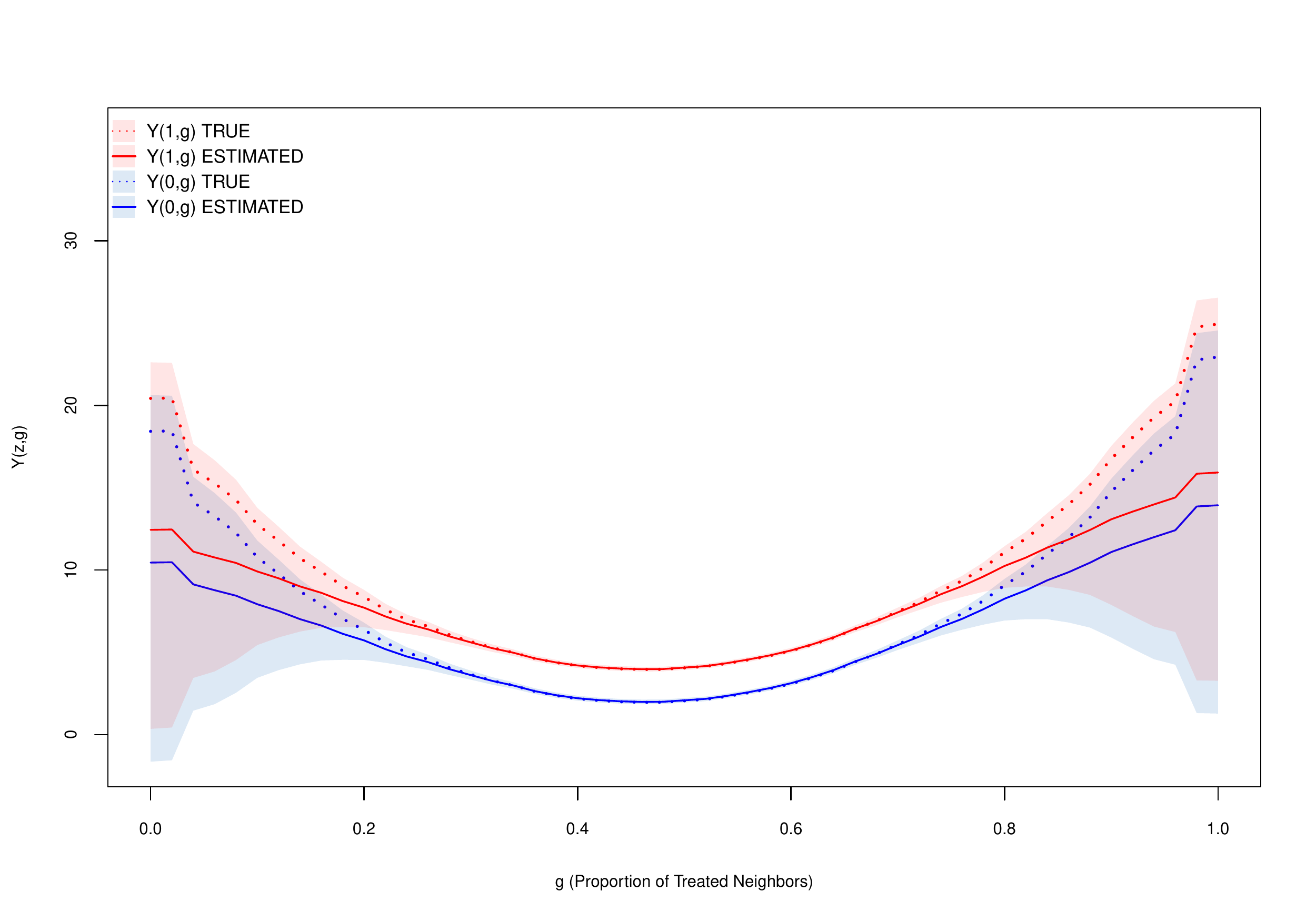}
        \caption{Linear DGP - Splines \phantom{0.3cm}}
    \end{subfigure}
     \end{adjustwidth}
        \begin{adjustwidth}{-0.8cm}{0cm}

   \begin{subfigure}[t]{0.45\textwidth}
        \centering
        \includegraphics[width=1.25\textwidth]{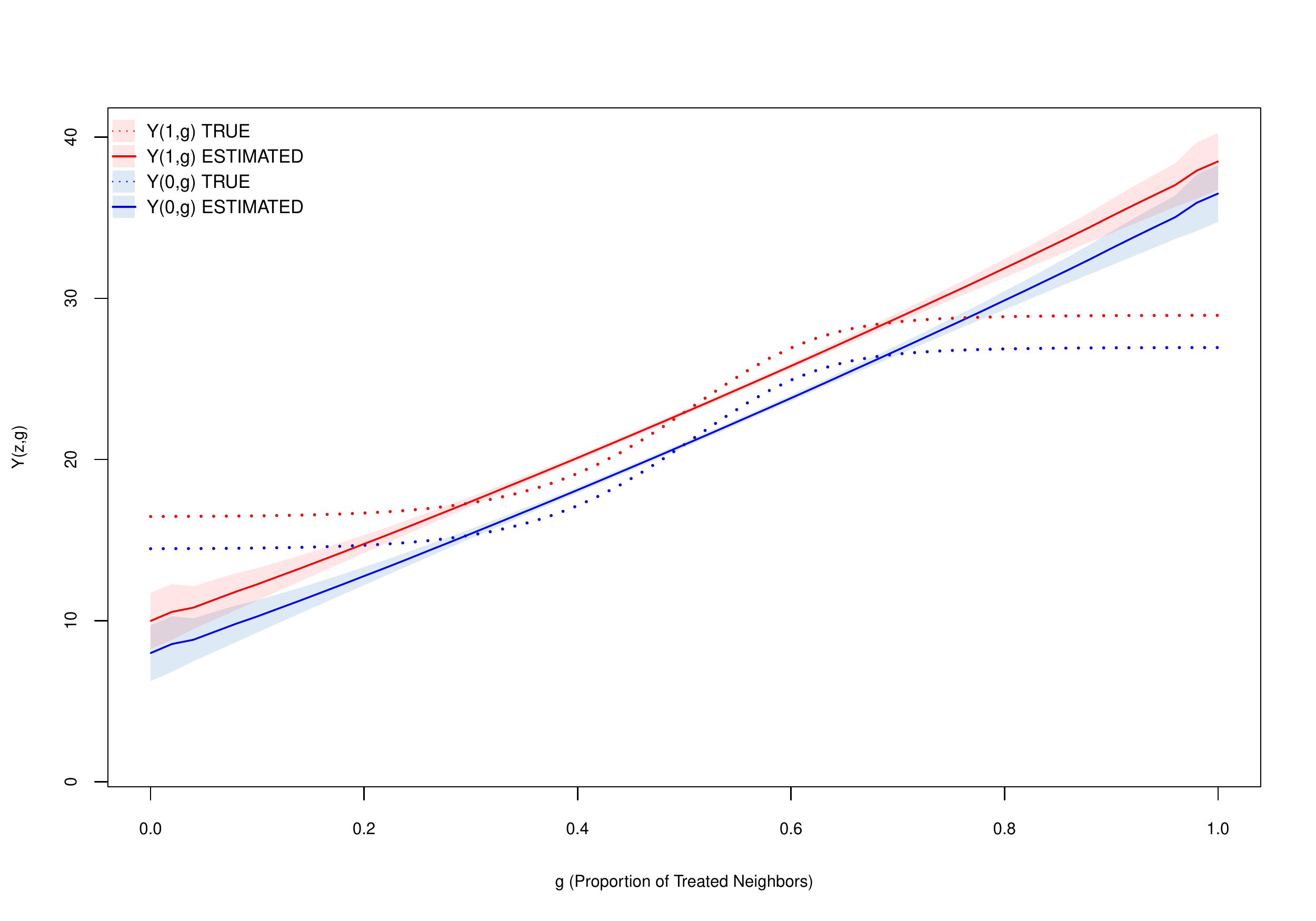}
        \caption{Non-Linear DGP - Linear Model }
    \end{subfigure}%
        ~~~~~~~~~~
    \begin{subfigure}[t]{0.45\textwidth}
        \centering
        \includegraphics[width=1.25\textwidth]{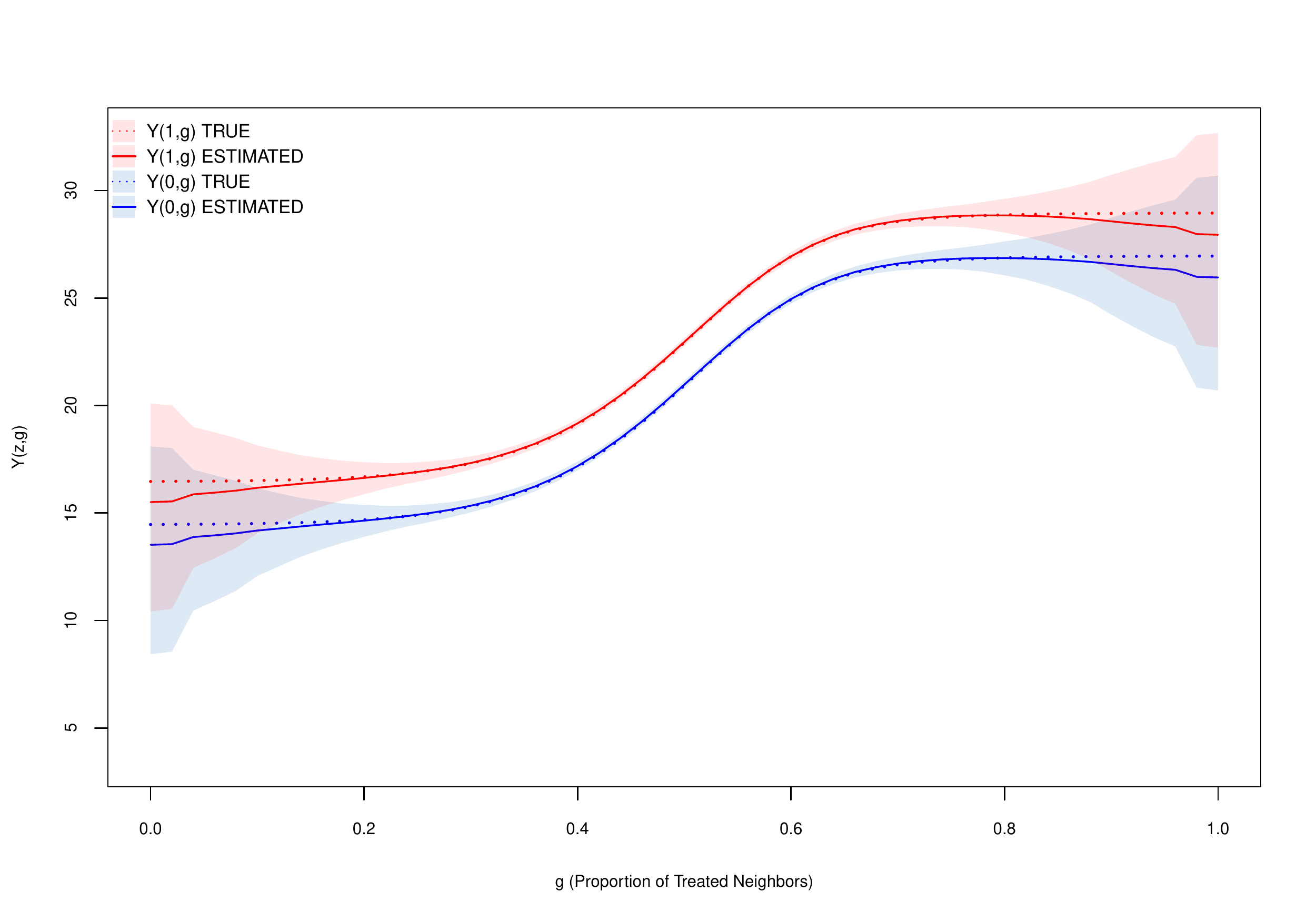}
        \caption{Non-Linear DGP - Splines}
    \end{subfigure}
     \end{adjustwidth}
  \caption{Estimated and true ()ADRF $\mu(z,g; V_g)$ under the scenario where the network is generated from a latent cluster model and the outcome is drawn from a linear model (top) or a non-linear model (bottom) with random effects. Average posterior mean (solid line) and 95 \% credible intervals of $\mu(0,g; V_g)$ (blue) and $\mu(1,g; V_g)$ (red) of linear (left) and splines-based (right) estimator with random effects are represented with the true ADRFs (dotted line). }
\label{fig: ls}
\end{figure}

\begin{table}[ht!]
\small
\def\firstrowcolor{}
\def\secondrowcolor{}
\begin{center}
\begin{adjustwidth}{0cm}{0cm}
\begin{tabular}{c<{\hspace{-\tabcolsep}\,\,\,\,}|>{\hspace{-\tabcolsep}}c<{\hspace{-\tabcolsep}}>{\hspace{-\tabcolsep}}c<{\hspace{-\tabcolsep}}>{\hspace{-\tabcolsep}\,}c<{\hspace{-\tabcolsep}}>{\hspace{-\tabcolsep}\,}c<{\hspace{-\tabcolsep}}>{\hspace{-\tabcolsep}\,}c<{\hspace{-\tabcolsep}}>{\hspace{-\tabcolsep}\,}c<{\hspace{-\tabcolsep}}>{\hspace{-\tabcolsep}\,}c<{\hspace{-\tabcolsep}}>{\hspace{-\tabcolsep}\,}c<{\hspace{-\tabcolsep}}>{\hspace{-\tabcolsep}\,}c<{\hspace{-\tabcolsep}}>{\hspace{-\tabcolsep}\,}c<{\hspace{-\tabcolsep}}>{\hspace{-\tabcolsep}}c<{\hspace{-\tabcolsep}}>{\hspace{-\tabcolsep}}c}
 &\multicolumn{6}{c}{{Linear}}&\multicolumn{6}{c}{{Splines}}\\
          &\multicolumn{3}{c}{{RE}}&\multicolumn{3}{c}{{NO RE}}&\multicolumn{3}{c}{{RE}}&\multicolumn{3}{c}{{NO RE}}\\
\cmidrule(lr){2-4}\cmidrule(lr){5-7}\cmidrule(lr){8-10}\cmidrule(lr){11-13}
          &Bias& RMSE&Coverage&Bias& RMSE &Coverage&Bias& RMSE&Coverage&Bias& RMSE&Coverage\\
\hline
$\tau(0, V_0)$&  0.002&0.080 &0.980&0.001&0.080&0.952&-0.010&0.082&0.974&-0.010&0.082&0.980\\
$\tau(0.1, V_{0.1})$&0.001  &0.080 &0.978&0.001&0.080&0.952&-0.010&0.082&0.980&-0.010&0.082&0.978\\
$\tau(0.2, V_{0.2})$&  0.001&0.080 &0.976&0.001&0.080&0.952&-0.010&0.082&0.970&-0.010&0.082&0.976\\
$\tau(0.3, V_{0.3})$&  0.001& 0.080&0.974&0.001&0.080&0.952&-0.010&0.082&0.976&-0.010&0.082&0.978\\
$\tau(0.4, V_{0.4})$&  0.002& 0.080&0.976&0.001&0.080&0.952&-0.010&0.082&0.976&-0.010&0.082&0.978\\
$\tau(0.5, V_{0.5})$&  0.002& 0.080&0.978&0.001&0.080&0.952&-0.010&0.082&0.972&-0.010&0.082&0.978\\
$\tau(0.6, V_{0.6})$&  0.001& 0.080&0.980&0.001&0.080&0.952&-0.010&0.082&0.976&-0.010&0.082&0.980\\
$\tau(0.7, V_{0.7})$&  0.001& 0.080&0.976&0.001&0.080&0.952&-0.010&0.082&0.980&-0.010&0.082&0.976\\
$\tau(0.8, V_{0.8})$&  0.001& 0.080&0.980&0.001&0.080&0.952&-0.010&0.082&0.982&-0.010&0.082&0.978\\
$\tau(0.9, V_{0.9})$&  0.002& 0.080&0.978&0.001&0.080&0.952&-0.010&0.082&0.972&-0.010&0.082&0.972\\
$\tau(1, V_{1})$&  0.001& 0.080&0.972&0.001&0.080&0.952&-0.010&0.082&0.976&-0.010&0.082&0.974\\[0.5cm]
\hline
$\delta(0.1,0,0)$&  0.002&0.274 &0.940&0.003&0.274&0.930&5.331&6.122&0.816&5.317&6.092&0.828\\
$\delta(0.2,0.1,0)$& 0.002& 0.165&0.942&0.003&0.265&0.928&2.303&2.739&0.818&2.301&2.731&0.822\\
$\delta(0.3,0.2,0)$&  0.002& 0.110&0.948&0.003&0.110&0.924&0.645&0.871&0.834&0.646&0.873&0.834\\
$\delta(0.4,0.3,0)$&  0.002& 0.069&0.970&0.003&0.069&0.928&0.076&0.153&0.926&0.077&0.154&0.938\\
$\delta(0.5,0.4,0)$& 0.002 &0.032&0.998&0.003&0.032&0.950&0.004&0.045&0.998&0.004&0.045&0.998\\
$\delta(0.6,0.5,0)$& 0.002 & 0.031&1.000&0.003&0.031&0.944&0.004&0.045&1.000&0.004&0.046&0.998\\
$\delta(0.7,0.6,0)$&  0.002& 0.067&0.982&0.003&0.068&0.960&-0.066&0.153&0.946&-0.065&0.152&0.954\\
$\delta(0.8,0.7,0)$&  0.002& 0.108&0.972&0.003&0.108&0.962&-0.794&1.108&0.784&-0.789&1.100&0.794\\
$\delta(0.9,0.8,0)$&  0.002& 0.175&0.960&0.003&0.175&0.950&-2.918&3.362&0.732&-2.903&3.339&0.734\\
$\delta(1,0.9,0)$&  0.002& 0.260&0.960&0.003&0.261&0.946&-5.543&6.272&0.736&-5.196&6.232&0.744\\[0.5cm]
\hline
$\delta(0.1,0,1)$&  0.002&0.274 &0.940&0.003&0.274&0.930&5.331&6.122&0.816&5.317&6.092&0.828\\
$\delta(0.2,0.1,1)$& 0.002& 0.165&0.942&0.003&0.265&0.928&2.303&2.739&0.818&2.301&2.731&0.822\\
$\delta(0.3,0.2,1)$&  0.002& 0.110&0.948&0.003&0.110&0.924&0.645&0.871&0.834&0.646&0.873&0.834\\
$\delta(0.4,0.3,1)$&  0.002& 0.069&0.970&0.003&0.069&0.928&0.076&0.153&0.926&0.077&0.154&0.938\\
$\delta(0.5,0.4,1)$& 0.002 &0.032&0.998&0.003&0.032&0.950&0.004&0.045&0.998&0.004&0.045&0.998\\
$\delta(0.6,0.5,1)$& 0.002 & 0.031&1.000&0.003&0.031&0.944&0.004&0.045&1.000&0.004&0.046&0.998\\
$\delta(0.7,0.6,1)$&  0.002& 0.067&0.982&0.003&0.068&0.960&-0.066&0.153&0.946&-0.065&0.152&0.954\\
$\delta(0.8,0.7,1)$&  0.002& 0.108&0.972&0.003&0.108&0.962&-0.794&1.108&0.784&-0.789&1.100&0.794\\
$\delta(0.9,0.8,1)$&  0.002& 0.175&0.960&0.003&0.175&0.950&-2.918&3.362&0.732&-2.903&3.339&0.734\\
$\delta(1,0.9,1)$&  0.002& 0.260&0.960&0.003&0.261&0.946&-5.543&6.272&0.736&-5.196&6.232&0.744\\[0.5cm]
\end{tabular}
\end{adjustwidth}
\end{center}
\caption{Performance of Bayesian GPS estimator for treatment effects $\tau(g, V_g)$, with $g\in\{0,0.1, \ldots, 0.9,1\}$,  and spillover effects  $\delta(g, g-0.1, z, V_g \cup V_{g-0.1})$, with $g\in\{0.1, \ldots, 0.9,1\}$ and $z \in \{0,1\}$,  under the scenario where the network is generated from a latent cluster model and the outcome is drawn from a linear model without random effects.}
\label{tab:ls_lin_nore}
\end{table}%

\begin{table}[ht!]
\small
\def\firstrowcolor{}
\def\secondrowcolor{}
\begin{center}
\begin{adjustwidth}{0cm}{0cm}
\begin{tabular}{c<{\hspace{-\tabcolsep}\,\,\,\,}|>{\hspace{-\tabcolsep}}c<{\hspace{-\tabcolsep}}>{\hspace{-\tabcolsep}}c<{\hspace{-\tabcolsep}}>{\hspace{-\tabcolsep}\,}c<{\hspace{-\tabcolsep}}>{\hspace{-\tabcolsep}\,}c<{\hspace{-\tabcolsep}}>{\hspace{-\tabcolsep}\,}c<{\hspace{-\tabcolsep}}>{\hspace{-\tabcolsep}\,}c<{\hspace{-\tabcolsep}}>{\hspace{-\tabcolsep}\,}c<{\hspace{-\tabcolsep}}>{\hspace{-\tabcolsep}\,}c<{\hspace{-\tabcolsep}}>{\hspace{-\tabcolsep}\,}c<{\hspace{-\tabcolsep}}>{\hspace{-\tabcolsep}\,}c<{\hspace{-\tabcolsep}}>{\hspace{-\tabcolsep}}c<{\hspace{-\tabcolsep}}>{\hspace{-\tabcolsep}}c}
 &\multicolumn{6}{c}{{Linear}}&\multicolumn{6}{c}{{Splines}}\\
          &\multicolumn{3}{c}{{RE}}&\multicolumn{3}{c}{{NO RE}}&\multicolumn{3}{c}{{RE}}&\multicolumn{3}{c}{{NO RE}}\\
\cmidrule(lr){2-4}\cmidrule(lr){5-7}\cmidrule(lr){8-10}\cmidrule(lr){11-13}
          &Bias& RMSE&Coverage&Bias& RMSE &Coverage&Bias& RMSE&Coverage&Bias& RMSE&Coverage\\
\hline
$\tau(0, V_0)$&  -0.004&0.120 &0.962&-0.011&0.201&0.938&-0.013&0.086&0.964&-0.034&0.184&0.960\\
$\tau(0.1, V_{0.1})$&-0.004  &0.120 &0.962&-0.011&0.201&0.938&-0.013&0.086&0.968&-0.034&0.184&0.962\\
$\tau(0.2, V_{0.2})$& -0.004 &0.120 &0.956&-0.011&0.201&0.938&-0.013&0.086&0.966&-0.034&0.184&0.964\\
$\tau(0.3, V_{0.3})$& -0.004 & 0.120&0.960&-0.011&0.201&0.938&-0.013&0.086&0.958&-0.034&0.184&0.966\\
$\tau(0.4, V_{0.4})$& -0.004 & 0.120&0.956&-0.011&0.201&0.938&-0.013&0.086&0.962&-0.034&0.184&0.968\\
$\tau(0.5, V_{0.5})$& -0.004 & 0.120&0.960&-0.011&0.201&0.938&-0.013&0.086&0.966&-0.034&0.184&0.964\\
$\tau(0.6, V_{0.6})$& -0.004 & 0.120&0.960&-0.011&0.201&0.938&-0.013&0.086&0.956&-0.034&0.184&0.954\\
$\tau(0.7, V_{0.7})$& -0.004 & 0.120&0.960&-0.011&0.201&0.938&-0.013&0.086&0.964&-0.034&0.184&0.956\\
$\tau(0.8, V_{0.8})$& -0.004 & 0.120&0.962&-0.011&0.201&0.938&-0.013&0.086&0.968&-0.034&0.184&0.964\\
$\tau(0.9, V_{0.9})$& -0.004 & 0.120&0.964&-0.011&0.201&0.938&-0.013&0.086&0.954&-0.034&0.184&0.958\\
$\tau(1, V_{1})$& -0.004 & 0.120&0.962&-0.011&0.201&0.938&-0.013&0.086&0.962&-0.034&0.184&0.958\\[0.5cm]
\hline
$\delta(0.1,0,0)$&  2.231&2.450&0.064&2.255&2.522&0.198&0.621&1.052&0.996&0.535&1.323&1.000\\
$\delta(0.2,0.1,0)$&  2.331&2.404 &0.000&2.344&2.436&0.012&0.285&0.596&0.988&0.379&0.814&0.998\\
$\delta(0.3,0.2,0)$&  1.999&2.033 &0.000&2.007&2.051&0.000&0.069&0.269&0.986&0.081&0.445&0.990\\
$\delta(0.4,0.3,0)$&  0.886& 0.914&0.000&0.890&0.926&0.008&0.014&0.147&0.978&0.041&0.251&0.988\\
$\delta(0.5,0.4,0)$&  -0.982&0.986 &0.000&-0.982&0.988&0.000&-0.004&0.110&0.974&-0.011&0.195&0.994\\
$\delta(0.6,0.5,0)$&  -1.110&1.115 &0.000&-1.114&1.121&0.000&-0.020&0.108&0.978&-0.057&0.203&0.986\\
$\delta(0.7,0.6,0)$&  1.371& 1.400&0.000&1.362&1.397&0.000&0.031&0.155&0.970&0.046&0.380&0.986\\
$\delta(0.8,0.7,0)$&2.260 &2.797 &0.000&2.747&2.791&0.000&-0.060&0.282&0.990&0.006&0.456&0.990\\
$\delta(0.9,0.8,0)$&  3.155& 3.235&0.000&3.135&3.232&0.002&0.346&0.673&0.984&-0.255&0.888&0.996\\
$\delta(1,0.9,0)$&3.888 & 3.543&0.008&3.360&3.548&0.024&-0.641&1.053&0.994&-0.523&1.296&1.000\\[0.5cm]
\hline
$\delta(0.1,0,1)$&  2.231&2.450&0.064&2.255&2.522&0.198&0.621&1.052&0.996&0.535&1.323&1.000\\
$\delta(0.2,0.1,1)$&  2.331&2.404 &0.000&2.344&2.436&0.012&0.285&0.596&0.988&0.379&0.814&0.998\\
$\delta(0.3,0.2,1)$&  1.999&2.033 &0.000&2.007&2.051&0.000&0.069&0.269&0.986&0.081&0.445&0.990\\
$\delta(0.4,0.3,1)$&  0.886& 0.914&0.000&0.890&0.926&0.008&0.014&0.147&0.978&0.041&0.251&0.988\\
$\delta(0.5,0.4,1)$&  -0.982&0.986 &0.000&-0.982&0.988&0.000&-0.004&0.110&0.974&-0.011&0.195&0.994\\
$\delta(0.6,0.5,1)$&  -1.110&1.115 &0.000&-1.114&1.121&0.000&-0.020&0.108&0.978&-0.057&0.203&0.986\\
$\delta(0.7,0.6,1)$&  1.371& 1.400&0.000&1.362&1.397&0.000&0.031&0.155&0.970&0.046&0.380&0.986\\
$\delta(0.8,0.7,1)$&2.260 &2.797 &0.000&2.747&2.791&0.000&-0.060&0.282&0.990&0.006&0.456&0.990\\
$\delta(0.9,0.8,1)$&  3.155& 3.235&0.000&3.135&3.232&0.002&0.346&0.673&0.984&-0.255&0.888&0.996\\
$\delta(1,0.9,1)$&3.888 & 3.543&0.008&3.360&3.548&0.024&-0.641&1.053&0.994&-0.523&1.296&1.000\\
\end{tabular}
\end{adjustwidth}
\end{center}
\caption{Performance of Bayesian GPS estimator for treatment effects $\tau(g, V_g)$, with $g\in\{0,0.1, \ldots, 0.9,1\}$,  and spillover effects  $\delta(g, g-0.1, z, V_g \cup V_{g-0.1})$, with $g\in\{0.1, \ldots, 0.9,1\}$ and $z \in \{0,1\}$,  under the scenario where the network is generated from a latent cluster model and the outcome is drawn from a non-linear model with random effects.}
\label{tab:ls_nolin_re}
\end{table}%

\begin{table}[ht!]
\small
\def\firstrowcolor{}
\def\secondrowcolor{}
\begin{center}
\begin{adjustwidth}{0cm}{0cm}
\begin{tabular}{c<{\hspace{-\tabcolsep}\,\,\,\,}|>{\hspace{-\tabcolsep}}c<{\hspace{-\tabcolsep}}>{\hspace{-\tabcolsep}}c<{\hspace{-\tabcolsep}}>{\hspace{-\tabcolsep}\,}c<{\hspace{-\tabcolsep}}>{\hspace{-\tabcolsep}\,}c<{\hspace{-\tabcolsep}}>{\hspace{-\tabcolsep}\,}c<{\hspace{-\tabcolsep}}>{\hspace{-\tabcolsep}\,}c<{\hspace{-\tabcolsep}}>{\hspace{-\tabcolsep}\,}c<{\hspace{-\tabcolsep}}>{\hspace{-\tabcolsep}\,}c<{\hspace{-\tabcolsep}}>{\hspace{-\tabcolsep}\,}c<{\hspace{-\tabcolsep}}>{\hspace{-\tabcolsep}\,}c<{\hspace{-\tabcolsep}}>{\hspace{-\tabcolsep}}c<{\hspace{-\tabcolsep}}>{\hspace{-\tabcolsep}}c}
 &\multicolumn{6}{c}{{Linear}}&\multicolumn{6}{c}{{Splines}}\\
          &\multicolumn{3}{c}{{RE}}&\multicolumn{3}{c}{{NO RE}}&\multicolumn{3}{c}{{RE}}&\multicolumn{3}{c}{{NO RE}}\\
\cmidrule(lr){2-4}\cmidrule(lr){5-7}\cmidrule(lr){8-10}\cmidrule(lr){11-13}
          &Bias& RMSE&Coverage&Bias& RMSE &Coverage&Bias& RMSE&Coverage&Bias& RMSE&Coverage\\
\hline
$\tau(0, V_0)$&  -0.005&0.105 &0.982&-0.005&0.106&0.974&-0.008&0.077&0.972&-0.008&0.077&0.972\\
$\tau(0.1, V_{0.1})$&-0.005  &0.105 &0.976&-0.005&0.106&0.974&-0.008&0.077&0.970&-0.008&0.077&0.970\\
$\tau(0.2, V_{0.2})$& -0.005 &0.105 &0.978&-0.005&0.106&0.974&-0.008&0.077&0.974&-0.008&0.077&0.972\\
$\tau(0.3, V_{0.3})$& -0.005 & 0.105&0.982&-0.005&0.106&0.974&-0.008&0.077&0.972&-0.008&0.077&0.972\\
$\tau(0.4, V_{0.4})$& -0.005 & 0.105&0.978&-0.005&0.106&0.974&-0.008&0.077&0.972&-0.008&0.077&0.976\\
$\tau(0.5, V_{0.5})$& -0.005 & 0.105&0.978&-0.005&0.106&0.974&-0.008&0.077&0.972&-0.008&0.077&0.972\\
$\tau(0.6, V_{0.6})$& -0.005 & 0.105&0.984&-0.005&0.106&0.974&-0.008&0.077&0.972&-0.008&0.077&0.970\\
$\tau(0.7, V_{0.7})$& -0.005 & 0.105&0.976&-0.005&0.106&0.974&-0.008&0.077&0.974&-0.008&0.077&0.974\\
$\tau(0.8, V_{0.8})$& -0.005 & 0.105&0.982&-0.005&0.106&0.974&-0.008&0.077&0.974&-0.008&0.077&0.970\\
$\tau(0.9, V_{0.9})$& -0.005 & 0.105&0.974&-0.005&0.106&0.974&-0.008&0.077&0.976&-0.008&0.077&0.972\\
$\tau(1, V_{1})$& -0.005 & 0.105&0.980&-0.005&0.106&0.974&-0.008&0.077&0.968&-0.008&0.077&0.974\\[0.5cm]
\hline
$\delta(0.1,0,0)$&  2.210&2.448&0.064&2.209&2.450&0.090&0.621&0.960&0.998&0.629&0.965&0.998\\
$\delta(0.2,0.1,0)$&  2.320&2.401 &0.000&2.320&2.401&0.000&0.274&0.524&0.998&0.278&0.523&0.996\\
$\delta(0.3,0.2,0)$&  1.997&2.033 &0.000&1.996&2.033&0.000&0.075&0.274&0.976&0.076&0.274&0.982\\
$\delta(0.4,0.3,0)$&  0.889& 0.918&0.000&0.888&0.918&0.000&0.008&0.135&0.982&0.007&0.135&0.976\\
$\delta(0.5,0.4,0)$&  -0.975&0.979 &0.000&-0.975&0.979&0.000&-0.004&0.105&0.980&-0.004&0.106&0.980\\
$\delta(0.6,0.5,0)$&  -1.099&1.105 &0.000&-1.100&1.105&0.000&-0.018&0.104&0.974&-0.017&0.104&0.968\\
$\delta(0.7,0.6,0)$&  1.388& 1.421&0.000&1.388&1.421&0.000&0.023&0.146&0.978&0.022&0.145&0.976\\
$\delta(0.8,0.7,0)$&2.782 &2.823 &0.000&2.782&2.823&0.000&-0.058&0.275&0.986&-0.058&0.276&0.988\\
$\delta(0.9,0.8,0)$&  3.188& 3.276&0.000&3.188&3.277&0.000&0.329&0.610&1.000&-0.331&0.613&0.996\\
$\delta(1,0.9,0)$&3.426 & 3.593&0.008&3.426&3.595&0.012&-0.639&0.967&1.000&-0.6483&0.971&1.000\\[0.5cm]
\hline
$\delta(0.1,0,1)$&  2.210&2.448&0.064&2.209&2.450&0.090&0.621&0.960&0.998&0.629&0.965&0.998\\
$\delta(0.2,0.1,1)$&  2.320&2.401 &0.000&2.320&2.401&0.000&0.274&0.524&0.998&0.278&0.523&0.996\\
$\delta(0.3,0.2,1)$&  1.997&2.033 &0.000&1.996&2.033&0.000&0.075&0.274&0.976&0.076&0.274&0.982\\
$\delta(0.4,0.3,1)$&  0.889& 0.918&0.000&0.888&0.918&0.000&0.008&0.135&0.982&0.007&0.135&0.976\\
$\delta(0.5,0.4,1)$&  -0.975&0.979 &0.000&-0.975&0.979&0.000&-0.004&0.105&0.980&-0.004&0.106&0.980\\
$\delta(0.6,0.5,1)$&  -1.099&1.105 &0.000&-1.100&1.105&0.000&-0.018&0.104&0.974&-0.017&0.104&0.968\\
$\delta(0.7,0.6,1)$&  1.388& 1.421&0.000&1.388&1.421&0.000&0.023&0.146&0.978&0.022&0.145&0.976\\
$\delta(0.8,0.7,1)$&2.782 &2.823 &0.000&2.782&2.823&0.000&-0.058&0.275&0.986&-0.058&0.276&0.988\\
$\delta(0.9,0.8,1)$&  3.188& 3.276&0.000&3.188&3.277&0.000&0.329&0.610&1.000&-0.331&0.613&0.996\\
$\delta(1,0.9,1)$&3.426 & 3.593&0.008&3.426&3.595&0.012&-0.639&0.967&1.000&-0.6483&0.971&1.000\\[0.5cm]
\end{tabular}
\end{adjustwidth}
\end{center}
\caption{Performance of Bayesian GPS estimator for treatment effects $\tau(g, V_g)$, with $g\in\{0,0.1, \ldots, 0.9,1\}$,  and spillover effects  $\delta(g, g-0.1, z, V_g \cup V_{g-0.1})$, with $g\in\{0.1, \ldots, 0.9,1\}$ and $z \in \{0,1\}$,  under the scenario where the network is generated from a latent cluster model and the outcome is drawn from a non-linear model without random effects.}
\label{tab:ls_nolin_nore}
\end{table}%

\begin{figure}[ht!]
        \centering
        \includegraphics[width=0.8\textwidth]{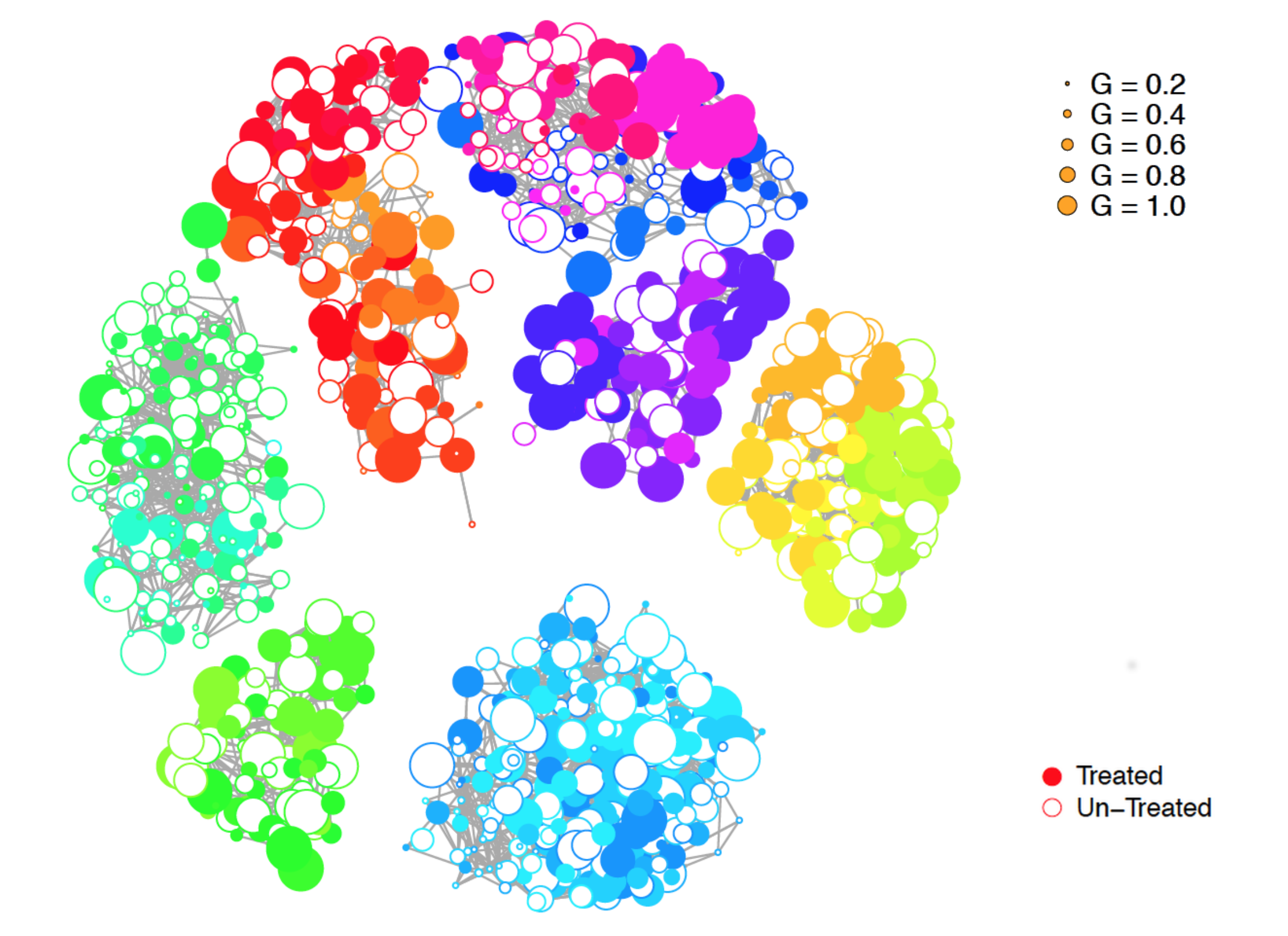}
        \caption{Add Health network}
        \label{fig: AH}
\end{figure}

Finally, Tables \ref{tab:ah_lin_re}, \ref{tab:ah_lin_nore},
\ref{tab:ah_nolin_re} and \ref{tab:ah_nolin_nore}  report the
results of the scenarios where the network is taken from the Add
Health data, which is likely affected by homophily. Here the
average degree is lower than the other two cases, resulting in a
distribution of the neighborhood treatment which is sparse and
concentrated around few values (see Appendix). Moreover, the size of the
communities used to generate outcome correlation is bigger, since
we define a community as the cluster of students in the same
school and with the same grade. 
In Figure \ref{fig: AH} we can see the Add Health social network, with colors representing the communities, the node symbol representing the individual treatment and the node size representing the neighborhood treatment.

In Tables \ref{tab:ah_lin_re} and
\ref{tab:ah_nolin_re} we can see that the mean square error of the
estimators with random effects is slightly higher than in the
previous scenarios, especially for treatment effects.
In addition, when outcome correlation is not
taken into account the MSE does not increase as much as in the previous cases. This is presumably due to the bigger size of the communities (the average size is 20.8 units).
Nevertheless, for the linear estimator the coverage is still smaller when random intercepts are not included.

\begin{table}[ht!]
\small
\def\firstrowcolor{}
\def\secondrowcolor{}
\begin{center}
\begin{adjustwidth}{0cm}{0cm}
\begin{tabular}{c<{\hspace{-\tabcolsep}\,\,\,\,}|>{\hspace{-\tabcolsep}}c<{\hspace{-\tabcolsep}}>{\hspace{-\tabcolsep}}c<{\hspace{-\tabcolsep}}>{\hspace{-\tabcolsep}\,}c<{\hspace{-\tabcolsep}}>{\hspace{-\tabcolsep}\,}c<{\hspace{-\tabcolsep}}>{\hspace{-\tabcolsep}\,}c<{\hspace{-\tabcolsep}}>{\hspace{-\tabcolsep}\,}c<{\hspace{-\tabcolsep}}>{\hspace{-\tabcolsep}\,}c<{\hspace{-\tabcolsep}}>{\hspace{-\tabcolsep}\,}c<{\hspace{-\tabcolsep}}>{\hspace{-\tabcolsep}\,}c<{\hspace{-\tabcolsep}}>{\hspace{-\tabcolsep}\,}c<{\hspace{-\tabcolsep}}>{\hspace{-\tabcolsep}}c<{\hspace{-\tabcolsep}}>{\hspace{-\tabcolsep}}c}
 &\multicolumn{6}{c}{{Linear}}&\multicolumn{6}{c}{{Splines}}\\
          &\multicolumn{3}{c}{{RE}}&\multicolumn{3}{c}{{NO RE}}&\multicolumn{3}{c}{{RE}}&\multicolumn{3}{c}{{NO RE}}\\
\cmidrule(lr){2-4}\cmidrule(lr){5-7}\cmidrule(lr){8-10}\cmidrule(lr){11-13}
          &Bias& RMSE&Coverage&Bias& RMSE &Coverage&Bias& RMSE&Coverage&Bias& RMSE&Coverage\\
\hline
$\tau(0, V_0)$&  0.007&0.132&0.988&0.006&0.141&0.958&-0.002&0.130&0.986&-0.002&0.140&0.992\\
$\tau(0.1, V_{0.1})$&0.007  &0.132 &0.986&0.006&0.141&0.958&-0.003&0.130&0.988&-0.001&0.140&0.990\\
$\tau(0.2, V_{0.2})$&  0.007&0.132 &0.986&0.006&0.141&0.958&-0.002&0.130&0.984&-0.001&0.140&0.990\\
$\tau(0.3, V_{0.3})$&  0.007& 0.132&0.990&0.006&0.141&0.958&-0.002&0.130&0.988&-0.002&0.140&0.992\\
$\tau(0.4, V_{0.4})$&  0.007& 0.132&0.988&0.006&0.141&0.958&-0.002&0.130&0.988&-0.002&0.140&0.984\\
$\tau(0.5, V_{0.5})$&  0.007& 0.132&0.988&0.006&0.141&0.958&-0.002&0.130&0.988&-0.002&0.140&0.992\\
$\tau(0.6, V_{0.6})$&  0.007& 0.132&0.988&0.006&0.141&0.958&-0.002&0.130&0.986&-0.002&0.140&0.996\\
$\tau(0.7, V_{0.7})$&  0.007& 0.132&0.988&0.006&0.141&0.958&-0.002&0.130&0.984&-0.002&0.140&0.992\\
$\tau(0.8, V_{0.8})$&  0.007& 0.132&0.988&0.006&0.141&0.958&-0.002&0.130&0.988&-0.002&0.140&0.990\\
$\tau(0.9, V_{0.9})$&  0.007& 0.132&0.986&0.006&0.141&0.958&-0.002&0.130&0.984&-0.001&0.140&0.988\\
$\tau(1, V_{1})$&  0.007& 0.132&0.986&0.006&0.141&0.958&-0.002&0.130&0.986&-0.002&0.140&0.988\\[0.5cm]
\hline
$\delta(0.1,0,0)$&  -0.005&0.131 &0.970&-0.009&0.150&0.888&0.845&1.019&0.896&0.841&1.008&0.916\\
$\delta(0.2,0.1,0)$& -0.004& 0.106&0.982&-0.008&0.121&0.886&0.175&0.287&0.938&0.172&0.288&0.968\\
$\delta(0.3,0.2,0)$&  -0.003& 0.069&0.998&-0.005&0.081&0.884&0.020&0.099&1.000&0.019&0.110&0.996\\
$\delta(0.4,0.3,0)$&  -0.002& 0.045&1.000&-0.004&0.056&0.852&0.002&0.069&0.998&0.001&0.087&1.000\\
$\delta(0.5,0.4,0)$& -0.001&0.032&1.000&-0.003&0.042&0.792&0.001&0.055&1.000&0.001&0.077&1.000\\
$\delta(0.6,0.5,0)$& 0.000& 0.031&1.000&-0.002&0.042&0.794&0.002&0.053&1.000&0.000&0.069&0.998\\
$\delta(0.7,0.6,0)$&  0.000& 0.050&1.000&0.000&0.061&0.854&0.002&0.071&1.000&0.002&0.084&1.000\\
$\delta(0.8,0.7,0)$&  0.001& 0.058&0.996&0.001&0.070&0.866&-0.030&0.090&1.000&-0.027&0.098&0.996\\
$\delta(0.9,0.8,0)$&  0.002& 0.080&0.992&0.002&0.094&0.880&-0.134&0.191&0.966&-0.135&0.195&0.972\\
$\delta(1,0.9,0)$&  0.004& 0.151&0.966&0.006&0.174&0.888&-0.742&0.898&0.836&-0.748&0.901&0.860\\[0.5cm]
\hline
$\delta(0.1,0,1)$&  -0.005&0.131 &0.970&-0.009&0.150&0.888&0.845&1.019&0.896&0.841&1.008&0.916\\
$\delta(0.2,0.1,1)$& -0.004& 0.106&0.982&-0.008&0.121&0.886&0.175&0.287&0.938&0.172&0.288&0.968\\
$\delta(0.3,0.2,1)$&  -0.003& 0.069&0.998&-0.005&0.081&0.884&0.020&0.099&1.000&0.019&0.110&0.996\\
$\delta(0.4,0.3,1)$&  -0.002& 0.045&1.000&-0.004&0.056&0.852&0.002&0.069&0.998&0.001&0.087&1.000\\
$\delta(0.5,0.4,1)$& -0.001&0.032&1.000&-0.003&0.042&0.792&0.001&0.055&1.000&0.001&0.077&1.000\\
$\delta(0.6,0.5,1)$& 0.000& 0.031&1.000&-0.002&0.042&0.794&0.002&0.053&1.000&0.000&0.069&0.998\\
$\delta(0.7,0.6,1)$&  0.000& 0.050&1.000&0.000&0.061&0.854&0.002&0.071&1.000&0.002&0.084&1.000\\
$\delta(0.8,0.7,1)$&  0.001& 0.058&0.996&0.001&0.070&0.866&-0.030&0.090&1.000&-0.027&0.098&0.996\\
$\delta(0.9,0.8,1)$&  0.002& 0.080&0.992&0.002&0.094&0.880&-0.134&0.191&0.966&-0.135&0.195&0.972\\
$\delta(1,0.9,1)$&  0.004& 0.151&0.966&0.006&0.174&0.888&-0.742&0.898&0.836&-0.748&0.901&0.860\\[0.5cm]
\end{tabular}
\end{adjustwidth}
\end{center}
\caption{Performance of Bayesian GPS estimator for treatment effects $\tau(g, V_g)$, with $g\in\{0,0.1, \ldots, 0.9,1\}$,  and spillover effects  $\delta(g, g-0.1, z, V_g \cup V_{g-0.1})$, with $g\in\{0.1, \ldots, 0.9,1\}$ and $z \in \{0,1\}$,  under the scenario where the network is taken from the Add Health data and the outcome is drawn from a linear model with random effects.}
\label{tab:ah_lin_re}
\end{table}%

\begin{table}[ht!]
\small
\def\firstrowcolor{}
\def\secondrowcolor{}
\begin{center}
\begin{adjustwidth}{0cm}{0cm}
\begin{tabular}{c<{\hspace{-\tabcolsep}\,\,\,\,}|>{\hspace{-\tabcolsep}}c<{\hspace{-\tabcolsep}}>{\hspace{-\tabcolsep}}c<{\hspace{-\tabcolsep}}>{\hspace{-\tabcolsep}\,}c<{\hspace{-\tabcolsep}}>{\hspace{-\tabcolsep}\,}c<{\hspace{-\tabcolsep}}>{\hspace{-\tabcolsep}\,}c<{\hspace{-\tabcolsep}}>{\hspace{-\tabcolsep}\,}c<{\hspace{-\tabcolsep}}>{\hspace{-\tabcolsep}\,}c<{\hspace{-\tabcolsep}}>{\hspace{-\tabcolsep}\,}c<{\hspace{-\tabcolsep}}>{\hspace{-\tabcolsep}\,}c<{\hspace{-\tabcolsep}}>{\hspace{-\tabcolsep}\,}c<{\hspace{-\tabcolsep}}>{\hspace{-\tabcolsep}}c<{\hspace{-\tabcolsep}}>{\hspace{-\tabcolsep}}c}
 &\multicolumn{6}{c}{{Linear}}&\multicolumn{6}{c}{{Splines}}\\
          &\multicolumn{3}{c}{{RE}}&\multicolumn{3}{c}{{NO RE}}&\multicolumn{3}{c}{{RE}}&\multicolumn{3}{c}{{NO RE}}\\
\cmidrule(lr){2-4}\cmidrule(lr){5-7}\cmidrule(lr){8-10}\cmidrule(lr){11-13}
          &Bias& RMSE&Coverage&Bias& RMSE &Coverage&Bias& RMSE&Coverage&Bias& RMSE&Coverage\\
\hline
$\tau(0, V_0)$&  0.000&0.066&0.984&0.000&0.066&0.952&-0.006&0.066&0.980&-0.005&0.066&0.980\\
$\tau(0.1, V_{0.1})$&0.000  &0.066 &0.984&0.000&0.066&0.952&-0.006&0.066&0.980&-0.006&0.066&0.978\\
$\tau(0.2, V_{0.2})$&  0.000&0.066 &0.984&0.000&0.066&0.952&-0.006&0.066&0.974&-0.005&0.066&0.978\\
$\tau(0.3, V_{0.3})$&  0.000& 0.066&0.984&0.000&0.066&0.952&-0.006&0.066&0.978&-0.006&0.066&0.976\\
$\tau(0.4, V_{0.4})$&  0.000& 0.066&0.984&0.000&0.066&0.952&-0.006&0.066&0.976&-0.005&0.066&0.978\\
$\tau(0.5, V_{0.5})$&  0.000& 0.066&0.984&0.000&0.066&0.952&-0.006&0.066&0.982&-0.006&0.066&0.976\\
$\tau(0.6, V_{0.6})$&  0.000& 0.066&0.984&0.000&0.066&0.952&-0.006&0.066&0.972&-0.006&0.066&0.978\\
$\tau(0.7, V_{0.7})$&  0.000& 0.066&0.984&0.000&0.066&0.952&-0.006&0.066&0.980&-0.006&0.066&0.978\\
$\tau(0.8, V_{0.8})$&  0.000& 0.066&0.984&0.000&0.066&0.952&-0.006&0.066&0.982&-0.006&0.066&0.978\\
$\tau(0.9, V_{0.9})$&  0.000& 0.066&0.984&0.000&0.066&0.952&-0.006&0.066&0.982&-0.006&0.066&0.976\\
$\tau(1, V_{1})$&  0.000& 0.066&0.984&0.000&0.066&0.952&-0.006&0.066&0.982&-0.005&0.066&0.978\\[0.5cm]
\hline
$\delta(0.1,0,0)$&  0.000&0.057 &0.986&0.001&0.057&0.940&0.555&0.737&0.894&0.553&0.740&0.882\\
$\delta(0.2,0.1,0)$& 0.000& 0.046&0.992&0.001&0.046&0.944&0.095&0.169&0.950&0.095&0.170&0.942\\
$\delta(0.3,0.2,0)$&  0.000& 0.030&1.000&0.000&0.030&0.946&0.009&0.045&0.998&0.009&0.044&0.994\\
$\delta(0.4,0.3,0)$&  0.000& 0.020&1.000&0.000&0.020&0.942&0.000&0.028&1.000&0.001&0.029&1.000\\
$\delta(0.5,0.4,0)$& 0.000&0.014&1.000&0.000&0.013&0.944&0.000&0.024&1.000&0.000&0.024&1.000\\
$\delta(0.6,0.5,0)$& 0.000& 0.013&1.000&0.000&0.013&0.936&0.000&0.024&1.000&0.000&0.024&1.000\\
$\delta(0.7,0.6,0)$&  0.000& 0.021&1.000&0.000&0.021&0.946&0.002&0.031&1.000&0.002&0.031&1.000\\
$\delta(0.8,0.7,0)$&  0.000& 0.024&1.000&0.000&0.024&0.948&-0.016&0.047&0.996&-0.016&0.047&0.994\\
$\delta(0.9,0.8,0)$&  0.000& 0.034&1.000&0.000&0.034&0.948&-0.077&0.115&0.928&-0.079&0.115&0.926\\
$\delta(1,0.9,0)$&  0.000& 0.064&0.986&0.000&0.064&0.948&-0.444&0.578&0.852&-0.442&0.580&0.846\\[0.5cm]
\hline
$\delta(0.1,0,1)$&  0.000&0.057 &0.986&0.001&0.057&0.940&0.555&0.737&0.894&0.553&0.740&0.882\\
$\delta(0.2,0.1,1)$& 0.000& 0.046&0.992&0.001&0.046&0.944&0.095&0.169&0.950&0.095&0.170&0.942\\
$\delta(0.3,0.2,1)$&  0.000& 0.030&1.000&0.000&0.030&0.946&0.009&0.045&0.998&0.009&0.044&0.994\\
$\delta(0.4,0.3,1)$&  0.000& 0.020&1.000&0.000&0.020&0.942&0.000&0.028&1.000&0.001&0.029&1.000\\
$\delta(0.5,0.4,1)$& 0.000&0.014&1.000&0.000&0.013&0.944&0.000&0.024&1.000&0.000&0.024&1.000\\
$\delta(0.6,0.5,1)$& 0.000& 0.013&1.000&0.000&0.013&0.936&0.000&0.024&1.000&0.000&0.024&1.000\\
$\delta(0.7,0.6,1)$&  0.000& 0.021&1.000&0.000&0.021&0.946&0.002&0.031&1.000&0.002&0.031&1.000\\
$\delta(0.8,0.7,1)$&  0.000& 0.024&1.000&0.000&0.024&0.948&-0.016&0.047&0.996&-0.016&0.047&0.994\\
$\delta(0.9,0.8,1)$&  0.000& 0.034&1.000&0.000&0.034&0.948&-0.077&0.115&0.928&-0.079&0.115&0.926\\
$\delta(1,0.9,1)$&  0.000& 0.064&0.986&0.000&0.064&0.948&-0.444&0.578&0.852&-0.442&0.580&0.846\\[0.5cm]
\end{tabular}
\end{adjustwidth}
\end{center}
\caption{Performance of Bayesian GPS estimator for treatment effects $\tau(g, V_g)$, with $g\in\{0,0.1, \ldots, 0.9,1\}$,  and spillover effects  $\delta(g, g-0.1, z, V_g \cup V_{g-0.1})$, with $g\in\{0.1, \ldots, 0.9,1\}$ and $z \in \{0,1\}$,  under the scenario where the network is taken from the Add Health data and the outcome is drawn from a linear model without random effects.}
\label{tab:ah_lin_nore}
\end{table}%

\begin{table}[ht!]
\small
\def\firstrowcolor{}
\def\secondrowcolor{}
\begin{center}
\begin{adjustwidth}{0cm}{0cm}
\begin{tabular}{c<{\hspace{-\tabcolsep}\,\,\,\,}|>{\hspace{-\tabcolsep}}c<{\hspace{-\tabcolsep}}>{\hspace{-\tabcolsep}}c<{\hspace{-\tabcolsep}}>{\hspace{-\tabcolsep}\,}c<{\hspace{-\tabcolsep}}>{\hspace{-\tabcolsep}\,}c<{\hspace{-\tabcolsep}}>{\hspace{-\tabcolsep}\,}c<{\hspace{-\tabcolsep}}>{\hspace{-\tabcolsep}\,}c<{\hspace{-\tabcolsep}}>{\hspace{-\tabcolsep}\,}c<{\hspace{-\tabcolsep}}>{\hspace{-\tabcolsep}\,}c<{\hspace{-\tabcolsep}}>{\hspace{-\tabcolsep}\,}c<{\hspace{-\tabcolsep}}>{\hspace{-\tabcolsep}\,}c<{\hspace{-\tabcolsep}}>{\hspace{-\tabcolsep}}c<{\hspace{-\tabcolsep}}>{\hspace{-\tabcolsep}}c}
 &\multicolumn{6}{c}{{Linear}}&\multicolumn{6}{c}{{Splines}}\\
          &\multicolumn{3}{c}{{RE}}&\multicolumn{3}{c}{{NO RE}}&\multicolumn{3}{c}{{RE}}&\multicolumn{3}{c}{{NO RE}}\\
\cmidrule(lr){2-4}\cmidrule(lr){5-7}\cmidrule(lr){8-10}\cmidrule(lr){11-13}
          &Bias& RMSE&Coverage&Bias& RMSE &Coverage&Bias& RMSE&Coverage&Bias& RMSE&Coverage\\
\hline
$\tau(0, V_0)$&  -0.049&0.202 &0.962&0.001&0.204&0.938&-0.012&0.138&0.980&-0.009&0.145&0.984\\
$\tau(0.1, V_{0.1})$&-0.048  &0.202 &0.954&0.001&0.204&0.938&-0.012&0.138&0.982&-0.009&0.145&0.982\\
$\tau(0.2, V_{0.2})$& -0.049 &0.202 &0.962&0.001&0.204&0.938&-0.012&0.138&0.980&-0.009&0.145&0.984\\
$\tau(0.3, V_{0.3})$& -0.049 & 0.202&0.964&0.001&0.204&0.938&-0.012&0.138&0.984&-0.009&0.145&0.986\\
$\tau(0.4, V_{0.4})$& -0.049 & 0.202&0.956&0.001&0.204&0.938&-0.012&0.138&0.982&-0.009&0.145&0.986\\
$\tau(0.5, V_{0.5})$& -0.049 & 0.202&0.952&0.001&0.204&0.938&-0.012&0.138&0.988&-0.009&0.145&0.982\\
$\tau(0.6, V_{0.6})$& -0.049 & 0.202&0.960&0.001&0.204&0.938&-0.012&0.138&0.982&-0.009&0.145&0.982\\
$\tau(0.7, V_{0.7})$& -0.049 & 0.202&0.962&0.001&0.204&0.938&-0.012&0.138&0.982&-0.009&0.145&0.980\\
$\tau(0.8, V_{0.8})$& -0.049 & 0.202&0.958&0.001&0.204&0.938&-0.012&0.138&0.984&-0.009&0.145&0.974\\
$\tau(0.9, V_{0.9})$& -0.049 & 0.202&0.960&0.001&0.204&0.938&-0.012&0.138&0.978&-0.009&0.145&0.980\\
$\tau(1, V_{1})$& -0.048 & 0.202&0.960&0.001&0.204&0.938&-0.012&0.138&0.978&-0.009&0.145&0.984\\[0.5cm]
\hline
$\delta(0.1,0,0)$&  1.459&1.471&0.000&1.482&1.495&0.000&0.288&0.471&0.996&0.252&0.491&0.958\\
$\delta(0.2,0.1,0)$&  1.443&1.451 &0.000&1.466&1.474&0.000&0.055&0.276&0.982&0.050&0.282&0.984\\
$\delta(0.3,0.2,0)$&  0.976&0.981 &0.000&1.000&1.005&0.000&0.035&0.227&0.974&0.031&0.241&0.966\\
$\delta(0.4,0.3,0)$&  -0.285& 0.291&0.348&-0.260&0.269&0.030&0.026&0.197&0.976&0.034&0.206&0.980\\
$\delta(0.5,0.4,0)$&  -2.173&2.173 &0.000&-2.148&2.148&0.000&-0.039&0.178&0.964&-0.040&0.199&0.966\\
$\delta(0.6,0.5,0)$&  -2.259&2.259 &0.000&-2.233&2.234&0.000&-0.135&0.221&0.932&-0.140&0.237&0.934\\
$\delta(0.7,0.6,0)$&  0.150& 0.166&0.922&0.176&0.192&0.212&0.155&0.251&0.912&0.168&0.268&0.910\\
$\delta(0.8,0.7,0)$&1.409 &1.412 &0.000&1.435&1.438&0.000&-0.015&0.211&0.988&-0.006&0.232&0.966\\
$\delta(0.9,0.8,0)$&  1.612& 1.617&0.000&1.638&1.644&0.000&-0.049&0.222&0.992&-0.053&0.236&0.984\\
$\delta(1,0.9,0)$&1.788 & 1.806&0.000&1.817&1.834&0.000&-0.238&0.437&0.972&-0.236&0.464&0.958\\[0.5cm]
\hline
$\delta(0.1,0,1)$&  1.459&1.471&0.000&1.482&1.495&0.000&0.288&0.471&0.996&0.252&0.491&0.958\\
$\delta(0.2,0.1,1)$&  1.443&1.451 &0.000&1.466&1.474&0.000&0.055&0.276&0.982&0.050&0.282&0.984\\
$\delta(0.3,0.2,1)$&  0.976&0.981 &0.000&1.000&1.005&0.000&0.035&0.227&0.974&0.031&0.241&0.966\\
$\delta(0.4,0.3,1)$&  -0.285& 0.291&0.348&-0.260&0.269&0.030&0.026&0.197&0.976&0.034&0.206&0.980\\
$\delta(0.5,0.4,1)$&  -2.173&2.173 &0.000&-2.148&2.148&0.000&-0.039&0.178&0.964&-0.040&0.199&0.966\\
$\delta(0.6,0.5,1)$&  -2.259&2.259 &0.000&-2.233&2.234&0.000&-0.135&0.221&0.932&-0.140&0.237&0.934\\
$\delta(0.7,0.6,1)$&  0.150& 0.166&0.922&0.176&0.192&0.212&0.155&0.251&0.912&0.168&0.268&0.910\\
$\delta(0.8,0.7,1)$&1.409 &1.412 &0.000&1.435&1.438&0.000&-0.015&0.211&0.988&-0.006&0.232&0.966\\
$\delta(0.9,0.8,1)$&  1.612& 1.617&0.000&1.638&1.644&0.000&-0.049&0.222&0.992&-0.053&0.236&0.984\\
$\delta(1,0.9,1)$&1.788 & 1.806&0.000&1.817&1.834&0.000&-0.238&0.437&0.972&-0.236&0.464&0.958\\[0.5cm]
\end{tabular}
\end{adjustwidth}
\end{center}
\caption{Performance of Bayesian GPS estimator for treatment effects $\tau(g, V_g)$, with $g\in\{0,0.1, \ldots, 0.9,1\}$,  and spillover effects  $\delta(g, g-0.1, z, V_g \cup V_{g-0.1})$, with $g\in\{0.1, \ldots, 0.9,1\}$ and $z \in \{0,1\}$,  under the scenario where the network is taken from the Add Health data and the outcome is drawn from a non-linear model with random effects.}
\label{tab:ah_nolin_re}
\end{table}%

\begin{figure}[t!]
        \begin{adjustwidth}{-0.8cm}{0cm}
    \begin{subfigure}[t]{0.45\textwidth}
        \centering
        \includegraphics[width=1.25\textwidth]{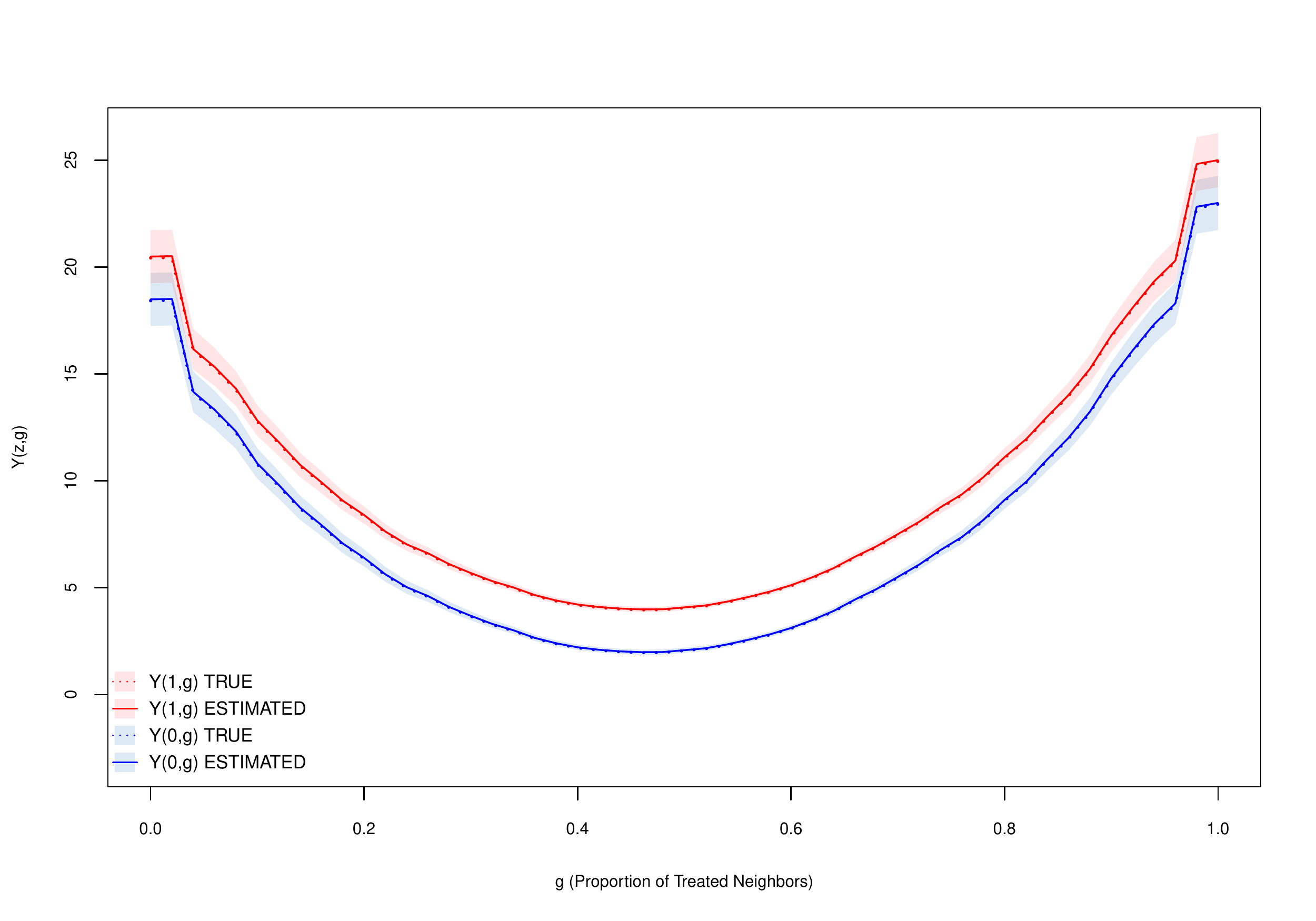}
        \caption{Linear DGP - Linear Model \phantom{0.3cm}}
    \end{subfigure}%
        ~~~~~~~~~~
    \begin{subfigure}[t]{0.45\textwidth}
        \centering
        \includegraphics[width=1.25\textwidth]{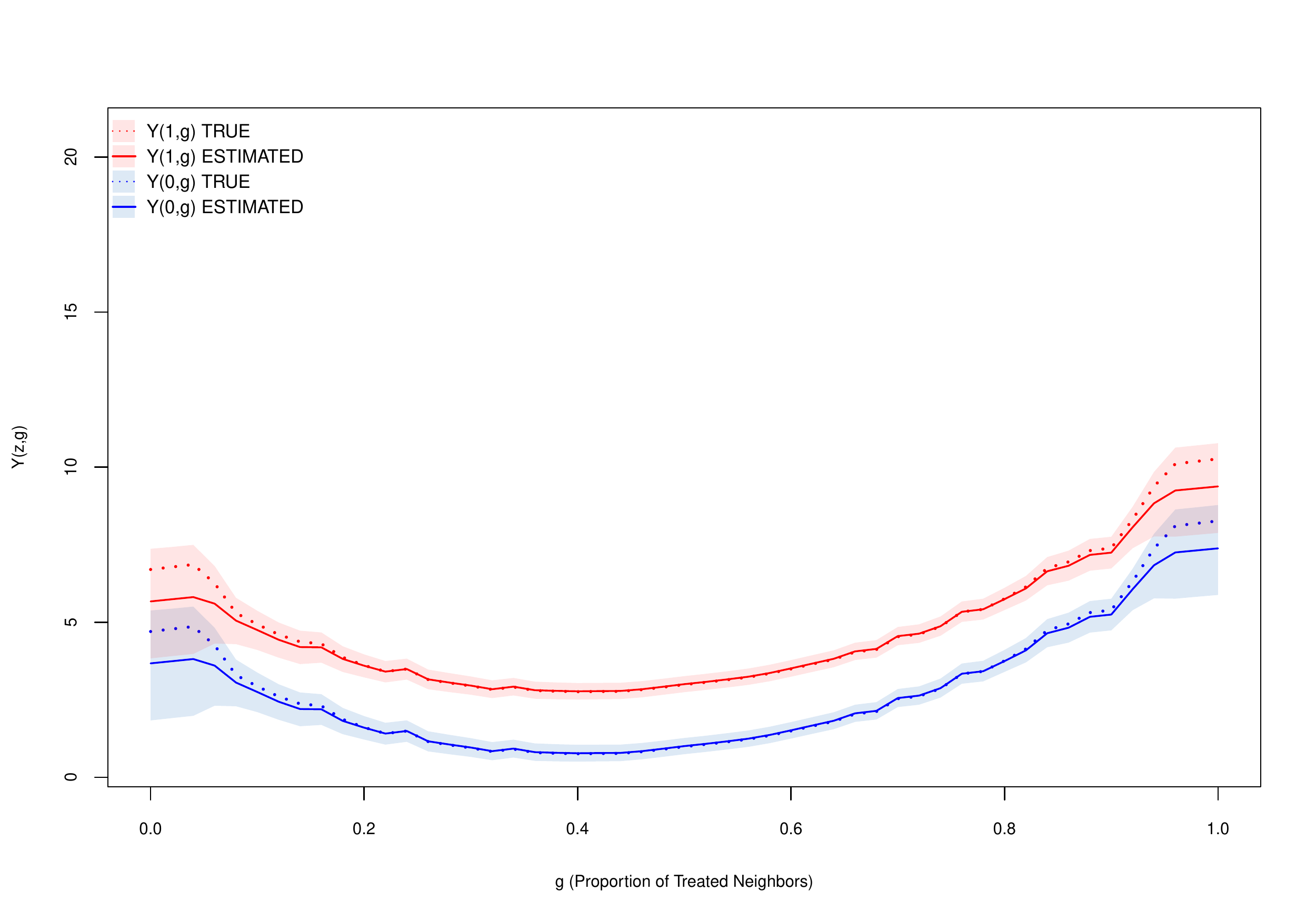}
        \caption{Linear DGP - Splines \phantom{0.3cm}}
    \end{subfigure}
     \end{adjustwidth}
        \begin{adjustwidth}{-0.8cm}{0cm}

   \begin{subfigure}[t]{0.45\textwidth}
        \centering
        \includegraphics[width=1.25\textwidth]{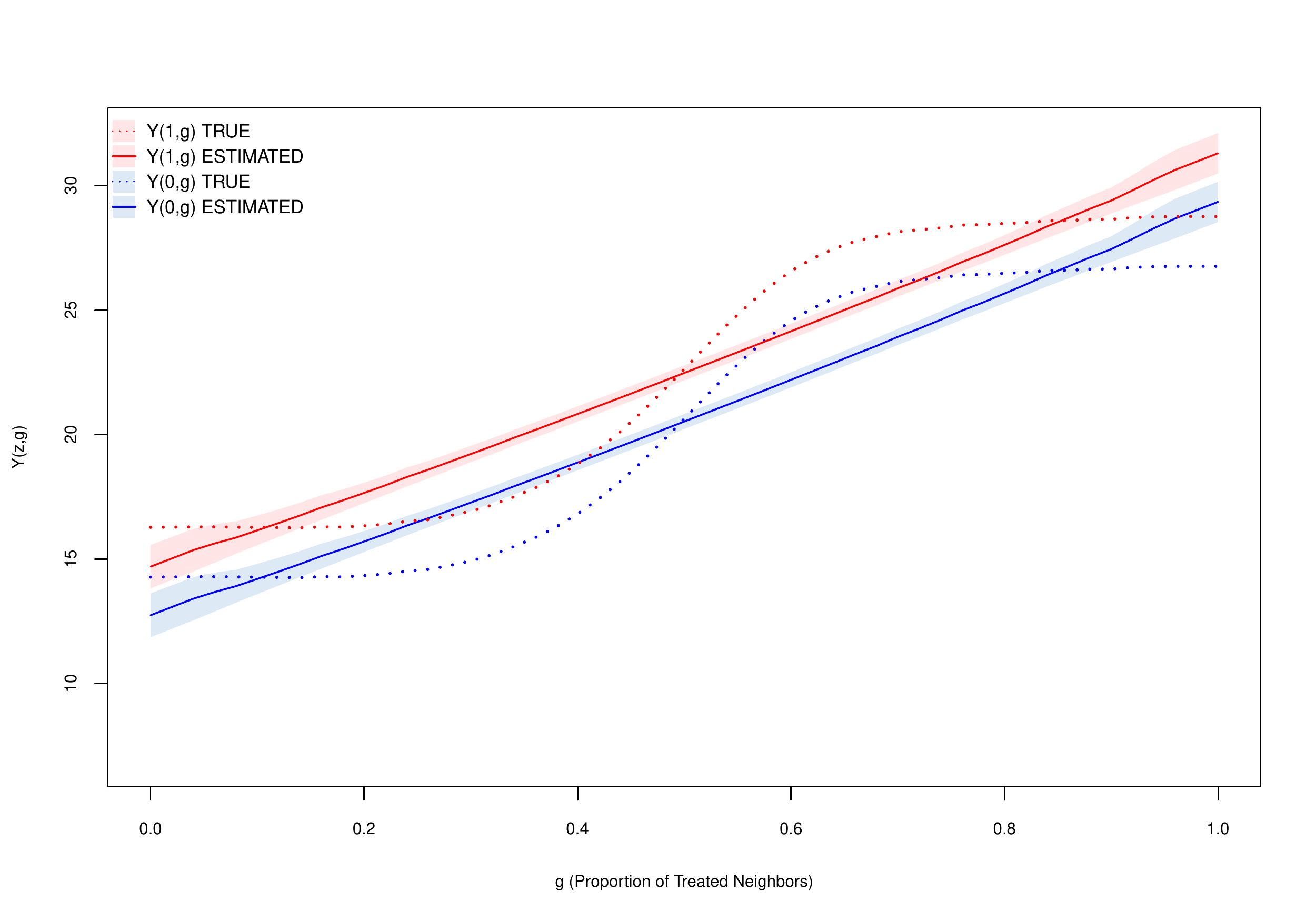}
        \caption{Non-Linear DGP - Linear Model }
    \end{subfigure}%
        ~~~~~~~~~~
    \begin{subfigure}[t]{0.45\textwidth}
        \centering
        \includegraphics[width=1.25\textwidth]{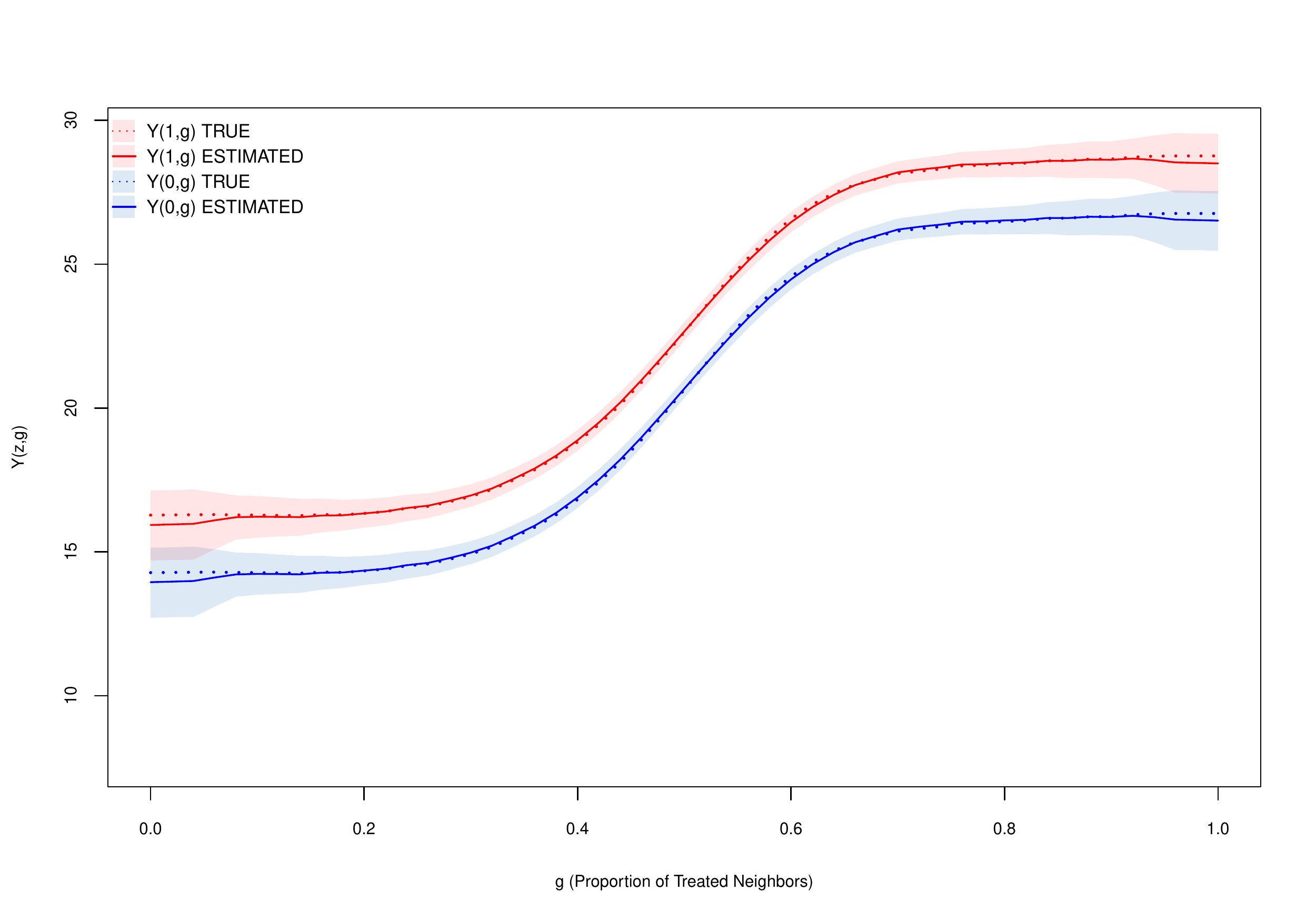}
        \caption{Non-Linear DGP - Splines}
    \end{subfigure}
     \end{adjustwidth}
  \caption{Estimated and true ()ADRF $\mu(z,g; V_g)$ under the scenario where the network is taken from the Add Health data and the outcome is drawn from a linear model (top) or a non-linear model (bottom) with random effects. Average posterior mean (solid line) and 95 \% credible intervals of $\mu(0,g; V_g)$ (blue) and $\mu(1,g; V_g)$ (red) of linear (left) and splines-based (right) estimator with random effects are represented with the true ADRFs (dotted line). }
\label{fig: ah}
\end{figure}

\begin{table}[ht!]
\small
\def\firstrowcolor{}
\def\secondrowcolor{}
\begin{center}
\begin{adjustwidth}{0cm}{0cm}
\begin{tabular}{c<{\hspace{-\tabcolsep}\,\,\,\,}|>{\hspace{-\tabcolsep}}c<{\hspace{-\tabcolsep}}>{\hspace{-\tabcolsep}}c<{\hspace{-\tabcolsep}}>{\hspace{-\tabcolsep}\,}c<{\hspace{-\tabcolsep}}>{\hspace{-\tabcolsep}\,}c<{\hspace{-\tabcolsep}}>{\hspace{-\tabcolsep}\,}c<{\hspace{-\tabcolsep}}>{\hspace{-\tabcolsep}\,}c<{\hspace{-\tabcolsep}}>{\hspace{-\tabcolsep}\,}c<{\hspace{-\tabcolsep}}>{\hspace{-\tabcolsep}\,}c<{\hspace{-\tabcolsep}}>{\hspace{-\tabcolsep}\,}c<{\hspace{-\tabcolsep}}>{\hspace{-\tabcolsep}\,}c<{\hspace{-\tabcolsep}}>{\hspace{-\tabcolsep}}c<{\hspace{-\tabcolsep}}>{\hspace{-\tabcolsep}}c}
 &\multicolumn{6}{c}{{Linear}}&\multicolumn{6}{c}{{Splines}}\\
          &\multicolumn{3}{c}{{RE}}&\multicolumn{3}{c}{{NO RE}}&\multicolumn{3}{c}{{RE}}&\multicolumn{3}{c}{{NO RE}}\\
\cmidrule(lr){2-4}\cmidrule(lr){5-7}\cmidrule(lr){8-10}\cmidrule(lr){11-13}
          &Bias& RMSE&Coverage&Bias& RMSE &Coverage&Bias& RMSE&Coverage&Bias& RMSE&Coverage\\
\hline
$\tau(0, V_0)$&  -0.026&0.148 &0.970&-0.001&0.148&0.958&-0.008&0.067&0.972&-0.008&0.067&0.974\\
$\tau(0.1, V_{0.1})$&-0.026  &0.149 &0.968&-0.001&0.148&0.958&-0.008&0.067&0.980&-0.008&0.067&0.974\\
$\tau(0.2, V_{0.2})$& -0.026 &0.149 &0.966&-0.001&0.148&0.958&-0.008&0.067&0.976&-0.008&0.067&0.972\\
$\tau(0.3, V_{0.3})$& -0.026 & 0.148&0.974&-0.001&0.148&0.958&-0.008&0.067&0.972&-0.008&0.067&0.980\\
$\tau(0.4, V_{0.4})$& -0.026 & 0.149&0.968&-0.001&0.148&0.958&-0.008&0.067&0.978&-0.008&0.067&0.978\\
$\tau(0.5, V_{0.5})$& -0.026 & 0.148&0.972&-0.001&0.148&0.958&-0.008&0.067&0.974&-0.008&0.067&0.972\\
$\tau(0.6, V_{0.6})$& -0.026 & 0.149&0.972&-0.001&0.148&0.958&-0.008&0.067&0.980&-0.008&0.067&0.980\\
$\tau(0.7, V_{0.7})$& -0.026 & 0.149&0.970&-0.001&0.148&0.958&-0.008&0.067&0.972&-0.008&0.067&0.974\\
$\tau(0.8, V_{0.8})$& -0.026 & 0.149&0.972&-0.001&0.148&0.958&-0.008&0.067&0.972&-0.008&0.067&0.98'\\
$\tau(0.9, V_{0.9})$& -0.026 & 0.149&0.968&-0.001&0.148&0.958&-0.008&0.067&0.972&-0.008&0.067&0.980\\
$\tau(1, V_{1})$& -0.026 & 0.149&0.964&-0.001&0.148&0.958&-0.008&0.067&0.970&-0.008&0.067&0.978\\[0.5cm]
\hline
$\delta(0.1,0,0)$&  1.465&1.473&0.000&1.477&1.485&0.000&0.359&0.436&0.874&0.359&0.435&0.858\\
$\delta(0.2,0.1,0)$&  1.450&1.455 &0.000&1.462&1.467&0.000&0.050&0.165&0.978&0.049&0.164&0.982\\
$\delta(0.3,0.2,0)$&  0.986&0.988 &0.000&0.998&1.001&0.000&0.027&0.128&0.968&0.028&0.129&0.968\\
$\delta(0.4,0.3,0)$&  -0.274& 0.278&0.068&-0.261&0.265&0.030&-0.017&0.113&0.964&-0.017&0.113&0.970\\
$\delta(0.5,0.4,0)$&  -2.161&2.161&0.000&-2.148&2.148&0.000&0.017&0.088&0.982&0.016&0.088&0.980\\
$\delta(0.6,0.5,0)$&  -2.246&2.247 &0.000&-2.233&2.233&0.000&-0.076&0.116&0.924&-0.075&0.116&0.920\\
$\delta(0.7,0.6,0)$&  0.164& 0.175&0.730&0.177&0.187&0.212&0.081&0.135&0.912&0.081&0.135&0.912\\
$\delta(0.8,0.7,0)$&1.423 &1.426 &0.000&1.438&1.440&0.000&-0.043&0.130&0.950&-0.042&0.130&0.964\\
$\delta(0.9,0.8,0)$&  1.628& 1.632&0.000&1.642&1.646&0.000&-0.014&0.145&0.976&-0.014&0.145&0.972\\
$\delta(1,0.9,0)$&1.809 & 1.822&0.000&1.824&1.837&0.000&-0.283&0.372&0.882&-0.283&0.373&0.868\\[0.5cm]
\hline
$\delta(0.1,0,1)$&  1.465&1.473&0.000&1.477&1.485&0.000&0.359&0.436&0.874&0.359&0.435&0.858\\
$\delta(0.2,0.1,1)$&  1.450&1.455 &0.000&1.462&1.467&0.000&0.050&0.165&0.978&0.049&0.164&0.980\\
$\delta(0.3,0.2,1)$&  0.986&0.988 &0.000&0.998&1.001&0.000&0.027&0.128&0.968&0.028&0.129&0.974\\
$\delta(0.4,0.3,1)$&  -0.274& 0.278&0.068&-0.261&0.265&0.030&-0.017&0.113&0.964&-0.017&0.113&0.968\\
$\delta(0.5,0.4,1)$&  -2.161&2.161&0.000&-2.148&2.148&0.000&0.017&0.088&0.982&0.016&0.088&0.976\\
$\delta(0.6,0.5,1)$&  -2.246&2.247 &0.000&-2.233&2.233&0.000&-0.076&0.116&0.924&-0.075&0.116&0.920\\
$\delta(0.7,0.6,1)$&  0.164& 0.175&0.730&0.177&0.187&0.212&0.081&0.135&0.912&0.081&0.135&0.912\\
$\delta(0.8,0.7,1)$&1.423 &1.426 &0.000&1.438&1.440&0.000&-0.043&0.130&0.950&-0.042&0.130&0.958\\
$\delta(0.9,0.8,1)$&  1.628& 1.632&0.000&1.642&1.646&0.000&-0.014&0.145&0.976&-0.014&0.145&0.972\\
$\delta(1,0.9,1)$&1.809 & 1.822&0.000&1.824&1.837&0.000&-0.283&0.372&0.882&-0.283&0.373&0.872\\
\end{tabular}
\end{adjustwidth}
\end{center}
\caption{Performance of Bayesian GPS estimator for treatment effects $\tau(g, V_g)$, with $g\in\{0,0.1, \ldots, 0.9,1\}$,  and spillover effects  $\delta(g, g-0.1, z, V_g \cup V_{g-0.1})$, with $g\in\{0.1, \ldots, 0.9,1\}$ and $z \in \{0,1\}$,  under the scenario where the network is taken from the Add Health data and the outcome is drawn from a non-linear model without random effects.}
\label{tab:ah_nolin_nore}
\end{table}%

\section{Concluding Remarks}
\label{sec:conc}
We have discussed definition, identification and estimation issues
of causal effects of interventions in contexts with interconnected
and interfering units. We have introduced assumptions,
neighborhood interference, on the way and the extent to which
spillover effects occur along the observed network are required.
For observational studies, where the treatment assignment is not
under the control of the investigator, we have introduced an
unconfoundedness of the individual and neighborhood treatment,
which rules out the presence of unmeasured confounding variables,
including those driving homophily. Under these assumptions we have
developed a new covariate-adjustment estimator for treatment and
spillover effects in observational studies on networks. Estimation
is based on a generalized propensity score that balances
individual and neighborhood covariates across units under
different levels of individual treatment and of exposure to
neighbors' treatment.
Adjustment for propensity score is performed using a penalized
spline regression. Inference capitalizes on a three-step Bayesian
procedure which allows taking into account the uncertainty in the
propensity score estimation and avoiding model feedback. Finally,
correlation of interacting units is taken into account using a
community detection algorithm and incorporating random effects in
the outcome model. We conducted a simulation study  to assess,
under different data generating scenarios, the  performances in
terms of bias, MSE, and coverage of our proposed estimator and its
variants -- fully model-based versus matching adjustment for
individual propensity score, linear versus non-linear outcome
model, with versus without community random intercept. The
simulation study has shown promising performance of the proposed
methods. The splines-based estimator is superior to the linear
estimator when the outcome does not linearly depend on the two
treatments and the two propensity scores. The drawback of the use
of splines is the large performance deterioration  in regions
where there are few observations. This, however, is not only
specific to use of splines but it is a common problem due to lack
of overlap. The ability to estimate the entire average
dose-response function and the spillover effects at different
levels of the neighborhood treatment clearly depends on both the
sparsity of the graph and the distribution of the individual
treatment and in turn of the neighborhood treatment. A sparse
graph with most units with low degree would result in a
distribution of the neighborhood treatment concentrated around few
values and the dose-response function would be poorly estimated
for the other values.  On the other hand a dense graph with many
units with a very large degree would result in a very large range
of the neighborhood treatment, with few observations for every
value. In this case a discretization of the neighborhood treatment
along with an approximation of the dose-response function might be
required, with obvious consequences in terms of causal effects'
interpretation.

A possible correlation between neighbors has been a major concern
in the field of causal inference on networks. The inclusion of a
random intercept defined on non-overlapping communities,
identified  using a community detection algorithm, allows us to
capture an outcome correlation due to latent characteristics. The
ability to estimate and take into account the outcome correlation
with a good approximation depends on the graph structure and on
the performances of the community detection algorithm and its
effectiveness in identifying  communities such that connections
between the nodes are denser than connections with the rest of the
network. In the simulation study we have defined random intercepts
on communities originating directly from the generating model or
pre-specified in the Add Health data. We have then used the same
known communities in the estimation stage. In this way, we were
able to assess the performances of our proposed estimator besides
the community identification. The use of community detection
algorithms would affect the performance of our estimator depending
on their specific performance. Moreover, the inclusion of
community random intercepts would succeed in taking into account
the presence of outcome correlation to the extent to which the
identified communities match the correlation structure with good
approximation.

\newpage




\clearpage
\newpage
\appendix
\section{Data Generating Models}
\label{app: DGP}
\subsection*{Network Model: Stochastic Block Model}
\label{app: sbm}
For each unit $i=1, \ldots, 1000$ we first generated a continuous covariate $X_{i1}$ and a binary covariate $X_{i2}$ from the following distributions:
\[X_{i1}\sim \mathrm{Gamma}(0.5,1) \quad X_{i2}\sim \textrm{Ber}(0.5)\]
We then generated the network from a stochastic block model, where the probability of link only depends on the community membership.
Let $\mathbf{M}$ be the stochastic block matrix, where each element $M_{lk}$ is the probability of link between a unit of community i and a unit of community k. In a stochastic block model the probability of link between two units i and j with community memberships $C_i$ and $C_j$ is
\[Pr(A_{ij}=1|  C_i, C_j)=M_{C_i C_j}\]
In our simulation study we created 100 communities of 10 units each and we set the diagonal elements of $\mathbf{M}$ to 0.08 and the off-diagonal elements to 0.02.

\subsection*{Network Model: Latent Cluster Model}
In scenarios with the network drawn from a latent cluster model, we first sampled covariates $X_{i1}$ and $X_{i2}$ as in the scenarios based on the stochastic block model. We created again 100 communities of 10 units each, with $C_i$ being the community membership indicator. The adjacency matrix is then generated using the following model:
\[Pr(A_{ij}=1 | \vX_{i}, \vX_{j}, C_i, C_j)=\textrm{exp}\big(\beta_0+ \bm{\beta}_X^T|\vX_{i}- \vX_{j}| +  \beta_C|C_{i}- C_{j}|\big)\]
where $\beta_0=-4.6$, $\bm{\beta}_X=[0.05, 0.005]^T$, and $\beta_C=0.18$.
The first term gives rise to homophily along the observed characteristics, that is, nodes with similar characteristics are more likely to be connected. The second term creates a cluster structure in the network, which can be interpreted as some homophily due to latent characteristics shared by group of units.


\subsection*{Network Model: Add Health Data}
The Add Health Study is a large survey of adolescents attending schools in the United States who were listed on 7th through 12th grade enrollment rosters during the 1994 -1995
academic year. The Wave 1 survey collected data on socio-demographic characteristics of the respondents, education and occupation of parents, household structure, risk behaviors, health status, and friendships. As part of the in-home survey, students selected their five best male and five best female friends from a complete school roster.
We used data on this friendship network together with 3 characteristics: sex, grade, race.  We selected a sample of 1000 students in 8 schools. We used the directed friendship graph, meaning that only the students who were nominated by student i were considered his friends.  Units with degree equal to 0 were not excluded a priori.  They were in fact considered as alters of a student i who nominated them and their treatment and characteristics were used to compute the neighborhood treatment and neighborhood characteristics of student i.

In the simulation study communities are defined as clusters of students in the same school and same grade.


\begin{table}[ht!]
\begin{center}
\begin{tabular}{c|cccc}
 &\multicolumn{2}{c}{{Individual Characteristics}}&\multicolumn{2}{c}{{Neighborhood Characteristics}}\\
\cmidrule(lr){2-3}\cmidrule(lr){4-5}
          &Mean& SD&Mean& SD\\[0.1cm]
  \cmidrule(lr){1-2} \cmidrule(lr){2-3}\cmidrule(lr){4-5}
Sex&0.49&0.50&0.46&0.30\\[0.3cm]
Race&0.71&0.46&0.65&0.42\\[0.3cm]
Grade&9.46&1.68&8.60&3.13\\[0.5cm]
Degree&5.07&2.92&$\cdot$&$\cdot$\\
\end{tabular}
\end{center}
\caption{Summary Statistics of the Add Health dataset.}
\label{tab:}
\end{table}%

 \subsection*{Individual Treatment Model}
In the scenarios based on the stochastic block model and the latent cluster model, the individual treatment $Z_i$ was drawn for each unit using the following individual propensity score:
\begin{equation}
\label{eq: ips1}
\textrm{logit}(Pr(Z_i=1| \vX_i))=2.6X_{1i}-2.2X_{2i}
\end{equation}

In the simulations based on the Add Health data we used all three characteristics and generated the individual treatment based on the following propensity score:
\begin{equation}
\label{eq: ips2}
\textrm{logit}(Pr(Z_i=1| \vX_i))=0.7\textrm{sex}_{i}-0.11\textrm{grade}_{i}+\textrm{race}_i
\end{equation}

\begin{figure}[htbp]
\begin{center}
\includegraphics[width=\textwidth]{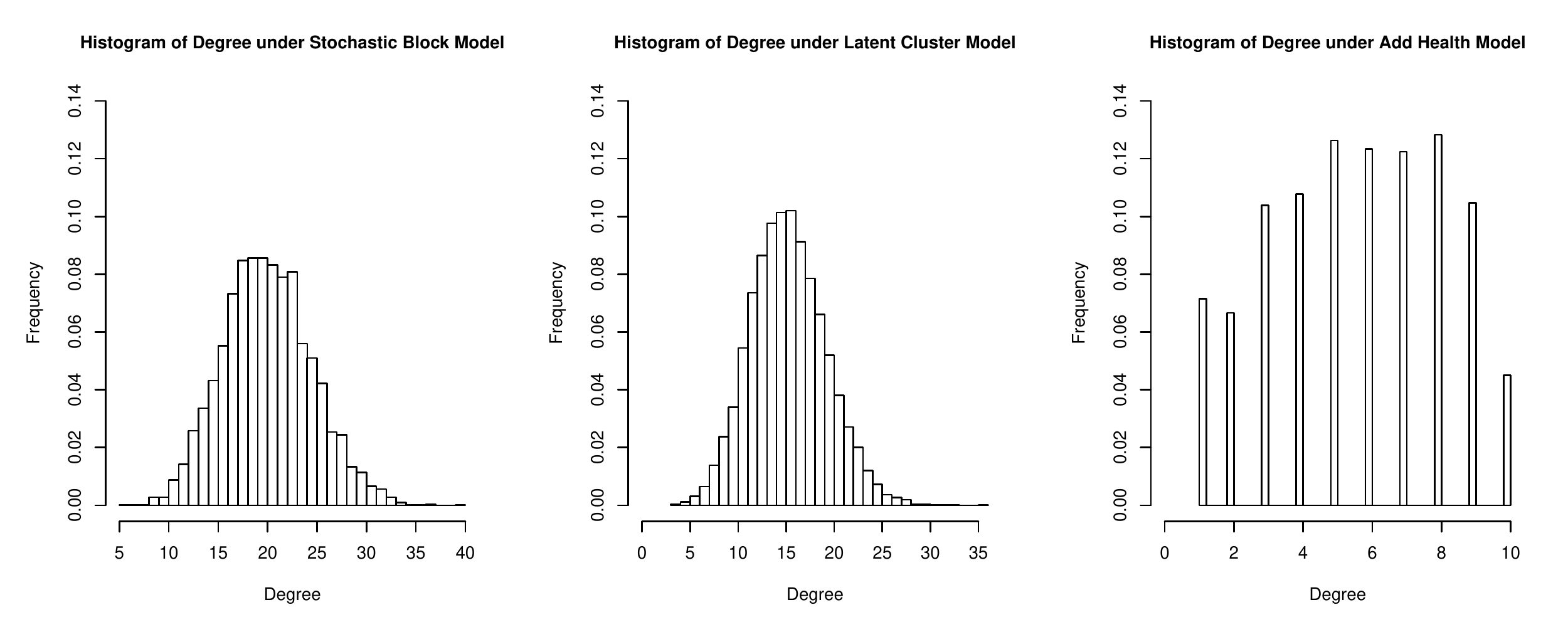}
\caption{Histograms of the degree $N_i$ in the 3 network types.}
\label{fig: HistDegree}
\end{center}
\end{figure}

\begin{figure}[htbp]
\begin{center}
\includegraphics[width=\textwidth]{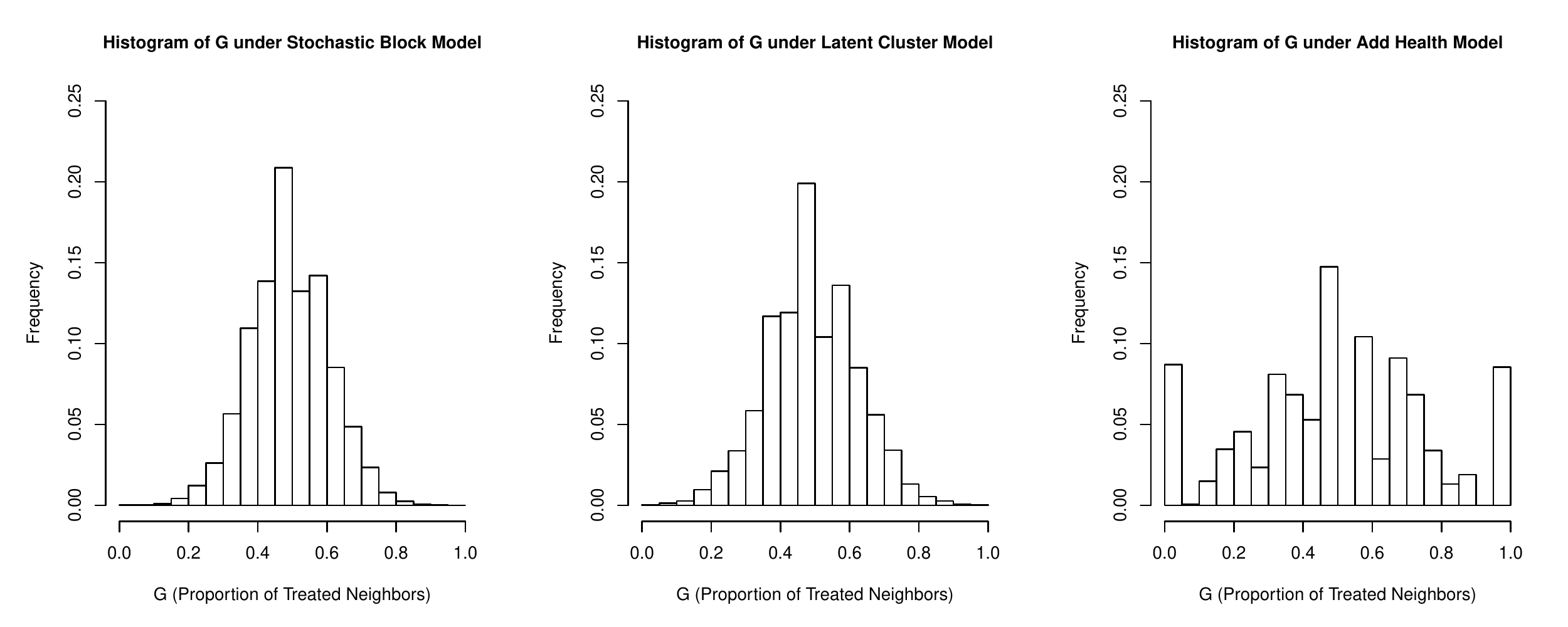}
\caption{Histograms of the neighborhood treatment $G_i$ in the 3 network types.}
\label{fig: HistG}
\end{center}
\end{figure}

The neighborhood treatment $G_i$ was then computed as  the proportion of treated neighbors.
The propensity scores in Equations \eqref{eq: ips1} \eqref{eq: ips2} applied on the 3 network structures, with degree distribution represented in Figure \ref{fig: HistDegree},  has given rise to the distributions of the neighborhood treatment $G_i$ shown in Figure \ref{fig: HistG}.

The neighborhood propensity score model was estimated using a binomial regression and the neighborhood propensity score $\Lambda_i=\lambda(G_i; Z_i; \vX_i^g)$ was then predicted for each unit corresponding to the actual $G_i$ affecting each unit, the individual treatment $Z_i$, the individual and neighborhood characteristics.

\subsection*{Potential Outcome Models}
In scenarios where the outcome was drawn from a linear model, we used the following distribution:
 \begin{equation}
\label{eq:  linY}
  Y_i \sim \mathcal{N}(-3+2Z_i +4G_i- \Phi_i-2\textrm{log}(\Lambda_i) +u_j , 1)
\end{equation}
with $u_j\sim \mathcal{N}(0, \Sigma_u)$. In scenarios with random effects $\Sigma_u=2I$, whereas is scenarios without random effects $\Sigma_u=0I$.

In scenarios where the outcome was drawn from a non-linear model, we instead used the following distribution:

 \begin{equation}
\label{eq:  nolinY}
  Y_i \sim \mathcal{N}(3+2Z_i +25\,\textrm{sigmoid}\,(10\,\textrm{exp}\,(-\frac{(G_i-1)^2}{0.12}))-2\Phi_i-2.5\Lambda_i +u_j , 1)
\end{equation}
where $\textrm{sigmoid}(x)=\frac{1}{1+\textrm{e}^{-x}}$ and again $u_j\sim \mathcal{N}(0, \Sigma_u)$ with $\Sigma_u=2I$ or $\Sigma_u=0I$.



\end{document}